\documentclass[12pt]{article}
\usepackage{epsfig}
\usepackage{amssymb}
\usepackage{amsmath}
\usepackage{amsfonts}
\usepackage{graphicx}
\usepackage{mathrsfs}
\usepackage[dvips]{color}
\usepackage{multirow}
\usepackage{graphicx}
\usepackage{subfig}
\usepackage{float}

\newcommand{\R}{\mathbb{R}}
\newcommand{\C}{\mathbb{C}}

\newcommand{\ff}{\mathfrak{f}}

\newcommand{\fK}{\mathfrak{K}}

\newcommand{\bba}{\boldsymbol{a}}

\newcommand{\bc}{\boldsymbol{c}}

\newcommand{\bg}{{\boldsymbol{g}}}

\newcommand{\bk}{\boldsymbol{k}}

\newcommand{\bp}{\boldsymbol{p}}
\newcommand{\bq}{\boldsymbol{q}}

\newcommand{\bx}{\boldsymbol{x}}

\newcommand{\bA}{\mathbf{A}}

\newcommand{\cL}{\mathcal{L}}

\newcommand{\cO}{\mathcal{O}}

\newcommand{\be}{\begin{equation}}
\newcommand{\ee}{\end{equation}}
\newcommand{\bea}{\begin{eqnarray}}
\newcommand{\eea}{\end{eqnarray}}
\newcommand{\nn}{\nonumber}
\newcommand{\kt}{\rangle}
\newcommand{\br}{\langle}

\newcommand{\ed}{\end{document}}

\newcommand{\bi}{\begin{itemize}}
\newcommand{\ei}{\end{itemize}}

\newcommand{\bce}{\begin{center}}
\newcommand{\ece}{\end{center}}

\newcommand{\sD}{\mathscr{D}}

\newcommand{\brho}{{\boldsymbol{\rho}}}
\newcommand{\bfK}{{\boldsymbol{\fK}}}

\newcommand{\ket}[1]{\mid\!\! \, #1\rangle}                              %
\usepackage{mathtools}
\DeclarePairedDelimiterX\MeijerM[3]{\lparen}{\rparen}%
{\begin{smallmatrix}#1 \\ #2\end{smallmatrix}\delimsize\vert\,#3}

\begin{document}

\title{Geometric scattering of a scalar particle moving on a curved surface in the presence of point defects}

\author{Hai Viet Bui\thanks{Email: hbuit@ku.edu.tr}~  and Ali Mostafazadeh\thanks{Corresponding author, Email: amostafazadeh@ku.edu.tr}\\[6pt]
    Departments of Physics and Mathematics,  Ko\c{c} University, \\ 34450 Sar{\i}yer,
    Istanbul, Turkey}
\date{ }
\maketitle

\begin{abstract}
A nonrelativistic scalar particle that is constrained to move on an asymptotically flat curved surface undergoes a geometric scattering that is sensitive to the mean and Gaussian curvatures of the surface. A careful study of possible realizations of this phenomenon in typical condensed matter systems requires dealing with the presence of defects. We examine the effect of delta-function point defects residing on a curved surface ${S}$. In particular, we solve the scattering problem for a multi-delta-function potential in plane, which requires a proper regularization of divergent terms entering its scattering amplitude, and include the effects of nontrivial geometry of ${S}$ by treating it as a perturbation of the plane. This allows us to obtain analytic expressions for the geometric scattering amplitude for a surface consisting of one or more Gaussian bumps. In general the presence of the delta-function defects enhances the geometric scattering effects. 

\vspace{2mm}



\end{abstract}

\section{Introduction}

The study of quantum mechanics on a curved surface may be considered as the first step towards developing a quantum theory of gravity. This has provided the motivation for the early investigations of the subject \cite{dewitt-1952,dewitt-1957}. Among the basic problems one encounters in trying to quantize a classical particle moving in a curved space $M$ is the non-uniqueness of the resulting quantum system. This manifests itself in the form of a factor-ordering problem in the canonical quantization of the system, the ambiguity in the choice of the measure in its path-integral quantization, and the existence of different constraint functions in employing Dirac's formulation of quantizing constrained systems \cite{golovnev}. In canonical quantization program this problem reduces to the lack of a basic prescription for determining the coefficients of the curvature terms appearing in the Hamiltonian operator. For a particle of mass $m$ moving on a surface ${S}$ embedded in a three-dimensional Euclidean space, the latter takes the form:
    \be
        H=-\frac{\hbar^2}{2m}\Delta_g+\frac{\hbar^2}{m}(\lambda_1K+\lambda_2 M^2),
        \label{H-gen}
        \ee
where $\Delta_g$ is the Laplace-Beltrami operator, which acts on complex-valued functions $\psi:{S}\to\C$ according to
     \be
        (\Delta_g\psi)(x):=g(x)^{-1/2}\partial_i \left[g^{ij}(x)g(x)^{1/2}\partial_j\psi(x)\right],
        \label{Delta}
        \ee
$g:=\det(\bg)$, $\bg=[g_{ij}]$ is the $2\times 2$ matrix representing the metric tensor induced on $S$ by the Euclidean metric of the embedding three-dimensional space, $g^{ij}$ are components of $\bg^{-1}$, the repeated indices are summed, $x=(x_1,x_2)$ label local coordinates of the points on $S$, $K$ and $M$ respectively stand for the Gaussian and mean curvatures of $S$, and $\lambda_1$ and $\lambda_2$ are a pair of dimensionless real coupling constants.

Ref.~\cite{dacosta} proposes a resolution for this problem in which the particle is assumed to move in a three-dimensional Euclidean space in the presence of a certain constraining potential that keeps the particle inside a thin shell of the form $S\times [-\epsilon,\epsilon]$. By performing a particular limiting procedure where the strength of the confining potential becomes infinitely large while the thickness of the shell tends to zero, one can decouple the motion of the particle along the normal and tangential directions and show that the tangential motion of the particle is described by an effective Hamiltonian of the form (\ref{H-gen}) with
    \be
    \lambda_1=-\lambda_2=\frac{1}{2}.
    \label{da-costa}
    \ee
The main problem with this so-called thin-shell quantization scheme is that in reality the strength of every constraining potential has a finite upper bound. Taking this into account leads to additional contributions to the curvature terms in the Hamiltonian that depend on the particular constraining potential one employs \cite{kaplan}. In other words, a careful treatment of the problem seems to indicate that the choice of the curvature coefficients depends on the details of the system. This supports the idea of fixing them using experimental data.

The first step in this direction is taken in Ref.~\cite{pra-1996} where the scattering of a particle moving in an asymptotically flat surface is examined and the contribution of $\lambda_1$ to the scattering cross-section is calculated. The results of this investigation apply to genuine two-dimensional scattering setups where the surface is void of an extrinsic geometric, i.e., it is not embedded in a particular Euclidean space. This limits the practical importance of the results of \cite{pra-1996}, because motion in such a surface is experimentally out of reach.

A thorough study of the geometric scattering of a particle moving in an embedded curved surface has been carried out in \cite{pra-2018}. This line of research is also motivated by the recent interest in the study of condensed matter systems involving electrons moving in an effectively curved surface \cite{Ono-2010,Onoe-2012,Ortix-2011,Silva-2013,Vadakkumbatt-2014,Pahlavani-2015}. In order to put the results on geometric scattering to an experimental test involving such systems, one needs to take into account the unavoidable presence of defects. The purpose of the present article is to explore the geometric scattering effects associated with an embedded surface involving delta-function point defects. We use the following generalization the Hamiltonian operator (\ref{H-gen}) to model the motion of a scalar particle on such a surface.
    \be
        H=-\frac{\hbar^2}{2m}\Delta_g+
        \frac{\hbar^2}{m}(\lambda_1K+\lambda_2 M^2)+
        \sum_{j=1}^N\xi_j\delta(\bx-\bba_j),
        \label{H-gen-defect}
        \ee
where $\xi_j$ are real or complex coupling constants, $\bx:=(x_1,x_2)$ marks the coordinates of the points of the surface $S$ in a local Cartesian coordinate system, $\delta(\bx)$ denotes the Dirac delta-function in two dimensions, and $\bba_j$ label the position of the point defects on $S$. 

The presence of the delta function(s) on the right-hand side of (\ref{H-gen-defect}) makes the solution of the corresponding scattering problem highly nontrivial even for the case where $S$ is a plane, i.e., the Hamiltonian has the form
    \be
    H'=-\frac{\hbar^2}{2m}\nabla^2+\sum_{j=1}^N\xi_j\delta(\bx-\bba_j).
        \label{H-defect}
        \ee
This provides a remarkable example of a nonrelativistic quantum system whose treatment leads to divergent terms requiring renormalization. The case of a single delta-function potential, which corresponds to taking $N=1$ in (\ref{H-defect}), has been thoroughly investigated in  \cite{mead-1991,manuel,Adhikari1,Adhikari2,Henderson,Mitra,Nyeo, Camblong}. See also \cite{teo} and references therein. The mathematical reason for the emergence of divergences in the treatment of the Hamiltonian operator (\ref{H-defect}) is that it fails to be a genuine self-adjoint operator \cite{albaverio}. The renormalization schemes developed for dealing with these divergences are known to correspond to self-adjoint extensions of this operator. Refs.~\cite{pra-2016,jpa-2018} outline an alternative solution of the scattering problem for (\ref{H-defect}) that avoids the divergences of the standard approaches.

Following the strategy pursued in Ref.~\cite{pra-2018}, we address the scattering problem for (\ref{H-gen-defect}) in two steps:
	\begin{enumerate}
	\item Consider the case that $S$ is a plane, i.e., determine the scattering solutions of the Schr\"odinger equation for the Hamiltonian~$H'$; 
	\item Include the nontrivial geometry of $S$ as a perturbation of the plane, i.e., use the first Born approximation to solve the scattering problem for $H$, with $H'$ and $H-H'$ respectively playing the role of the unperturbed Hamiltonian and the perturbation.
	\end{enumerate}

The organization of the article is as follows. In Sec.~\ref{sec2}, we review some basic results on Lippmann-Schwinger equation and its standard (Born) series solution. In Sec.~\ref{sec3}, we use these to solve the scattering problem for the Hamiltonian (\ref{H-defect}). Here we provide a detailed discussion of the emerging singularities and their regularization. In Sec.~\ref{sec4}, we use the first Born approximation to compute the effects of nontrivial geometry of $S$ on the scattering amplitude. In Sec.~\ref{sec5}, we examine the utility of our general results for cylindrically symmetric surfaces. In Sec.~\ref{sec6}, we examine the consequences of our findings for surfaces involving one or more separated Gaussian bumps, and in Sec.~\ref{sec7} we present our concluding remarks.

\section{Potential scattering and Born series in two dimensions}
\label{sec2}

Consider the scattering problem defined in a plane by the Schr\"odinger equation,
    \begin{equation}
    \label{sch-eq}
    H |\psi\kt=E |\psi\kt,
    \end{equation}
where $E$ is a real and positive value of energy. Suppose that the Hamiltonian operator admits the decomposition:
    \[H=H_0+V_0+\zeta\, V_1=H'+\zeta\, V_1,\]
where $H_0:=\bp^2/2m$ is the Hamiltonian for a free particle, $V_0$ and $V_1$ are scattering potentials, $\zeta$ is a real perturbation parameter
, and $H':=H_0+V_0$.

Clearly, for any wavevector $\bk$ with wavenumber $k:=|\bk|=\sqrt{2mE}/\hbar$, we have $H_0 |\bk\kt=E |\bk\kt$, where $\br\bx|\bk\kt=e^{i\bk\cdot\bx}/2\pi$. Solutions of the Schr\"odinger equation \eqref{sch-eq} that tend to $|\bk\kt$ in the absence of the potential $V:=V_0+\zeta\, V_1$ are linear combinations of a pair of solutions $|\psi^\pm(\bk)\kt$ fulfilling the Lippmann-Schwinger equation \cite{sakurai},
    \begin{equation}
    \label{ls}
    |\psi^\pm(\bk)\kt=\ket{{\bk}}+
    G_0^\pm (E) (V_0+\zeta\, V_1)|\psi^\pm(\bk)\kt,
    \end{equation}
where 
    \begin{equation}
    G_0^\pm (E):=\lim_{\epsilon\to 0^+}\frac{1}{E-H_0\pm i\epsilon}
    =\lim_{\epsilon\to 0^+}
    \int d^2{\bk'} \frac{|\bk'\kt\br\bk'|}{E-\hbar^2k^{\prime 2}/2m\pm i\epsilon},
    \nn
    \end{equation}
is the resolvent operator whose integral kernel yields the Green's functions,
    \bea
    \label{green} 
    G^{+}_0(\bx,\bx')&:=&\br\bx|G_0^{+}(E)|\bx'\kt
    =-\frac{ im}{2\hbar^2} H_0^{(1)}(k|\bx-\bx'|),\\
    G^{-}_0(\bx,\bx')&:=&\br\bx|G_0^-(E)|\bx'\kt
    =\frac{ im}{2\hbar^2} H_0^{(2)}(k|\bx-\bx'|),
    \eea
and $H_0^{(1)}(x)$ and $H_0^{(2)}(x)$ are respectively the zero-order Hankel function of the first and second kind \cite{pra-1996}.

Next, we recall that
    \begin{align}
    &H_0^{(1)}(x)\to \sqrt{\frac{2}{\pi x}}\; e^{i(x-\pi/4)}~~{\rm for}~~x\to\infty,
    \nn\\
    &H_0^{(2)}(x)\to \sqrt{\frac{2}{\pi x}}\; e^{-i(x-\pi/4)}~~{\rm for}~~x\to\infty,
    \nn\\
    &|\bx-\bx'|\to r-\frac{\bx\cdot\bx'}{r}~~{\rm for}~~r\to\infty,
    \nn
    \end{align}
where $r:=|\bx|$, \cite{watson}. These in turn imply
	\begin{align}
    	&H_0^{(1)}(k|\bx-\bx'|)\to \sqrt{\frac{2}{\pi k r}}\:
	e^{-i(\bk'\cdot\bx'+\pi/4)}e^{ikr}
	~~{\rm for}~~r\to\infty,
	\label{asym1}\\
	&H_0^{(2)}(k|\bx-\bx'|)\to \sqrt{\frac{2}{\pi k r}}\:
	e^{i(\bk'\cdot\bx'+\pi/4)}e^{-ikr}
	~~{\rm for}~~r\to\infty,
	\label{asym2}
	\end{align}
where $\bk':=k\,\bx/r$. In view of (\ref{ls}), (\ref{green}), (\ref{asym1}), and (\ref{asym2}),
    \be
    \br\bx|\psi^\pm(\bk)\kt\to
    \frac{1}{2\pi}\left[ e^{ i \textbf{k}\cdot \textbf{x} }+\ff^\pm({\bk}',{\bk})\,\frac{e^{\pm i k r }}{\sqrt{r}}\right]\quad{\rm for}\quad r:=|\bx|\to\infty,
    \label{asymp}
    \ee
where 
    \bea
    \ff^\pm({\bk}',{\bk})&:=&\frac{\mp im e^{\mp\pi i/4}}{\hbar^2}\sqrt{\frac{2\pi}{k}}
    \int_{\R^2}d^2\bx' e^{\mp i\bk'\cdot\bx'}\br\bx'|(V_0+\zeta\, V_1)|\psi^{\pm}(\bk)\kt,\nn\\
    &=&\frac{-2\pi m}{\hbar^2}\sqrt{\frac{\pm 2\pi i}{k}}\;
    \br\pm\bk'|(V_0+\zeta\, V_1)|\psi^{\pm}(\bk)\kt.
    \label{f=def}
    \eea
According to (\ref{asymp}), $|\psi^{+}(\bk)\kt$ and $|\psi^{-}(\bk)\kt$ are respectively the solutions of the Schr\"odinger equation \eqref{sch-eq} that correspond to the outgoing and incoming waves. It is the former that we identify with the scattering solutions of this equation. With the choice of the plus sign, the second term on the right-hand side of (\ref{asymp}) represents the asymptotic form of the scattered wave. Its amplitude $\ff^+({\bk}',{\bk})$ is called the scattering amplitude for the potential $V$. In the following we drop the superscript + and label the the scattering amplitude by $\ff({\bk}',{\bk})$ for brevity.

We can use (\ref{ls}) to obtain a series expansion for the scattering solutions of  \eqref{sch-eq} in powers of the perturbation parameter $\zeta$. To do this we substitute the ansatz
    \begin{equation}
    \label{expand}
    |\psi^+(\bk)\kt=\sum_{n=0}^{\infty} \zeta^n
    |\psi^+_n(\bk)\kt
    \end{equation}
    in both sides of (\ref{ls}) and demand that it holds at each order of $\zeta$ separately. This yields a set of equations for the unknowns $|\psi^+_n(\bk)\kt$ that admit the following solution.
    \bea
    \label{unpertured wave}
    |\psi^+_n(\bk)\kt&=&\left\{\begin{array}{ccc}
    \left(1-G_0^+(E) V_0 \right)^{-1} |\bk\kt &{\rm for}& n=0,\\[6pt]
    \left(1-G_0^+(E) V_0 \right)^{-1} G_0^+(E)
    V_1 |\psi_{n-1}^+(\bk)\kt &{\rm for}& n\geq 1,
    \end{array}\right.\nn\\[6pt]
    &=&\left\{\begin{array}{ccc}
    |\bk\kt+G^{\prime +}(E) V_0  |\bk\kt &{\rm for}& n=0,\\[6pt]
    G^{\prime +}(E)V_1 |\psi_{n-1}^+(\bk)\kt &{\rm for}& n\geq 1,
    \end{array}\right.
    \label{psi-n=}\nn\\[6pt]
    &=&\left\{\begin{array}{ccc}
    |\bk\kt+G^{\prime +}(E) V_0  |\bk\kt &{\rm for}& n=0,\\[6pt]
    [G^{\prime +}(E)V_1]^n |\psi_0^+(\bk)\kt &{\rm for}& n\geq 1,
    \end{array}\right.\nn
    \eea
where
    \begin{equation}
    G^{\prime +} (E):=
    \lim_{\epsilon\to 0^+}\frac{1}{E-H'+ i\epsilon}.
    \nn
    \end{equation}

Next, we substitute (\ref{expand}) in (\ref{f=def}) to obtain a series expansion of the scattering amplitude in powers of $\zeta$. This gives the Born series,
    \be
    \ff(\bk',\bk)=\sum_{n=0}^\infty \,\zeta^n \ff_n(\bk',\bk),
    \label{f-expand}
    \ee
where
    \be
    \ff_n(\bk',\bk):=\frac{-2\pi m}{\hbar^2}\sqrt{\frac{2\pi i}{k}}\times
    \left\{\begin{array}{ccc}
    \br\bk'|V_0|\psi^+_0(\bk)\kt & {\rm for} & n=0,\\[6pt]
    \br\bk'|V_0|\psi^+_n(\bk)\kt+\br\bk'|V_1|\psi^+_{n-1}(\bk)\kt
    & {\rm for} & n\geq 1.\end{array}\right.
    \label{fn-def}
     \ee
Ignoring terms of order $N+1$ and higher in (\ref{f-expand}), we arrive at the $N$-th order Born approximation: $f(\bk',\bk)\approx\sum_{n=0}^N \,\zeta^n \ff_n(\bk',\bk)$. In particular, the first Born approximation gives
    \bea
    \ff(\bk',\bk)&\approx& \ff_0(\bk',\bk)+\zeta\, \ff_1(\bk',\bk),
    \label{first-BA}
    \eea
where
	\bea    
	\ff_0(\bk',\bk)&=&\frac{-2\pi m}{\hbar^2}\sqrt{\frac{2\pi i}{k}}\br\bk'|V_0|\psi^+_0(\bk)\kt,
	\label{f-zero}\\
	\ff_1(\bk',\bk)&=&\frac{-2\pi m}{\hbar^2}\sqrt{\frac{2\pi i}{k}}
	\left[\left(\br\bk'|V_0|\psi^+_1(\bk)\kt+\br\bk'|V_1|\psi^+_{0}(\bk)\kt\right)\right].
	\label{f-one}
	\eea
Note that $\ff_0(\bk',\bk)$ is the exact scattering amplitude for the potential $V_0$, and we can use (\ref{unpertured wave}) to show that
	\be
	\left(\br\bk'|V_0|\psi^+_1(\bk)\kt+\br\bk'|V_1|\psi^+_{0}(\bk)\kt\right)=
	\br\psi^+_0(\bk')|[1-V_0^\dagger G_0^-(E)][1- G_0^+(E)V_0]^{-1}V_1
	|\psi^+_0(\bk)\kt.
	\label{f1-term}
	\ee
For $V_0=0$, the right-hand side of this equation becomes $\br\bk'|V_1|\bk\kt$, and (\ref{first-BA}) --(\ref{f-one}) yield the familiar expression for the first Born approximation \cite{sakurai}, namely 
	\[\ff(\bk',\bk)\approx\zeta\, \ff_1(\bk',\bk)=\frac{-2\pi m}{\hbar^2}\sqrt{\frac{2\pi i}{k}}\br\bk'|V_1|\bk\kt.\]

In order to explore the scattering of a particle moving in a curved surface and interacting with $N$ point defects, we employ (\ref{first-BA}) with 
	\bea
    	V_0&:=&\sum_{j=0}^N \xi_j\, \delta(\bx-\bba_j),
	\label{delta}\\
	\zeta\,V_1&:=&H-H'=-\frac{\hbar^2}{2m}(\Delta_g-\nabla^2)+
	\frac{\hbar^2}{m}(\lambda_1K+\lambda_2 M^2),
	\label{V1=}
   	 \eea
where $\xi_j$ are real or complex coupling constants, $\bba_j$ label the position of the defects, and $H$ and $H'$ are respectively given by (\ref{H-gen-defect}) and (\ref{H-defect}).

\section{Scattering by delta-function potentials in two dimensions}
\label{sec3}

Consider the scattering process in two dimensions that is defined by the delta-function potential (\ref{delta}). Because for every state vector $|\psi\kt$,
    \[\br\bx|\bba_j\kt\br\bba|\psi\kt=\delta(\bx-\bba_j)\psi(\bba_j)=\delta(\bx-\bba_j)\psi(\bx),\]
we can express this potential in the form:
    \begin{equation}
    \label{modi delta}
    V_0=\sum_{j=0}^N \xi_j\,|\bba_j\kt\br\bba_j|.
    \end{equation}
Using this relation in (\ref{f-zero}), we find 
    \be
    \ff_0(\bk',\bk)=\frac{-1}{2}\sqrt{\frac{2\pi i}{k}}
    \;\sum_{j=1}^N X_j(\bk)\:e^{-i\bba_j\cdot\bk'}.
    \label{f-for-delta}
    \ee
where
    \be
    X_j(\bk):=\frac{2m \xi_j }{\hbar^2}\:\br\bba_j|\psi_0^+(\bk)\kt,
    \label{Xj-def}
    \ee

According to (\ref{unpertured wave}), $|\psi_0^+(\bk)\kt$ is the solution of the Lippmann-Schwinger equation,
    \[(1-G_0^+(E) V_0)|\psi_0^+(\bk)\kt=|\bk\kt.\]
This relation together with (\ref{green}) and (\ref{modi delta}) imply
    \be
    \br\bx|\psi_0^+(\bk)\kt=\br\bx|\bk\kt-\frac{im}{2\hbar^2}
    \sum_{j=1}^N \xi_j \br\bba_j|\psi_0^+(\bk)\kt\, H_0^{(1)}(k|\bx-\bba_j|).
    \label{eq001}
    \ee
Setting $\bx=\bba_i$ in this equation, with $i=1,2,\cdots, N$, and using (\ref{Xj-def}), we find the following system of $N$ equations for the coefficients $X_j(\bk)$.
    \be
    \sum_{j=1}^N\left[\frac{\hbar^2}{2m\xi_i}\,\delta_{ij}
    +\frac{i}{4}H_0^{(1)}(k|\bba_i-\bba_j|)\right]X_j(\bk)=\br\bba_i|\bk\kt,
    \label{eq002}
    \ee
where $\delta_{ij}$ is the Kronecker delta symbol. The main problem with this procedure for computing $X_j(\bk)$ is that the terms $H_0^{(1)}(k|\bba_i-\bba_j|)$ entering (\ref{eq002}) diverge for $i=j$. The standard treatment of this problem involves removing these divergences via a coupling-constant renormalization \cite{manuel}. As we explain below, this in turn introduces a length scale for the problem.

A simple physical interpretation for this length scale is provided by the observation that a point interaction modeled using (\ref{delta}) is an idealization of a scattering interaction whose range is much smaller than the wavelength of the scattered wave, $\lambda:=2\pi/k$. More specifically, it vanishes outside the discs $\sD_n$ with center $\bba_j$ and radius $\rho\ll \lambda$. Furthermore, Eq.~(\ref{eq001}), which determines the scattering solutions of the Schr\"odinger equation, is valid away from these discs. Therefore, we can at best set $\bx=\bba_i+\brho$, where $\brho\in\R^2$ is any vector with magnitude $\rho$. The emergence of divergent terms in setting $\bx=\bba_i$ in (\ref{eq001}) is a manifestation of the fact that we are actually not allowed to set $\bx=\bba_i$. Rather we must use the information about the behavior of the potential inside the discs $\sD_n$ to determine the solution of the Schr\"odinger equation that satisfies the outgoing boundary conditions at their boundary.

Now, consider a scattering experiment where a plane wave with wavevector $\bk$ is incident up on $N$ point scatterers located at $\bba_j$ with an interaction range $\rho$ much smaller than $k^{-1}$. Away from the discs $\sD_n$, the expression (\ref{f-for-delta}) for the scattering amplitude is valid. We need the details of the potential inside these discs to compute the terms $\br\bba_j|\psi_0^+(\bk)\kt$ and hence $X_j(\bk)$. Alternatively, we can measure the latter scattering amplitude for $N$ different values of $\bk'$ and insert the result in (\ref{f-for-delta}) to obtain a system of equations which we can solve to determine $X_j(\bk)$ and hence the scattering amplitude for all $\bk'$.

Motivated by the above argument on the need for a length scale $\rho$, we set $\bx=\bba_i+\brho$ in (\ref{eq001}), and suppose that the $\rho\to 0$ limit of both sides of the resulting equation exists. This is possible, only if we allow the coupling constants $\xi_j$ to depend on $\rho$. Assuming that this is the case and using (\ref{Xj-def}), we find the following regularized analog of (\ref{eq002}).
    \be
    \sum_{j=1}^N A_{ij}X_j(\bk)=\br\bba_i|\bk\kt,
    \label{eq-003}
    \ee
where
    \begin{align}
    &A_{\ell j}:=\frac{1}{4}\times\left\{\begin{array}{ccc}
    2\hbar^2/m\tilde\xi_j+i&{\rm for}&j=\ell,\\[6pt]
    H_0^{(1)}(k|\bba_\ell-\bba_j |)&{\rm for}&j\neq \ell,
    \end{array}\right.
    \label{Aij-def}\\
    &\tilde\xi_j:=\frac{1}{\xi_j^{-1}-\frac{m}{\pi\hbar^2}\left[\ln(k\rho/2)+\gamma\right]},
    \label{txi-def}
    \end{align}
$\gamma$ is Euler's constant, and we have made use of the fact that
    \[H_0^{(1)}(x)=\frac{2i[\ln(x/2)+\gamma]}{\pi}+1+\cO(x^2).\]

Letting $\bA$ denote the matrix of coefficients $A_{ij}$ and using $A^{-1}_{ij}$ to label the entries of $\bA^{-1}$, we can express the solution of (\ref{eq-003}) as
    \be
    X_i(\bk)=\sum_{j=1}^N A^{-1}_{ij}\br\bba_j|\bk\kt=\frac{1}{2\pi}
    \sum_{j=1}^N A^{-1}_{ij} e^{i\bba_j\cdot\bk}.
    \label{Xj=}
    \ee
Substituting this relation in (\ref{f-for-delta}), we find
    \be
    \ff_0(\bk',\bk)=\frac{-1}{2}\sqrt{\frac{i}{2\pi k}}
    \;\sum_{i,j=1}^N A^{-1}_{ij}\:e^{i(\bba_j\cdot\bk-\bba_i\cdot\bk')}.
    \label{f-for-delta=}
    \ee
Because the scattering amplitude is a physical quantity, it should not depend on the choice of the running renormalization scale $\rho$. This implies that the same should be true for the renormalized coupling constants $\tilde\xi_j$, i.e., the $\rho$-dependence of the bare coupling constants $\xi_j$ should be such that $\tilde\xi_j$ do not depend on $\rho$.

For $N=1$, i.e., a single-delta-function potential,
    \be
    V_0(\bx)=\xi_1\delta(\bx-\bba_1),
    \label{delta-1}
    \ee
Eqs.~(\ref{Aij-def}) and (\ref{f-for-delta=}) imply
    \be
    \ff_0(\bk',\bk)=-\sqrt{\frac{2i}{\pi k}}\:
    \frac{\tilde\xi_1\: e^{i\bba_1\cdot(\bk-\bk')}}{2\hbar^2/m+i\tilde\xi_1}.
    \label{f-for-single-delta}
    \ee
This agrees with the results of performing alternative renormalization schemes that are discussed in Ref.~\cite{manuel} as well as the result obtained by the transfer-matrix approach of Refs.~\cite{pra-2016,jpa-2018}. The latter, which avoids divergent terms, yield (\ref{f-for-single-delta}) with renormalized coupling constant $\tilde\xi$ replaced with the original (bare) coupling constant $\xi$. This is equivalent to setting $\rho=2 e^{-\gamma}/k$.

For $N=2$,
    \bea
    \bA&=&\frac{1}{4}\left[\begin{array}{cc}
    2\hbar^2/m\tilde\xi_1+i&
    iH_0^{(1)}(k|\bba_1-\bba_2|)
     \\
    iH_0^{(1)}(k|\bba_2-\bba_1|)&
    2\hbar^2/m\tilde\xi_2+i
    \end{array}\right],
    \label{A-N2=}\\
    \bA^{-1}&=&\frac{4 \tilde\xi_1\tilde\xi_2}{D}\left[\begin{array}{cc}
    2\hbar^2/m\tilde\xi_2+i&
    -iH_0^{(1)}(k|\bba_1-\bba_2|)
     \\
    -iH_0^{(1)}(k|\bba_2-\bba_1|)&
    2\hbar^2/m\tilde\xi_1+i
    \end{array}\right],
    \label{Ainv-N2=}\\
    D&=&[H_0^{(1)}(k|\bba_2-\bba_1|)^2-1]\tilde\xi_1\tilde\xi_2
    +2i\hbar^2(\tilde\xi_1+\tilde\xi_2)/m+4\hbar^4/m^2.
    \eea
Substituting (\ref{Ainv-N2=}) in (\ref{f-for-delta=}), we obtain
    \bea
    \ff_0(\bk',\bk)&=&-\sqrt{\frac{2i}{\pi k}}D^{-1}
    \Big\{\tilde\xi_1(2\hbar^2/m+i\tilde\xi_2)\: e^{i\bba_1\cdot(\bk-\bk')}+
    \tilde\xi_2(2\hbar^2/m+i\tilde\xi_1)\: e^{i\bba_2\cdot(\bk-\bk')}\nn\\
    &&\hspace{2.3cm}-i\tilde\xi_1\tilde\xi_2 H_0^{(1)}(k|\bba_2-\bba_1|)\left[
    e^{i(\bba_2\cdot\bk-\bba_1\cdot\bk')}+
    e^{i(\bba_1\cdot\bk-\bba_2\cdot\bk')}\right]\Big\}.
    \label{f-for-2delta=}
    \eea
The application of the transfer-matrix method for the double-delta-function potential \cite{jpa-2018}, which does not involve divergent terms, yields (\ref{f-for-2delta=}) with $H_0^{(1)}(k|\bba_2-\bba_1|)$ replaced with its real part, namely $J_0(k|\bba_2-\bba_1|)$.\footnote{Here and in what follows $J_n$ labels the $n$-th order Bessel function of the first kind.} This disagreement may be a sign that the multi-delta-function potential as treated in the transfer-matrix approach has a different physical interpretation as the one we have provided above, namely as a model for describing scatterers having a small but finite size.

\section{Effects of nontrivial geometry on scattering}
\label{sec4}

Consider the motion of a particle on an asymptotically flat surface ${S}$ that is embedded in a three-dimensional Euclidean space. Suppose that $S$ has the topology of a plane, i.e., ${S}$ is obtained by a local smooth deformation of a plane, and that there are $N$ point defects on $S$ whose effect on the particle's motion is described by a (multi-) delta-function potential of the form~(\ref{delta}). We are interested in determining the effect of the nontrivial geometry of ${S}$ on the scattering amplitude of this potential. We do this by considering the contribution of the geometry as a perturbation of the plane. Specifically, we use the first Born approximation (\ref{first-BA}) with unperturbed potential (\ref{delta}) subject to the perturbation (\ref{V1=}). In the coordinate representation, the latter takes the form
	\be
	\zeta\br\bx'|V_1|\bx\kt=\frac{\hbar^2}{2m}\,\cL_{x'}\,\delta(\bx'-\bx),	
	\label{geometric potential}
	\ee
where $\cL_{x}$ is the differential operator:
	\bea
	\label{Lx-def}
	\cL_{x}&:=&\left[g_0^{ij}(x)-g^{ij}(x)\right]\partial_i \partial_j 
	-\frac{\partial_i[\sqrt{g(x)} g^{ij}(x)]}{\sqrt{g(x)}}\partial_j 
	+ 2\lambda_1 K(x)+2 \lambda_2 M(x)^2,
	\eea
$g_0^{ij}(x)$ are the components of the inverse of the local matrix representation of the Euclidean metric, so that whenever $x=(x_1,x_2)$ are Cartesian coordinates $g_0^{ij}(x)=\delta_{ij}$. 

To determine the contribution of the perturbation to the scattering amplitude, we need to obtain an explicit expression for the right-hand side of (\ref{f-one}). This requires computing $\br\bk'|V_1|\psi_0^+(\bk)\kt$ and $\br\bk'|V_0|\psi_1^+(\bk)\kt$.

First, we observe that
	\bea
	\br\bk'|V_1|\psi_0^+(\bk)\kt&=&\frac{1}{2\pi}
	\int_{\R^2}d^2\bx'\int_{\R^2}d^2\bx\: e^{-i\bk'\cdot\bx'}\br\bx'|V_1|\bx\kt
	\br\bx|\psi_0^+(\bk)\kt\nn\\
	&=&\frac{\hbar^2}{4\pi m\zeta}
	\int_{\R^2}d^2\bx'\: e^{-i\bk'\cdot\bx'}\cL_{x'}\br\bx'|\psi_0^+(\bk)\kt.
	\label{eq005}
	\eea
Moreover, according to (\ref{Xj-def}), (\ref{eq001}), and (\ref{Xj=}),
	\bea
	\br\bx|\psi_0^+(\bk)\kt&=&\frac{e^{i\bk\cdot\bx}}{2\pi}-\frac{i}{4}
	\sum_{j=1}^N H_0^{(1)}(k|\bx-\bba_j|)X_j(\bk),\nn\\
	&=&\frac{1}{2\pi}\left[e^{i\bk\cdot\bx}-\frac{i}{4}
	\sum_{i,j=1}^N  e^{i\bk\cdot \bba_i} A_{ij}^{-1} H_0^{(1)}(k|\bx-\bba_j|) \right],
	\label{eq0011}
	\eea
where we have also benefited from the fact that $\bA$ and consequently $\bA^{-1}$ are symmetric matrices.

The calculation of $\br\bk'|V_0|\psi_1^+(\bk)\kt$ is slightly more involved. We begin using (\ref{unpertured wave}) to show that
	\be
	V_0|\psi_1^+(\bk)\kt-V_0G_0^+(E)V_0|\psi_1^+(\bk)\kt=
	V_0G_0^+(E)V_1|\psi_0^+(\bk)\kt.
	\ee
Substituting (\ref{modi delta}) in this equation, introducing
	\be
	Y_j(\bk):=\frac{2m}{\hbar^2}\,\xi_j\br\bba_j|\psi_1^+(\bk)\kt,
	\label{Yj-def}
	\ee
and noting that $|\bba_1\kt, |\bba_2\kt, \cdots, |\bba_N\kt$ are linearly independent, we obtain
	\be
	\sum_{i=1}^NA_{ij}Y_j(\bk)=B_i(\bk),
	\label{Y-eq}
	\ee
where $A_{ij}$ are given by (\ref{Aij-def}), 
	\bea
	B_i(\bk)&:=&\br\bba_i|G_0^+(E)V_1|\psi_0^+(\bk)\kt\nn\\
	&=&-\frac{i}{4\zeta}\int_{\R^2}d^2\bx\,H_0^{(1)}(k|\bx-\bba_i|)\cL_x
	\br\bx|\psi_0^+(\bk)\kt,
	\label{Bj-def}
	\eea
and we have made use of (\ref{green}). We can express the solution of (\ref{Y-eq}) as
	\be
	Y_i(\bk)=\sum_{j=1}^NA^{-1}_{ij}B_j(\bk).
	\label{Yj=}
	\ee
Furthermore, according to (\ref{modi delta}), (\ref{Yj-def}), and (\ref{Yj=}),
	\be
	\br\bk'|V_0|\psi_1^+(\bk)\kt=
	\frac{\hbar^2}{2m}\sum_{i=1}^N\br\bk'|\bba_i\kt Y_i(\bk)=
	\frac{\hbar^2}{4\pi m}
	\sum_{i,j=1}^N e^{-i\bba_i\cdot\bk'}A^{-1}_{ij}B_j(\bk).
	\label{eq006}
	\ee
	
Next, we substitute (\ref{eq005}), (\ref{Bj-def}), and (\ref{eq006}) in (\ref{f-one}) and make use of  (\ref{eq0011}) to identify the contribution of the geometry of the surface to the scattering amplitude with
	\bea
	\zeta \ff_1(\bk',\bk)&=&
	-\pi\sqrt{\frac{2\pi i}{k}}\int_{\R^2}d^2\bx'\;
	\br\bx'|\psi_0^+(-\bk')\kt \cL_{x'}\br\bx'|\psi_0^+(\bk)\kt.
	\label{eq0012}
	\eea
In view of (\ref{eq0011}), we can express this formula in the following more explicit form:
	\be
	\zeta \ff_1(\bk',\bk)=-\frac{1}{2}\sqrt{\frac{i}{2\pi k}}\left[I_{0}-\frac{i}{4}
	\sum_{i,j=1}^N A_{ij}^{-1}\, I_{ij}-\frac{1}{16}
	\sum_{i,j,i',j'=1}^NA_{ij}^{-1}A_{i'j'}^{-1}\,I_{iji'j'}\right],
	\label{f1=Is}
	\ee
where
	\bea
	I_{0}&=&\int_{\R^2}d^2\bx'\:e^{-i\bk'\cdot\bx'}\cL_{x'}\: e^{i\bk\cdot\bx'},
	\label{I-zero}\\
	I_{ij}&=&\int_{\R^2}d^2\bx'\left\{
	e^{-i\bk'\cdot\bba_i} H_0^{(1)}(k|\bx'-\bba_j|)\,\cL_{x'}\: e^{i\bk\cdot\bx'}+
	e^{i(\bk\cdot\bba_i-\bk'\cdot\bx')}\cL_{x'} H_0^{(1)}(k|\bx'-\bba_j|) \right\},~~~
	\label{I-ij}\\
	I_{iji'j'}&=&e^{i(\bk\cdot\bba_{i'}-\bk'\cdot\bba_i)}\!\! \int_{\R^2}d^2\bx'\:
	H_0^{(1)}(k|\bx'-\bba_j|)\, \cL_{x'}
	H_0^{(1)}(k|\bx'-\bba_{j'}|).
	\label{I-ijij}
	\eea
An important outcome of these equations is that the presence of the defects can in general boost the influence of the nontrivial geometry of the surface on the scattering of the particle. This is because depending on the value of $A_{ij}^{-1}$, the terms on the right hand side of (\ref{f1=Is}) that involve $A_{ij}^{-1}$ can dominate $I_0$ which signifies the effect of geometry in the absence of the defects.  

For  a given surface ${S}$, we can compute the coefficients of the differential operator (\ref{Lx-def}) and use the result to evaluate the right-hand side of (\ref{I-zero}) -- (\ref{I-ijij}). In general, this cannot be done analytically unless we impose certain simplifying conditions on the shape of the surface and the number and position of the defects. There is, in principle, no obstacle to compute $I_{0}$, $I_{ij}$, and $I_{iji'j'}$ numerically, but this requires the knowledge of the renormalized coupling constants $\tilde\xi_j$ which need to be determined experimentally. This manifests the importance of deriving analytic expressions for the scattering amplitude that would clarify how the coupling constants $\tilde\xi_i$ relate to the scattering data.

\section{Geometric scattering for cylindrically symmetric surfaces}
\label{sec5}

Let $(r,\theta,z)$ denote the cylindrical coordinates in $\R^3$, with $(r,\theta)$ being the polar coordinates in the $x$-$y$ plane, and suppose that the surface ${S}$ is the graph of a smooth function of $r$, i.e., there is a smooth function $f:[0,\infty)\to\R$ such that 
	\be
	{S}:=\{(r,\theta,z)\in\R^3~|~z=f(r)~\}.
	\label{S-def}
	\ee
This is an asymptotically flat differentiable surface, if
	\begin{equation}
	\label{cond}
	\lim_{r \to \infty} \dot{f}(r)=\dot{f}(0)=0,
	\end{equation}
where a dot denotes differentiation with respect to $r$. We will parameterize the points of ${S}$ by the polar coordinates $(r,\theta)$, i.e., set $x_1:=r$ and $x_2:=\theta$. This in turn yields the following expression for the metric induced on ${S}$ by the Euclidean metric on $\R^3$. 
	\begin{equation}
	\bg=[g_{ij}]=\begin{bmatrix}
	1+\dot{f}^2&0\\
	0& r^2
	\end{bmatrix}.
	\label{bg=}
	\end{equation}
As noted in Ref.~\cite{pra-2018}, the Gaussian curvature $K$ and mean curvature $M$ of the surface ${S}$  are respectively given by
	\begin{equation}
	K=\frac{G\dot{G}}{r} \qquad\qquad\qquad M=\frac{1}{2}\left(\frac{G}{r}+\dot{G}\right),
	\label{KM=}
	\end{equation}
where
	\begin{equation}
	G:=\frac{\dot{f}}{\sqrt{1+\dot{f}^2}}=r\left(M+\sqrt{M^2-K}\right).
	\label{G=MK}
	\end{equation}
To ensure that $K$ and $M$ are regular (nonsingular) functions of $r$, we demand that $f'(r)/r$ and $f''(r)$ tend to finite values as $r \rightarrow 0$, i.e., $\lim_{r\to 0}f'(r)/r$ and $\lim_{r\to 0}f''(r)$ exist.

In light of (\ref{Lx-def}), (\ref{bg=}), (\ref{KM=}), and (\ref{G=MK}) the differential operator $\cL_x$ associated with the surface (\ref{S-def}) takes the form
	\bea
	\label{geometric}
	\cL_x&=& G^2\left[\partial^2_{r}+\frac{1}{r}\left(1+\frac{r\dot{G}}{G}\right)\partial_{r}+
	\frac{2\lambda_1\dot{G}}{r\, G} +\frac{\lambda_2}{2r^2} \left(1+
	\frac{r\dot{G}}{G}\right)^{\!2}\right]\\
	&=&G^2 \partial^2_{r}+2MG\,\partial_r+2(\lambda_1K+\lambda_2M^2).
	\label{geometric2}
	\eea
Substituting (\ref{geometric})  in (\ref{I-zero}) we obtain an integral that we can partially evaluate. Following a procedure described in Ref.~\cite{pra-2018}, we identify the integration variables $\bx'=(x',y')$ with coordinates in a Cartesian coordinate system in which $\bk-\bk'$ is along the $x'$-axis. If we use $\theta_0$, $\theta$, and $\Theta$ to respectively denote the angles between $\bk$ and the $x'$-axis, $\bk'$ and the $x'$-axis, and $\bk$ and $\bk'$, we can show that
	\begin{align}
	&\theta_0=\frac{1}{2}(\pi-\Theta), &&\theta=\frac{1}{2}(\pi+\Theta),
	&&|\bk'-\bk|=2ks,
	\label{id1}
	\end{align}
where
	\[s:=\sin(\Theta/2).\]
Next, we transform to the polar coordinates $(r',\theta')$ associated with $(x',y')$, and perform the integration over the polar angle $\theta'$. To do this we use (\ref{id1}) to establish
	\be
	e^{-i\bk'\cdot\bx'}\partial_{r'}^\ell e^{i\bk\cdot\bx'}
	=(ik)^\ell\cos^\ell(\theta'-\theta_0)\,e^{2iksr'\cos\theta'},
	\label{id2}
	\ee
and recall that for all $a\in\R^+$ and $\ell=0,1,2$, 
	\begin{align}
	&\int_0^{2\pi}\!\! d\theta'  \cos^\ell(\theta'-\theta_0)e^{ia\cos\theta'}=\left\{
	\begin{array}{ccc}
	2\pi J_0(a) & {\rm for} & \ell=0,\\
	2\pi i s J_1(a) & {\rm for} & \ell=1,\\
	2\pi[s^2 J_0(a)-(\frac{2s^2-1}{a})J_1(a)]& {\rm for} & \ell=2,
	\end{array}\right.
	\nn
	\\
	&\dot{G}(r)G(r)J_0(ar)=\frac{1}{2}\left\{\frac{d}{dr}\left[G(r)^2J_0(ar)\right]+a
	G(r)^2 J_1(ar)]\right\},
	\nn
	\\
	&r\dot{G}(r)G(r)J_1(ar)=\frac{1}{2}\left\{\frac{d}{dr}\left[r G(r)^2J_1(ar)\right]
	-a r G(r)^2 J_0(ar)]\right\}.
	\nn
	\end{align}
With the help of these equations, we obtain 	
	\bea
	I_0= \pi \left[\left(2\lambda_1+\lambda_2-\frac{1}{2s^2} \right)g_1(2ks)
	+\lambda_2\, g_2(2ks)\right],
	\label{I-zero=}
	\eea
where we have introduced the functions:
	\bea
	g_1(\kappa)&:=&\kappa\int_0^\infty dr\: G(r)^2J_1(\kappa r),
	\label{g1-def}\\
	g_2(\kappa)&:=&\int_0^\infty dr\: r^{-1}
	\left[G(r)^2+r^2\dot{G}(r)^2\right]J_0(\kappa r),
	\label{g2-def}
	\eea
and assumed that
	\begin{equation}
	\label{constrain}
	\lim\limits_{r\rightarrow \infty} r J_1 (2 k s r) G(r)^2=0.
	\end{equation}
The latter holds whenever $|G(r)|$ decays to zero faster than $r^{-1/4}$ as $r\to\infty$, which is a rather mild condition on ${S}$. 

We can express the integrand on the right-hand side of (\ref{g1-def}) and (\ref{g2-def}) in terms of the mean and Gaussian curvatures of the surface. This requires the use of (\ref{G=MK}) and
	\begin{align}
	&G^2+r^2\dot G^2=2r^2(2M^2-K),
	\label{G-dG=MK}
	\end{align}
which follows from (\ref{KM=}).

Inserting (\ref{I-zero=}) in (\ref{f1=Is}) and neglecting terms involving $A_{ij}^{-1}$, we obtain the scattering amplitude for geometric scattering in the absence of delta-function defects \cite{pra-2018}.  In order to determine the effect of the defects, we need to compute $I_{ij}$ and $I_{iji'j'}$. It is not difficult to see that the application of the same procedure for separating and performing the angular integral appearing in the calculation of $I_{ij}$ and $I_{iji'j'}$ is intractable. Therefore, we focus on some special cases.

\subsection{A central point defect}

Suppose that there is a single delta-function defect located at the origin, i.e., $N=1$ and $\bba_1=(0,0)$. Then (\ref{f1=Is}), (\ref{I-ij}), and (\ref{I-ijij}) respectively read:
	\bea
	\zeta \ff_1(\bk',\bk)&=&-\frac{1}{2}\sqrt{\frac{i}{2\pi k}}
	\left(I_{0}-\frac{i  I_{11}}{4 A_{11}}-\frac{I_{1111}}{16 A_{11}^2}\right),
	\label{f1-central}\\
	I_{11}&=&\int_{\R^2}d^2\bx'\left[H_0^{(1)}(kr')\,\cL_{x'}\: e^{i\bk\cdot\bx'}
	+e^{-i\bk'\cdot\bx'}\cL_{x'}\, H_0^{(1)}(kr')\right],
	\label{I-11a}\\
	I_{1111}&=&\int_{\R^2}d^2\bx'\: H_0^{(1)}(kr')\, \cL_{x'}\, H_0^{(1)}(kr'),
	\label{I-1111a}
	\eea
where according to (\ref{Aij-def}), 
	\be
	A_{11}=\frac{\hbar^2}{2m\tilde\xi_1}+\frac{i}{4}.
	\label{A11=}
	\ee
Substituting (\ref{geometric}) in (\ref{I-11a}) and (\ref{I-1111a}), pursuing an approach similar to the one we have employed in the calculation of $I_0$, and using various properties of Bessel and Hankel functions, we obtain 
	\bea
	I_{11}&=&2\pi\int_0^\infty dr'\,r'
	\left\{\Big[4C(r')-k^2 G(r')^2\Big]J_0(kr')H_0^{(1)}(kr')
	-k^2 G(r')^2 J_1(kr')H_1^{(1)}(kr')\right\},
	\label{I-11}\\
	I_{1111}&=&\pi\int_0^\infty dr'\,r'
	\left\{\Big[4C(r')-k^2 G(r')^2\Big]H_0^{(1)}(kr')^2
	-k^2 G(r')^2 H_1^{(1)}(kr')^2\right\},
	\label{I-1111}
	\eea 
where $C:[0,\infty)\to\R$ is the function defined by
	\[C:=\frac{G^2}{r}\Big[\frac{\lambda_1\dot G}{G}+
	\frac{\lambda_2}{4r}\Big(1+\frac{r\dot{G}}{G}\Big)^2\Big]=\lambda_1 K+\lambda_2 M^2.\]
The latter equation together with (\ref{G=MK}) allow us to express the integrands on the right-hand side of (\ref{I-11}) and (\ref{I-1111}) in terms of the mean and Gaussian curvatures of the surface.

\subsection{Distant point defects}

Consider the situation where the particle interacts with the delta-function defects that are located sufficiently far from the origin so that in their vicinity the surface is nearly flat, i.e., there is a  neighborhood, $N_d:=\{\bx\in\R^2~|~|\bx|<d\}$, of the origin such that the mean and Gaussian curvatures of the surface take negligible values outside $N_d$ and the location of the defects $\bba_j$ satisfy $|\bba_j|\gg d$. 
Then, we can ignore the contribution of the integrands on the right-hand side of (\ref{I-ij}) and (\ref{I-ijij}) unless $|\bx'|\ll|\bba_j|$ for all $j$. In view of this observation and (\ref{asym1}), we can employ the approximation,
	\be
	H_0^{(1)}(k|\bx'-\bba_j|)\approx \sqrt{\frac{2}{\pi i k|\bba_j|}}\:
	e^{ik|\bba_j|} e^{-i\bfK_j\cdot\bx'},
	\ee
in (\ref{I-ij}) and (\ref{I-ijij}), where $\bfK_j:= k\bba_j/|\bba_j|$. This gives
	\bea
	I_{ij}&=&\sqrt{\frac{2}{\pi i k|\bba_j|}}\:e^{ik|\bba_j|}
	\left[e^{-i\bk'\cdot\bba_i} J(\bfK_j,\bk)+
	e^{i\bk\cdot\bba_i}J(\bk',-\bfK_j)\right],~~~
	\label{I-ij-far}\\
	I_{iji'j'}&=&\frac{2}{\pi i k\sqrt{|\bba_j||\bba_{j'}|}}\:\:e^{ik(|\bba_j|+|\bba_{j'}|)}
	e^{i(\bk\cdot\bba_{i'}-\bk'\cdot\bba_i)} J(\bfK_j,-\bfK_{j'}),
	\label{I-ijij-far}
	\eea
where
	\be
	J(\bp,\bq):=\int_{\R^2}\!\!d^2\bx'\,e^{-i\bp\cdot\bx'}\, \cL_{x'}\,e^{i\bq\cdot\bx'}.
	\nn
	\ee
Comparing this relation with (\ref{I-zero}), we see that $I_0=J(\bk',\bk)$. We can repeat the steps of the calculation of $I_0$ to show that
	\bea
	J(\bp,\bq)&=&\pi\Big[
	\Big(2\lambda_1+\lambda_2+
	\frac{2q^2\cos(2\theta_q)}{\kappa^2}-\frac{2q\cos\theta_q}{\kappa}\Big)
	g_1(\kappa)+\lambda_2g_2(\kappa)+\nn\\
	&&\hspace{1cm}
	\frac{q\cos\theta_q}{\kappa}
	\Big(1-\frac{2q\cos\theta_q}{\kappa}\Big)g_3(\kappa)\Big],
	\label{J=}
	\eea
where $q:=|\bq|$, $\theta_q$ is the angle between $\bq$ and the $x'$-axis, $\kappa:=|\bq-\bp|$, $g_1$ and $g_2$ are functions defined by (\ref{g1-def}) and (\ref{g2-def}), and 
	\be
	g_3(\kappa):=\kappa^2\int_0^\infty dr\, r\, G(r)^2J_0(\kappa r).
	\label{g3-deg}
	\ee
For $\bp=\bk'$ and $\bq=\bk$, $\cos\theta_q=s$ and $\kappa=2ks$. Substituting these equations in (\ref{J=}), we recover the expression (\ref{I-zero=}) for $I_0$.

\section{Geometric scattering due to Gaussian bumps}
\label{sec6}

Suppose that ${S}$ is a Gaussian bump characterized by 
	\begin{equation}
	z=f(r):=\delta\, e^{-r^2/2\sigma^2},
	\label{G-bump}
	\end{equation}
where $\delta$ and $\sigma$ are real parameters with the dimension of length. In order to obtain analytic formulas for the integrals appearing in the expressions for $I_0, I_{ij}$, and $I_{iji'j'}$, we introduce the dimensionless parameters,
	\be
	\eta:=\frac{\delta^2}{\sigma^2},\quad\quad\quad\quad\quad
	\fK:=k\sigma,
	\ee
and confine our attention to cases where $\eta$ is so small that we can expand the integrands in (\ref{g1-def}), (\ref{g2-def}), (\ref{I-11}), (\ref{I-1111}), and (\ref{g3-deg}) in powers of $\eta$ and neglect the quadratic and higher order terms. This allows for an analytic evaluation of the integrals entering these equations, and with the help of (\ref{I-zero=}) and (\ref{J=}) yields
	\bea
	I_0&=&\frac{\pi\,\eta\, e^{-s^2\fK^2}}{2} \Big[
	(4\lambda_1s^2-1)\fK^2+\lambda_2(s^4\fK^4+2)\Big]+\cO(\eta^2),
	\label{I-zero=1}\\
	I_{11}&=&\frac{\pi\,\eta\, e^{-\fK^2/2}}{2} \Big\{
	\Big[2(2\lambda_1-1)\fK^2+\lambda_2(\fK^4+4)\Big]
	\Big[I_0(\fK^2/2)-iK_0(\fK^2/2)/\pi\Big]\nn\\
	&&\hspace{2cm}-\fK^2
	\Big[4\lambda_1+\lambda_2(\fK^2+1)\Big]
	\Big[I_1(\fK^2/2)+iK_1(\fK^2/2)/\pi\Big]
	\Big\}+\cO(\eta^2)
	\label{I-11=1}\\
	I_{1111}&=&\eta\left\{\frac{-i\fK^2 e^{-\fK^2/2}}{2}\,
	\Big[
	\big(\fK^2-2+4(\lambda_1+\lambda_2)\big)K_0(\fK^2/2)+
	\big(\fK^2+4(\lambda_1+\lambda_2)\big)K_1(\fK^2/2)\Big]+\right.\nn\\
	&&\sqrt\pi\left[\big(\fK^2+4(\lambda_1+\lambda_2)\big)
	G^{3,1}_{3,4}\Big(\fK^2\Big|\begin{array}{c}
	-1,\frac{1}{2},\frac{1}{2}\\
	0,0,0,\frac{1}{2}\end{array}\Big)
	-4(\lambda_1+\lambda_2)
	G^{3,1}_{3,4}\Big(\fK^2\Big|\begin{array}{c}
	0,\frac{1}{2},\frac{1}{2}\\
	0,0,0,\frac{1}{2}\end{array}\Big)+\right. \nn\\
	&&\fK^2
	G^{3,1}_{3,4}\Big(\fK^2\Big|\begin{array}{c}
	-1,-\frac{1}{2},\frac{1}{2}\\
	-1,0,1,-\frac{1}{2}\end{array}\Big)+
	\lambda_2
	G^{3,1}_{3,4}\Big(\fK^2\Big|\begin{array}{c}
	-2,-\frac{1}{2},\frac{1}{2}\\
	0,0,0,\frac{1}{2}\end{array}\Big)+\nn\\
	&&\left.\left. i\fK^2 G^{2,0}_{1,2}\Big(\fK^2\Big|\begin{array}{c}
	\frac{1}{2}\\
	0,1\end{array}\Big)-i\lambda_2
	G^{2,1}_{2,3}\Big(\fK^2\Big|\begin{array}{c}
	-2,\frac{1}{2}\\
	0,0,0\end{array}\Big)\right]\right\}+\cO(\eta^2),
	\label{I-1111=1}\\
	J(\bp,\bq)&=&\frac{\pi\,\eta\, e^{-\kappa^2\sigma^2/4}}{32} \Big\{
	4q^2\sigma^2\Big[\kappa^2\sigma^2\cos\theta_q
	(2\cos\theta_q-\kappa/q)-4\Big]+\nn\\
	&&\hspace{2.5cm}
	16\lambda_1\kappa^2\sigma^2+\lambda_2(\kappa^4\sigma^4+32)
	\Big\}+\cO(\eta^2),
	\label{J=1}
	\eea
where $I_\ell(x)$ and $K_\ell(x)$ are respectively the $\ell$-th order modified Bessel functions of the first and second kind\footnote{For real $x$,  $I_\ell(x)=i^{-\ell}J_\ell(ix)$ and $K_\ell(x)=\pi i^{\ell+1}H^{(1)}_\ell(ix)/2$.}, $G^{m,n}_{p,q}\Big(x\Big|\begin{array}{c}
	a_1,\cdots,a_p\\
	b_1,\cdots,b_q\end{array}\Big)$ labels the Meijer G function \cite{GR-table}, and $\cO(\eta^2)$ stands for the quadratic and higher order terms in powers of $\eta$.

\begin{figure}[ht]
    	\begin{center}
    \includegraphics[scale=.65]{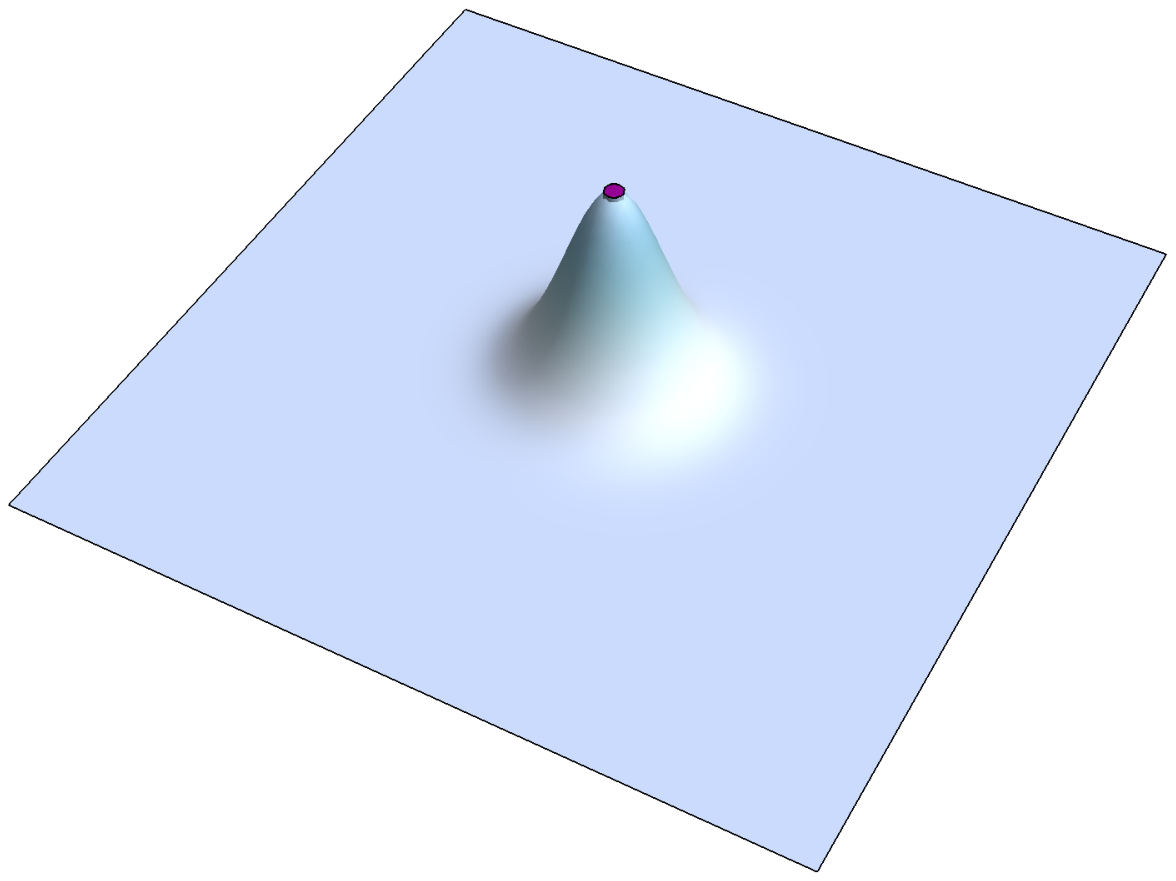}~~~~
    \includegraphics[scale=.65]{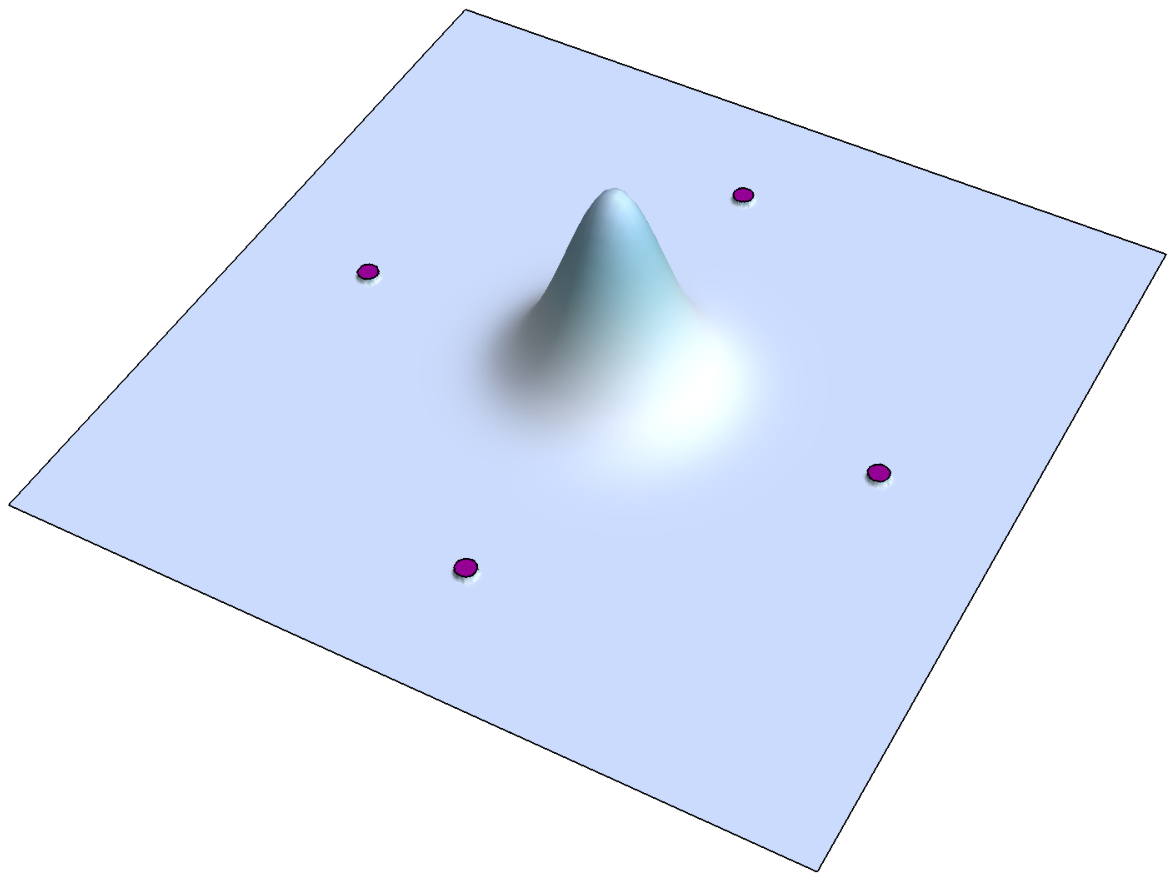}
    \caption{Schematic view of a Gaussian bump with a central point defect (on the left) and a Gaussian bump surrounded by four distant point defects (on the right).}
    \label{fig1}
    \end{center}
    \end{figure}

Substituting (\ref{I-zero=1}) -- (\ref{I-1111=1}) in (\ref{f1=Is}) and making use of (\ref{f-for-single-delta}) with $\bba_1=(0,0)$, we obtain the scattering amplitude for the Gaussian bump (\ref{G-bump}) with a point defect placed at its peak. Similarly, using (\ref{J=1}) in (\ref{I-ij-far}) and (\ref{I-ijij-far}) we can determine the scattering amplitude for the same Gaussian bump in the presence of a number of defects that are sufficiently far from its peak. See Fig.~\ref{fig1}. 
    
\begin{figure}[!ht]
	\begin{center}
	\includegraphics[scale=.6]{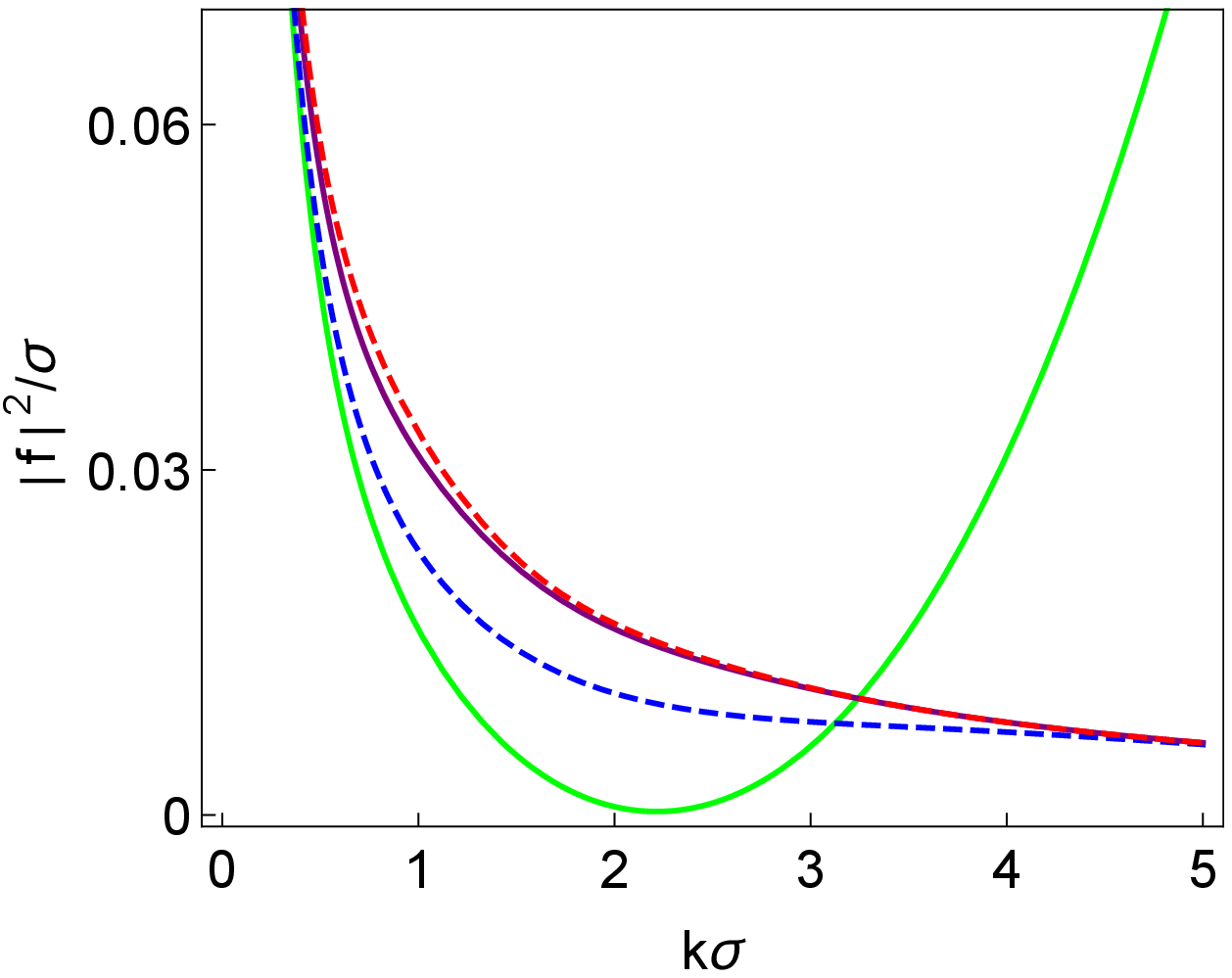}~~~~~~~~
	\includegraphics[scale=.62]{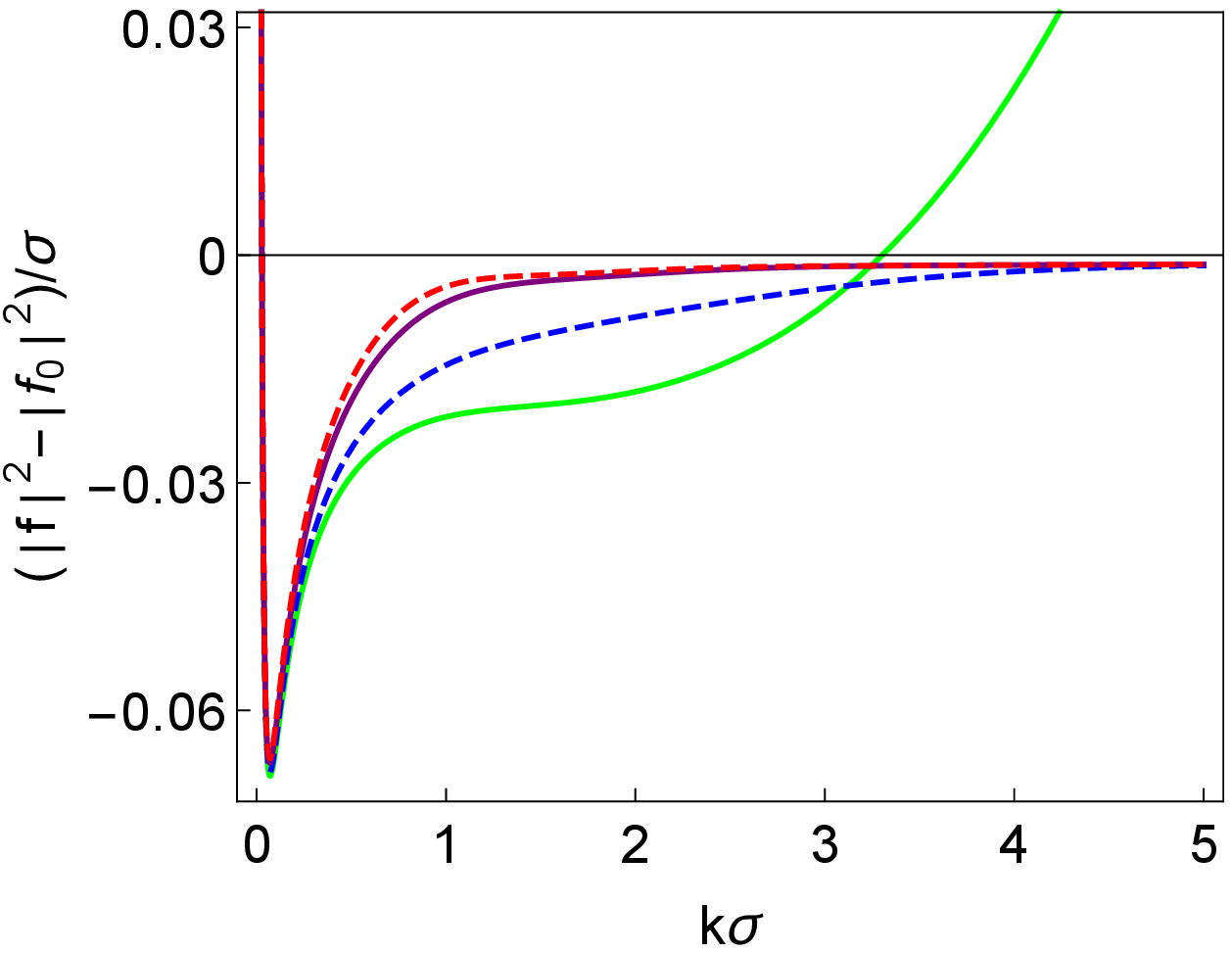}
	\caption{Plots of $|f(\boldsymbol{k}',\boldsymbol{k})|^2/\sigma$ and
	$\left[|f(\boldsymbol{k}',\boldsymbol{k})|^2-|f_0(\boldsymbol{k}',	
	\boldsymbol{k})|^2\right]/\sigma$ for a Gaussian bump (\ref{G-bump}) with a central defect as functions of $k\sigma$ for $\theta_0=0$, $\eta=0.1$, $\tilde\xi=\hbar^2/2m$, $\lambda_1=-\lambda_2=1/2$, and different values of $\theta$, namely $\theta=0$ (green), $\pi/3$ (dashed blue), $2\pi/3$ (purple), and $\pi$ (dashed red).}
	\label{fig2a}
	\end{center}
	\end{figure}

\begin{figure}[!ht]
	\begin{center}
	\includegraphics[scale=.60]{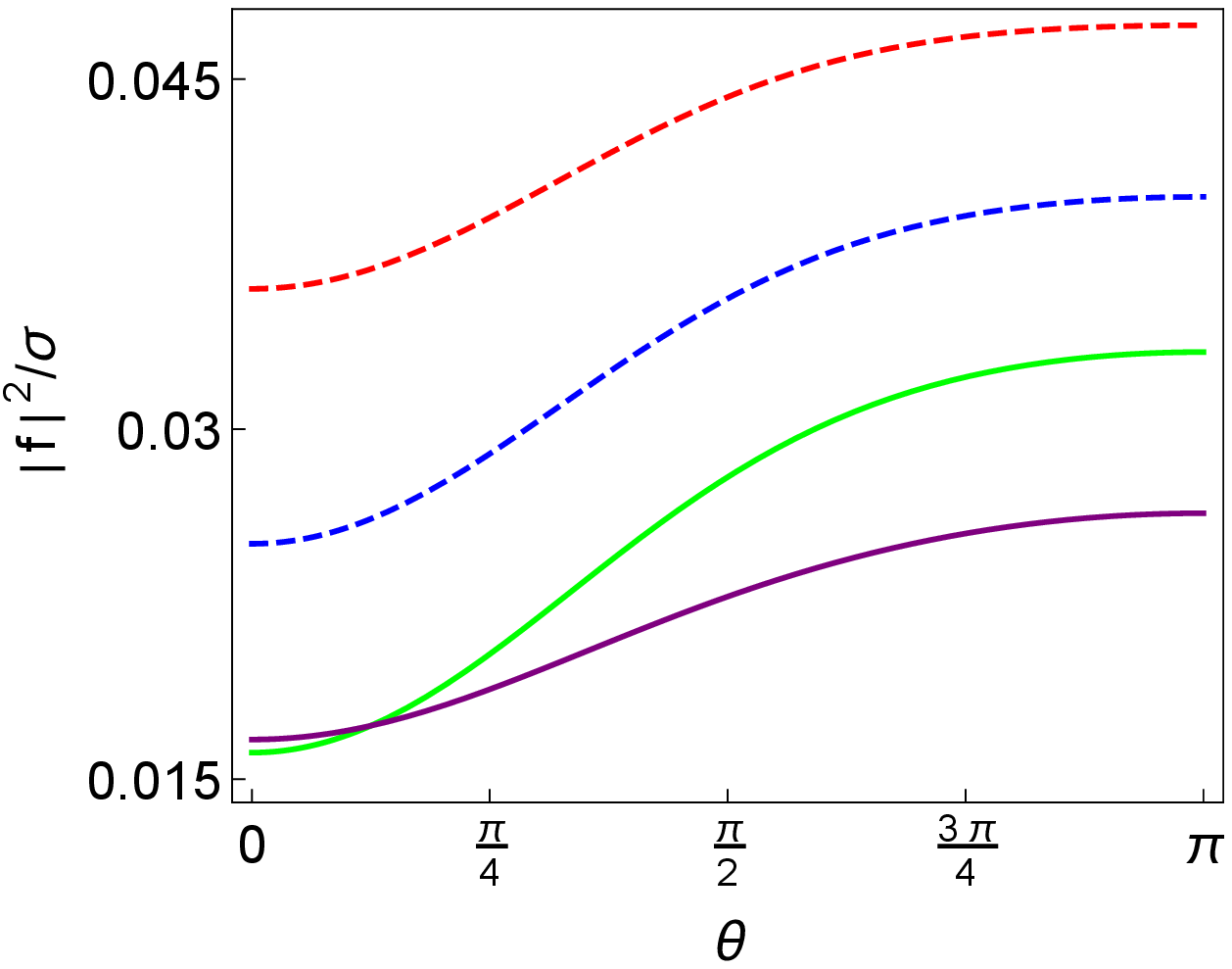}~~~~~~~~
	\includegraphics[scale=.62]{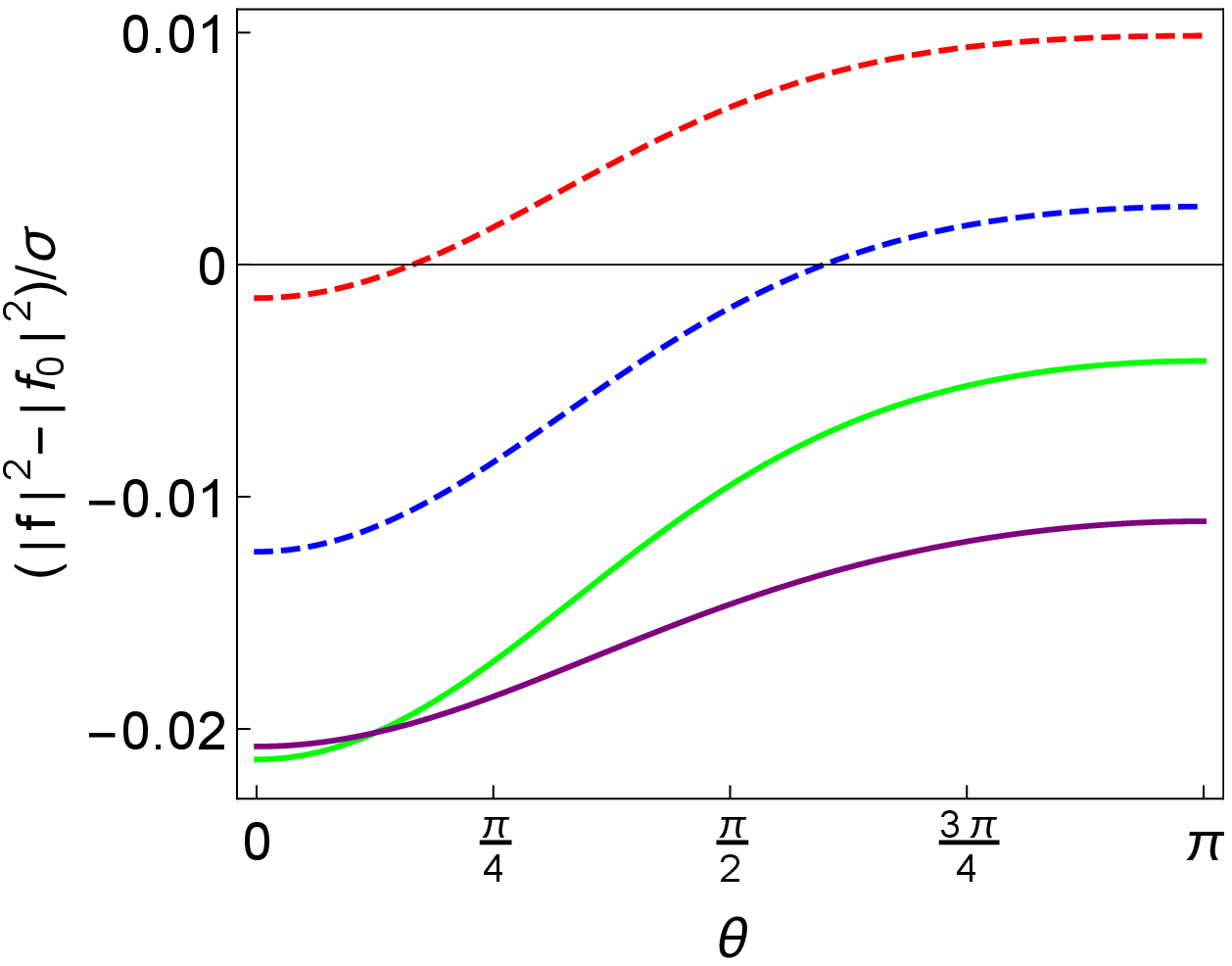}
	\caption{Plots of $|f(\boldsymbol{k}',\boldsymbol{k})|^2/\sigma$ and
	$\left[|f(\boldsymbol{k}',\boldsymbol{k})|^2-|f_0(\boldsymbol{k}',	
	\boldsymbol{k})|^2\right]/\sigma$ for a Gaussian bump (\ref{G-bump}) with a central defect as functions of $\theta$ for $\theta_0=0$, $\eta=0.1$, $\tilde\xi=\hbar^2/2m$, $k\sigma=1$, and different values of $\lambda_1$ and $\lambda_2$, namely $\lambda_1=-\lambda_2=1/2$ (green), $\lambda_1=0$ and $\lambda_2=-1/2$ (dashed blue), $\lambda_1=1/2$ and $\lambda_2=0$ (purple), and $\lambda_1=\lambda_2=1/2$ (dashed red).}
		\label{fig2b}
	\end{center}
	\end{figure}

\begin{figure}[!ht]
	\begin{center}
	\includegraphics[scale=.6]{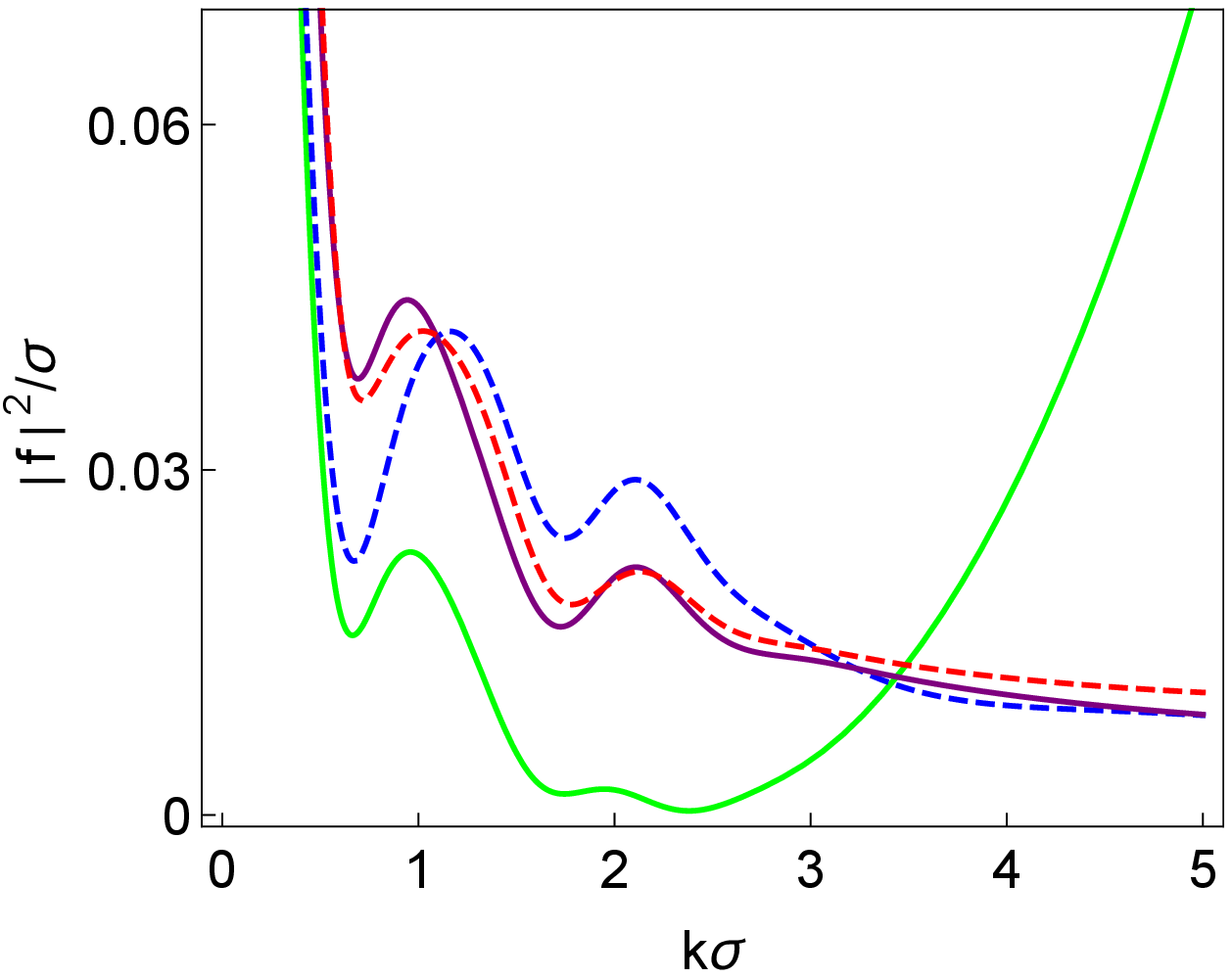}~~~~~~~~
	\includegraphics[scale=.62]{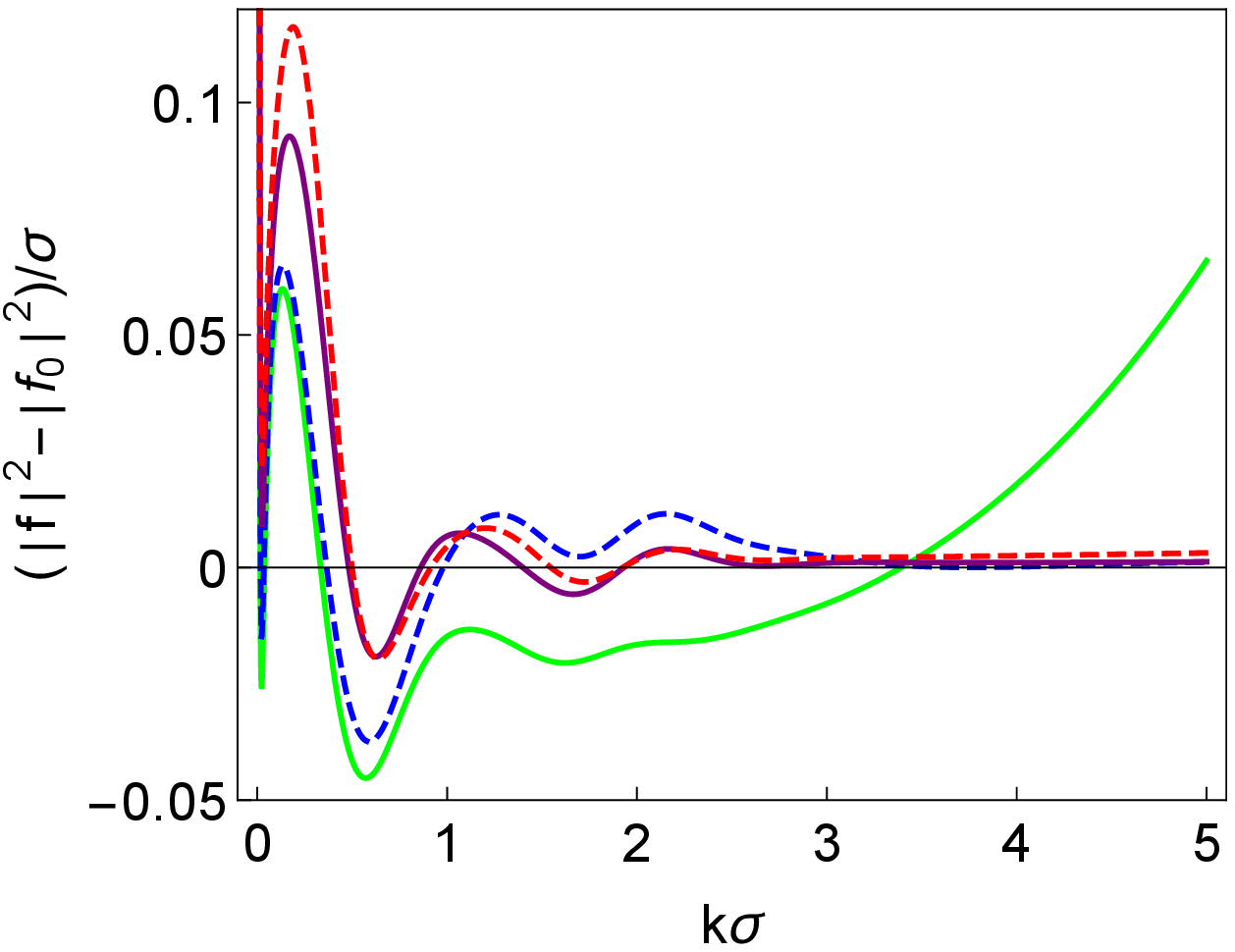}
	\caption{Plots of $|f(\boldsymbol{k}',\boldsymbol{k})|^2/\sigma$ and $\left[|f(\boldsymbol{k}',\boldsymbol{k})|^2-|f_0(\boldsymbol{k}',\boldsymbol{k})|^2\right]/\sigma$ for a Gaussian bump (\ref{G-bump}) with a 	distant defect located at $\bba=(3\sigma,0)$ as functions of $k\sigma$ for $	\theta_0=0$, $\eta=0.1$, $\tilde\xi=\hbar^2/2m$, and $\lambda_1=-\lambda_2 =1/2$, and different values of $\theta$, namely $\theta=0$ (green), $\pi/3$ (dashed blue), $2\pi/3$ (purple), and $\pi$ (dashed red).}
	\label{fig3a}
	\end{center}
	\end{figure}

\begin{figure}[!ht]
	\begin{center}
	\includegraphics[scale=.6]{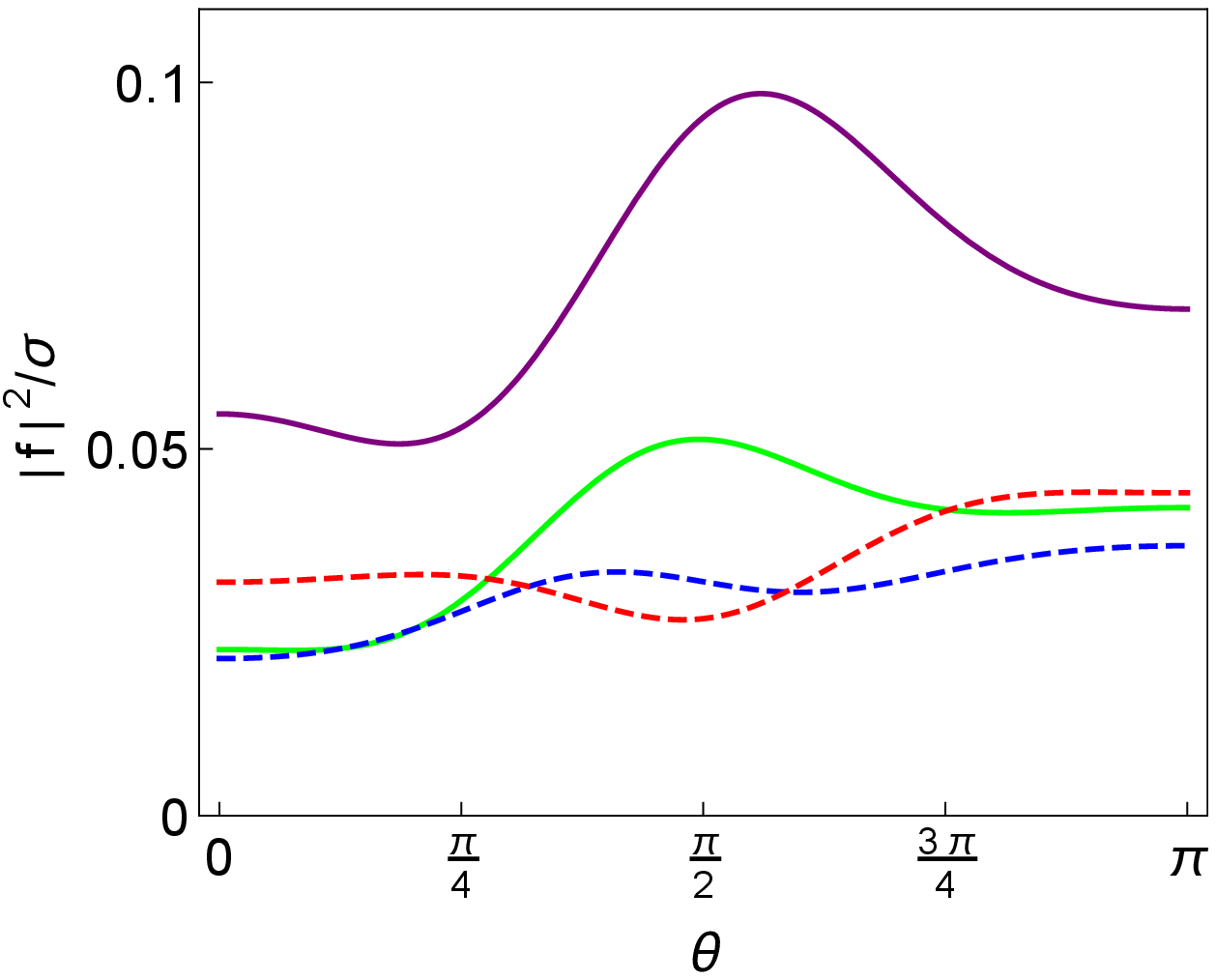}~~~~~~~~
	\includegraphics[scale=.62]{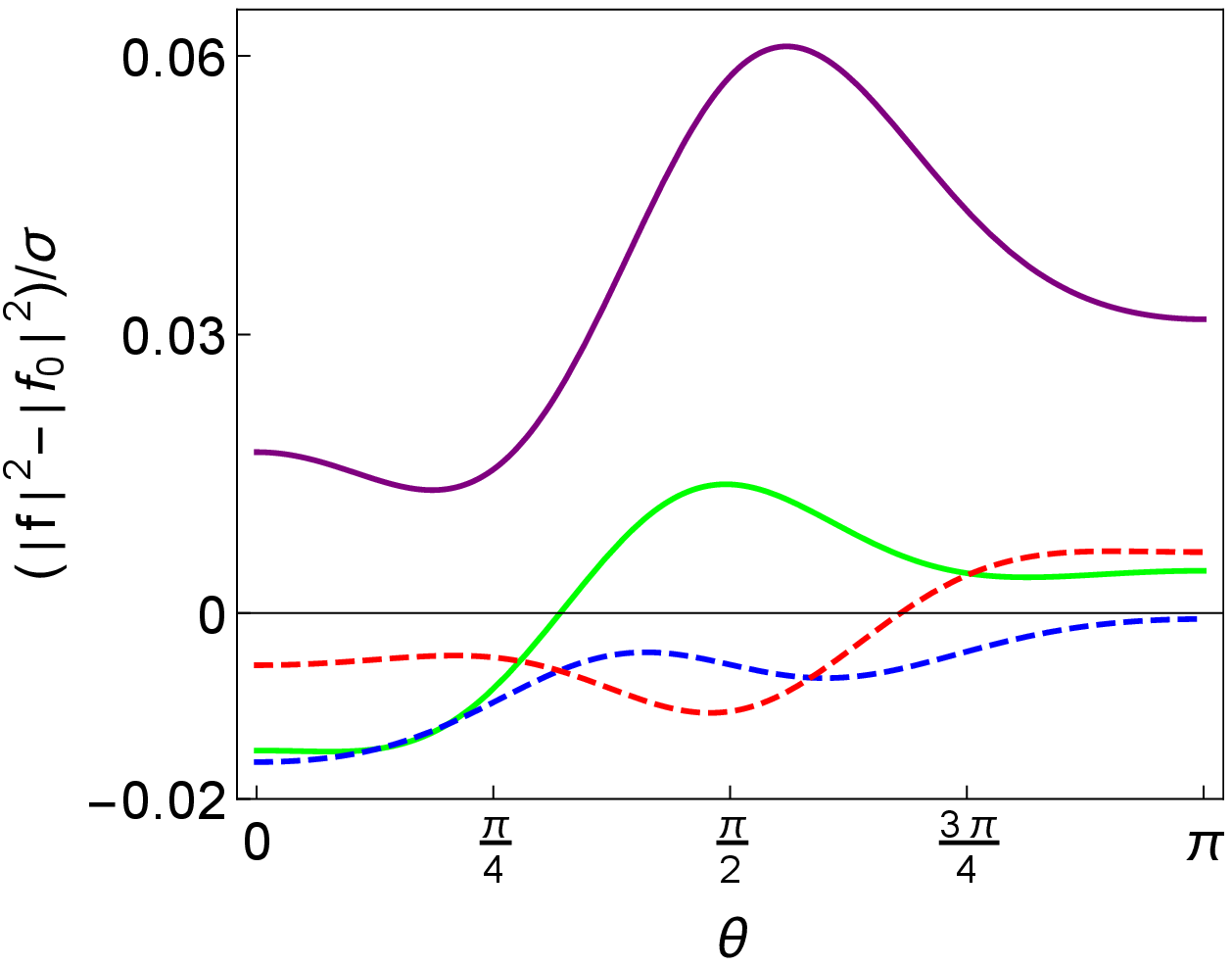}
	\caption{Plots of $|f(\boldsymbol{k}',\boldsymbol{k})|^2/\sigma$ and
	$\left[|f(\boldsymbol{k}',\boldsymbol{k})|^2-|f_0(\boldsymbol{k}',	
	\boldsymbol{k})|^2\right]/\sigma$ for a Gaussian bump (\ref{G-bump}) with a distant defect located at $\bba=(3\sigma,0)$ as functions of $\theta$ for $\theta_0=0$, $\eta=0.1$, $\tilde\xi=\hbar^2/2m$, $k\sigma=1$, and different values of $\lambda_1$ and $\lambda_2$, namely $\lambda_1=-\lambda_2=1/2$
	(green), $\lambda_1=0$ and $\lambda_2=-1/2$ (dashed blue), $\lambda_1=1/2$ and $\lambda_2=0$ (purple), and $\lambda_1=\lambda_2=1/2$ (dashed red).}
	\label{fig3b}
	\end{center}
	\end{figure}

\begin{figure}[!ht]
	\begin{center}
	\includegraphics[scale=.6]{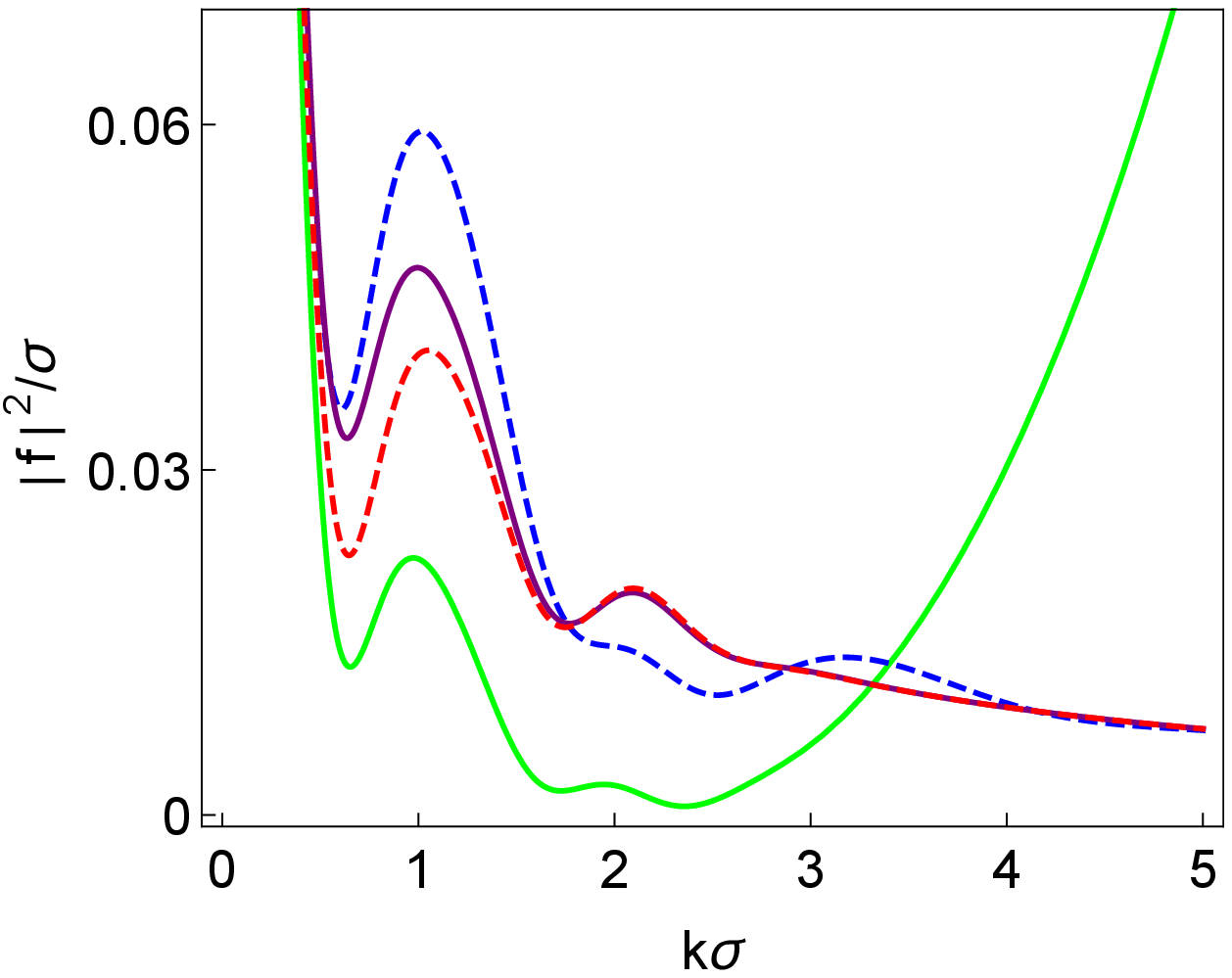}~~~~~~~~
	\includegraphics[scale=.62]{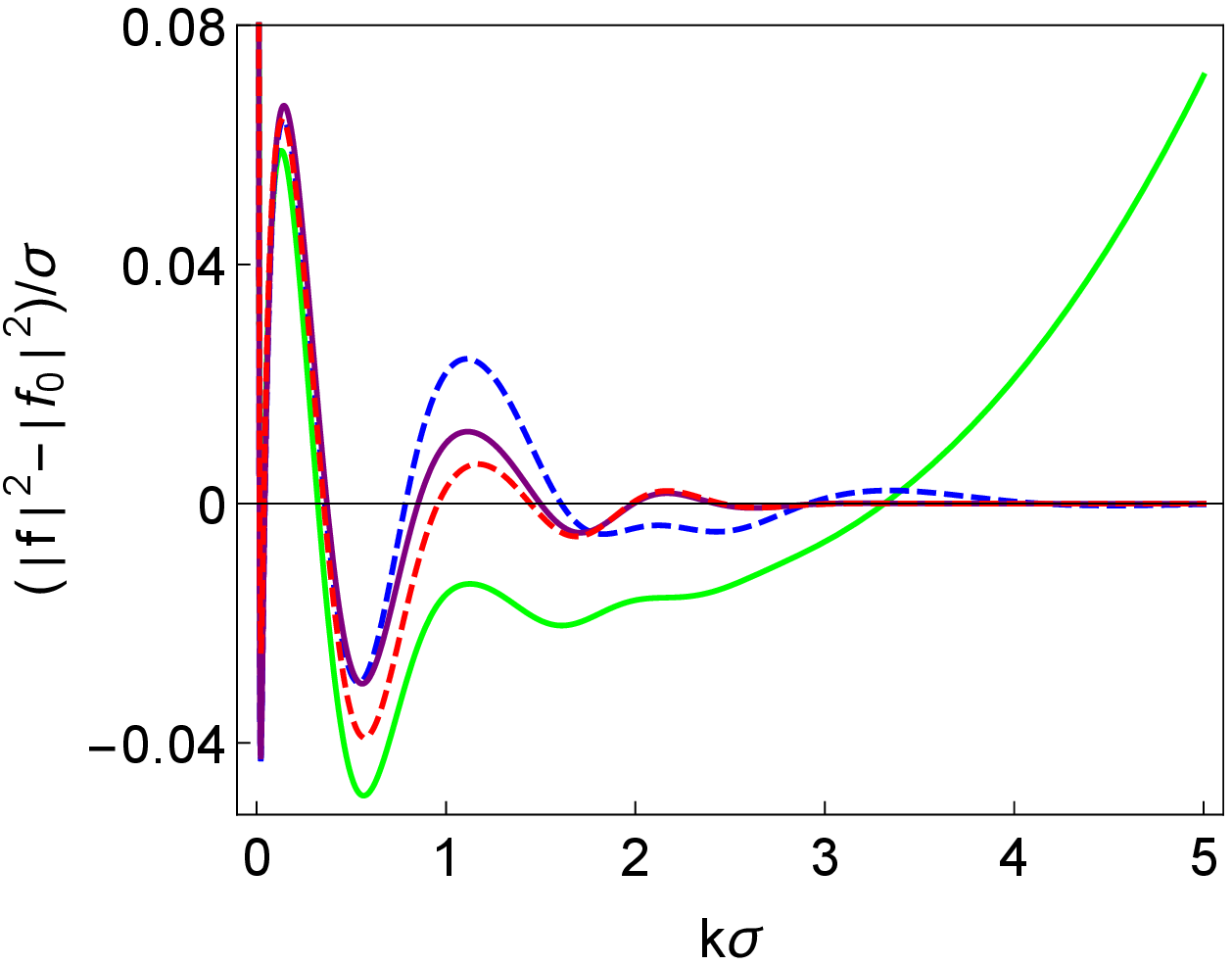}
	\caption{Plots of $|f(\boldsymbol{k}',\boldsymbol{k})|^2/\sigma$ and $\left[|f(\boldsymbol{k}',\boldsymbol{k})|^2-|f_0(\boldsymbol{k}',\boldsymbol{k})|^2\right]/\sigma$ for a Gaussian bump (\ref{G-bump}) with a distant defect located at $\bba=(0,3\sigma)$ as functions of $k\sigma$ for $	\theta_0=0$, $\eta=0.1$, $\tilde\xi=\hbar^2/2m$, and $\lambda_1=-\lambda_2 =1/2$, and different values of $\theta$, namely $\theta=0$ (green), $\pi/3$ (dashed blue), $2\pi/3$ (purple), and $\pi$ (dashed red)}
	\label{fig4a}
	\end{center}
\end{figure}

\begin{figure}[!ht]
	\begin{center}
		\includegraphics[scale=.6]{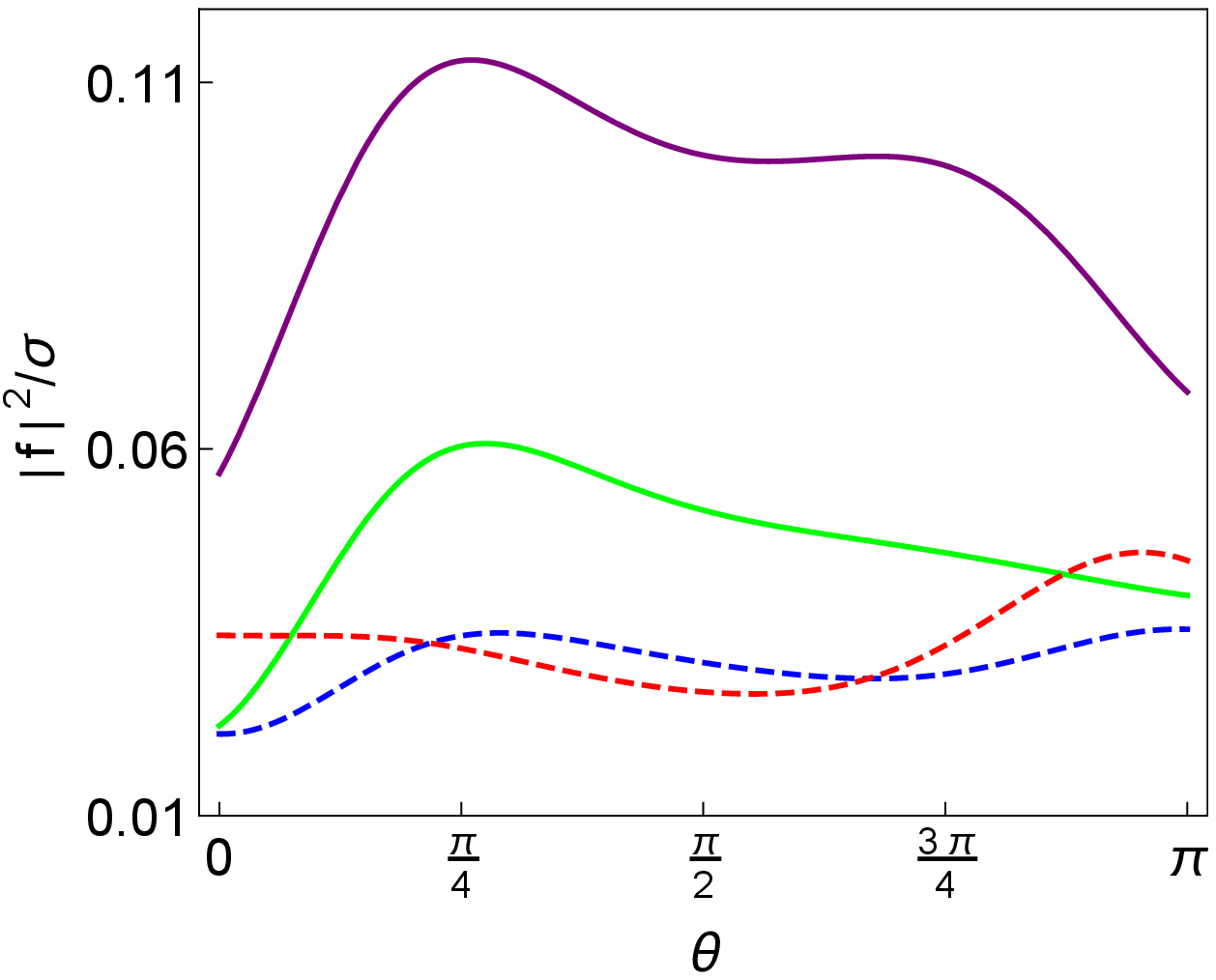}~~~~~~~~
		\includegraphics[scale=.62]{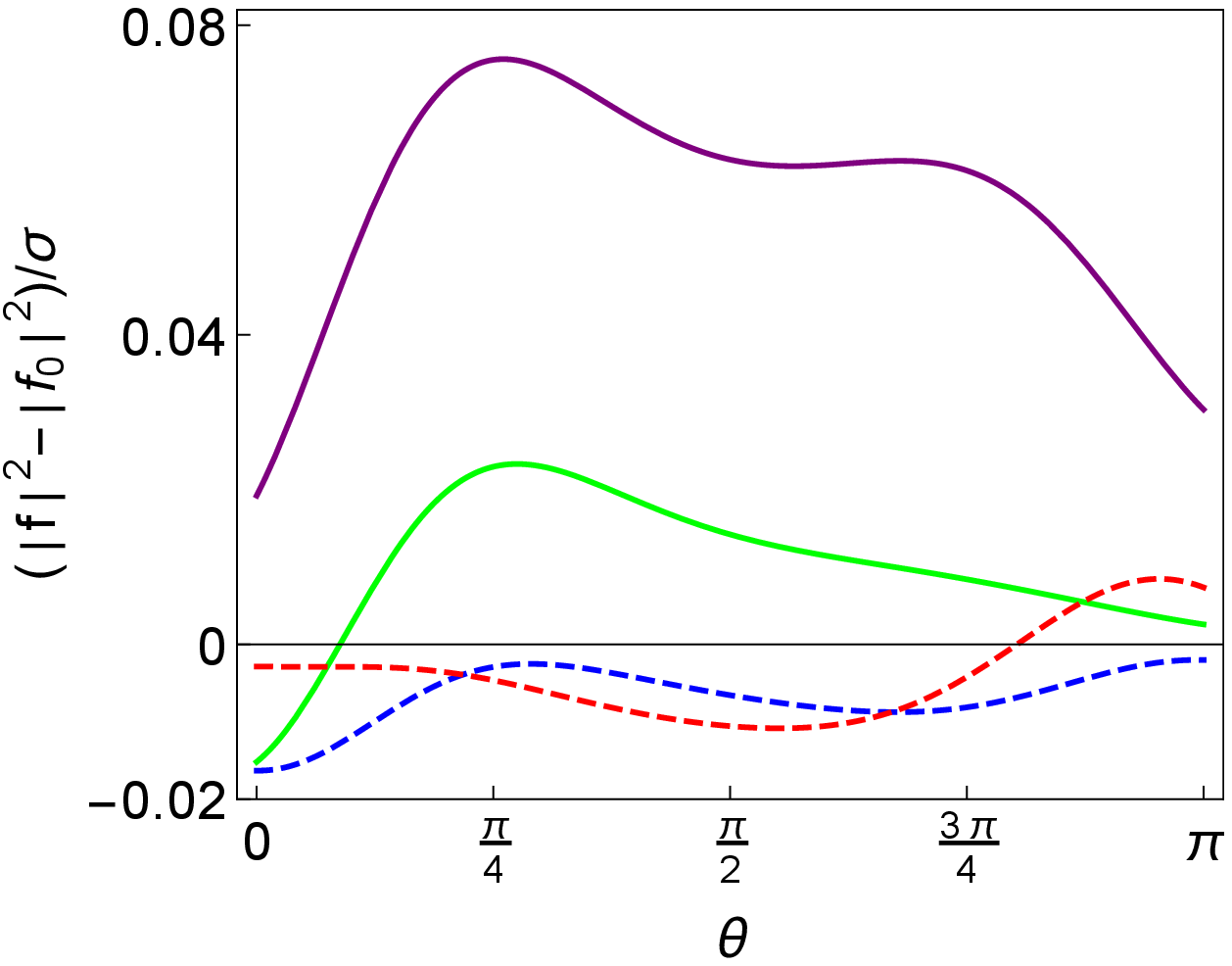}
		\caption{Plots of $|f(\boldsymbol{k}',\boldsymbol{k})|^2/\sigma$ and
	$\left[|f(\boldsymbol{k}',\boldsymbol{k})|^2-|f_0(\boldsymbol{k}',	
	\boldsymbol{k})|^2\right]/\sigma$ for a Gaussian bump (\ref{G-bump}) with a distant defect located at $\bba=(0,3\sigma)$ as functions of $\theta$ for $\theta_0=0$, $\eta=0.1$, $\tilde\xi=\hbar^2/2m$, $k\sigma=1$, and different values of $\lambda_1$ and $\lambda_2$, namely $\lambda_1=-\lambda_2=1/2$ 
	(green), $\lambda_1=0$ and $\lambda_2=-1/2$ (dashed blue), $\lambda_1=1/2$ and $\lambda_2=0$ (purple), and $\lambda_1=\lambda_2=1/2$ (dashed red).}
	\label{fig4b}
	\end{center}
	\end{figure}

Figures~\ref{fig2a}, \ref{fig2b}, \ref{fig3a}, \ref{fig3b}, \ref{fig4a}, and \ref{fig4b} show the plots of the differential cross section $|\ff(\bk',\bk)|^2$ and its difference with $|\ff_0(\bk',\bk)|^2$ for a Gaussian bump in the presence of a central or distant point defect. Here we have taken $\tilde\xi=\hbar^2/2m$. To see if this is a reasonable choice, we consider an electron gas and identify the defect with a barrier potential (respectively potential well) of height (respectively depth) $V_0:=|\tilde\xi|/\rho^2\approx~3~{\rm eV}$ and cross section $\rho^2\approx 1~{\rm nm}^2$. For an electron having an effective mass $m\approx10^{-2}m_{\rm e}$, this corresponds to taking $\tilde\xi_j\approx\pm\hbar^2/2m$. A delta-function potential provides a reliable approximation for such a barrier potential (resp.\ potential well) if the energy $E:=\hbar^2k^2/2m$ of the incident electron is much smaller than $V_0\approx 3~{\rm eV}$  and its de~Broglie wavelength $\lambda:=2\pi/k=2\pi\hbar/\sqrt{2mE}$ is much larger than $\rho\approx 1~{\rm nm}$. For $m\approx10^{-2}m_{\rm e}$, the latter condition is equivalent to $E\ll 100~{\rm eV}$, which agrees with the requirement that $E\ll V_0$. 
		
\begin{figure}[!ht]
	\begin{center}
	\includegraphics[scale=.6]{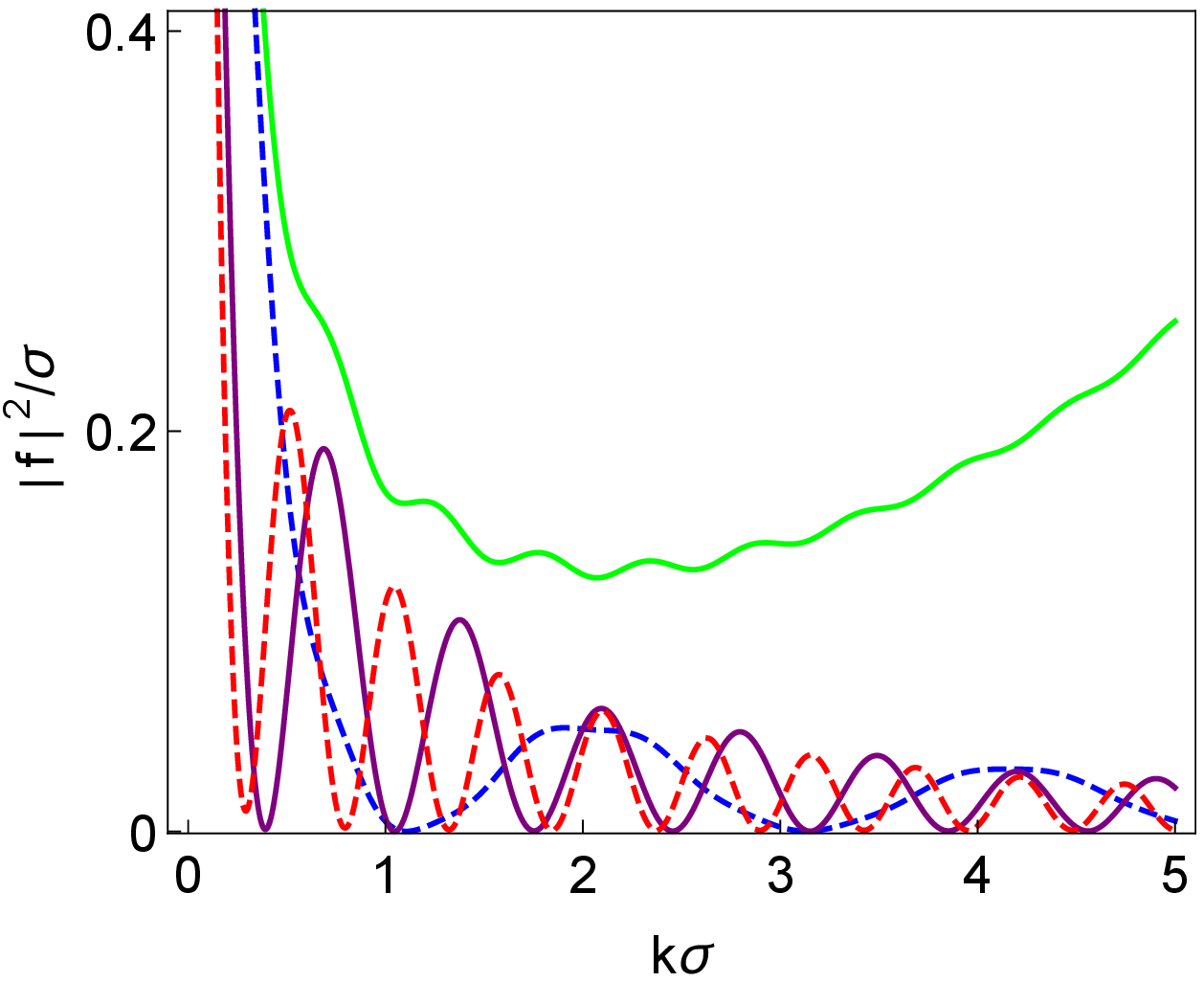}~~~~~~~~
	\includegraphics[scale=.64]{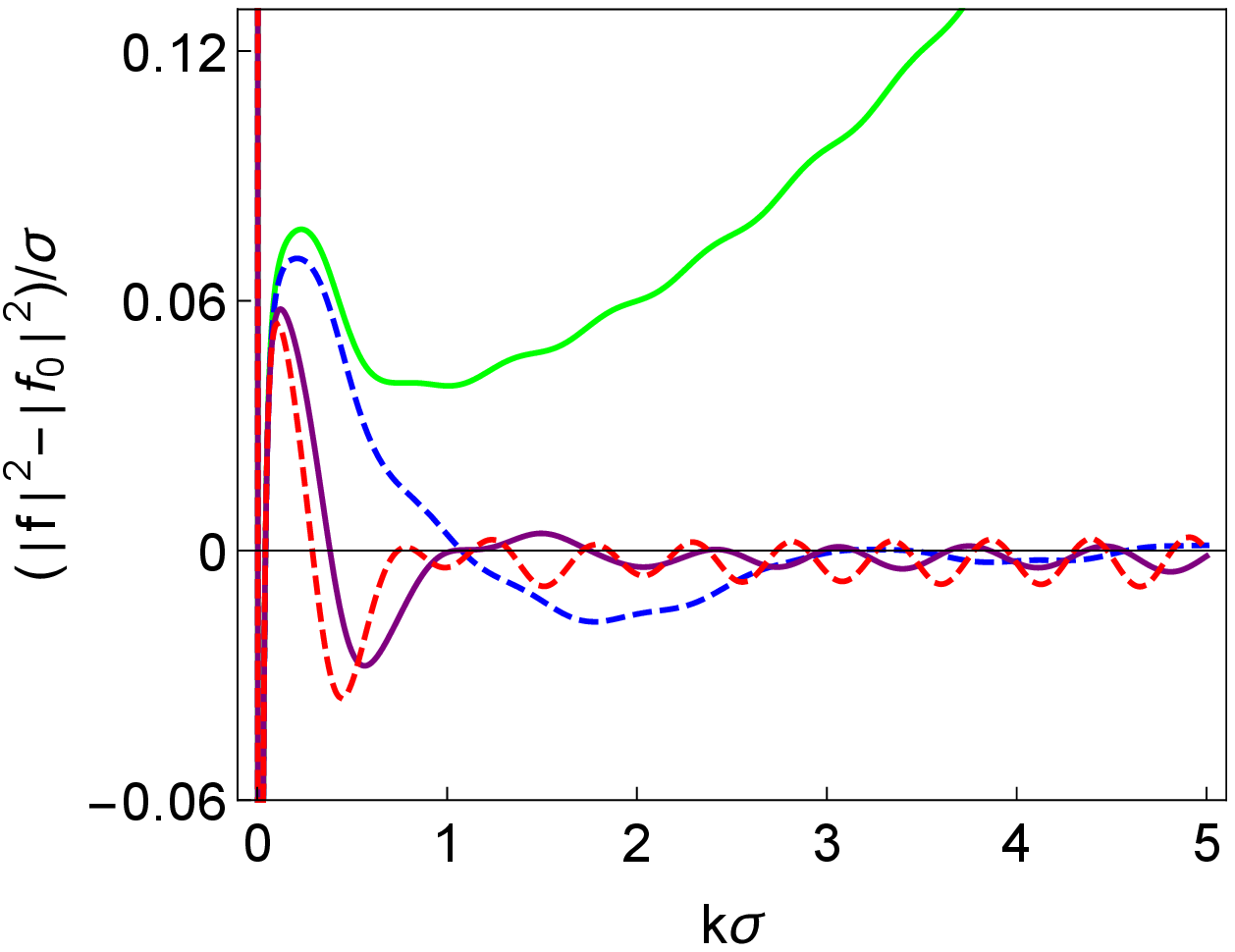}
	\caption{Plots of $|f(\boldsymbol{k}',\boldsymbol{k})|^2/\sigma$ and $\left[|f(\boldsymbol{k}',\boldsymbol{k})|^2-|f_0(\boldsymbol{k}',\boldsymbol{k})|^2\right]/\sigma$ for a Gaussian bump (\ref{G-bump}) with a pair of distant defects located at $\bba_1=(-3\sigma,0)$ and $\bba_2=(3\sigma,0)$ as functions of $k\sigma$ for $	\theta_0=0$, $\eta=0.1$, $\tilde\xi=\hbar^2/2m$, and $\lambda_1=-\lambda_2 =1/2$, and different values of $\theta$, namely $\theta=0$ (green), $\pi/3$ (dashed blue), $2\pi/3$ (purple), and $\pi$ (dashed red).}
	\label{fig5a}
	\end{center}
	\end{figure}

\begin{figure}[!ht]
	\begin{center}
	\includegraphics[scale=.6]{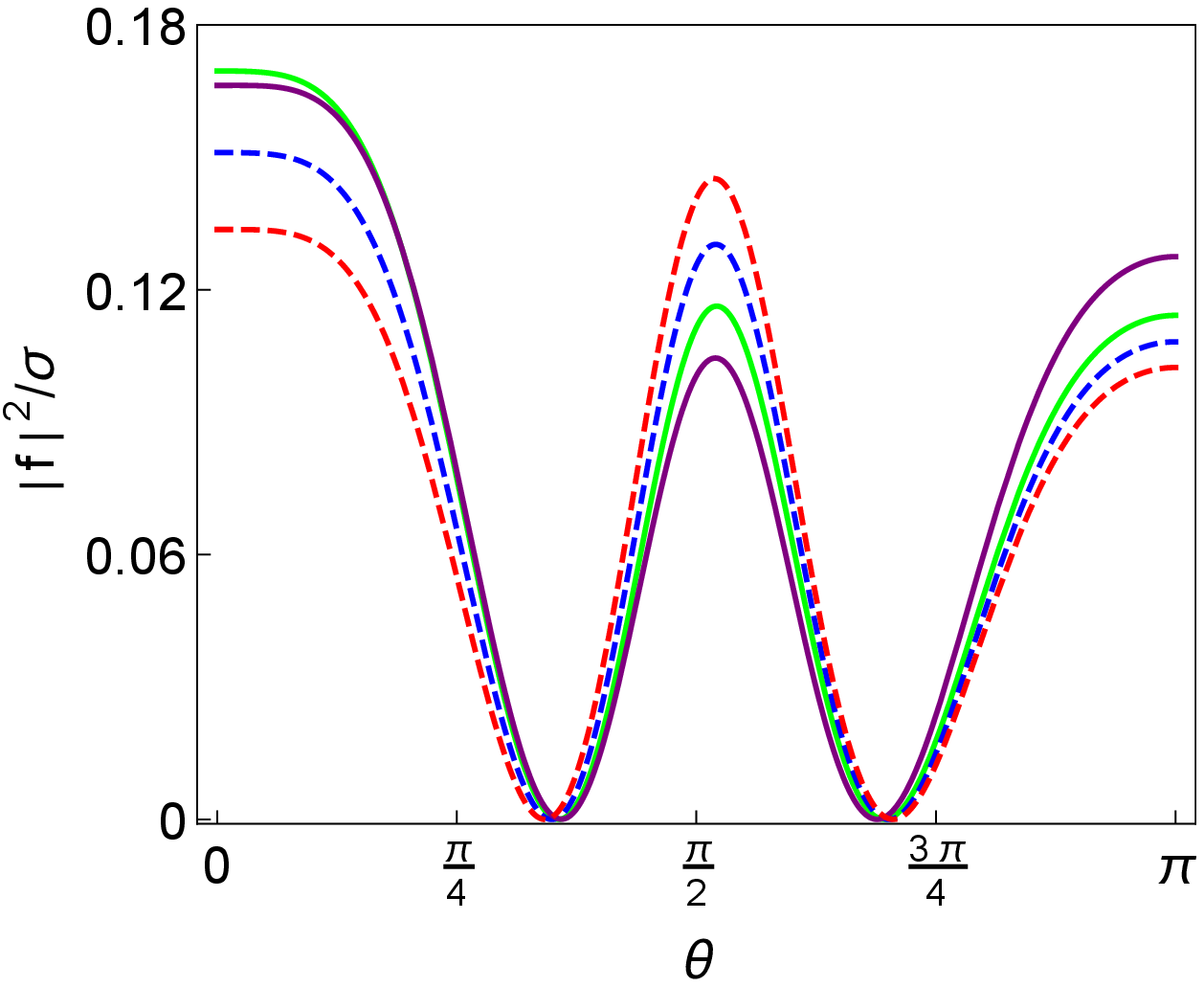}~~~~~~~~
	\includegraphics[scale=.64]{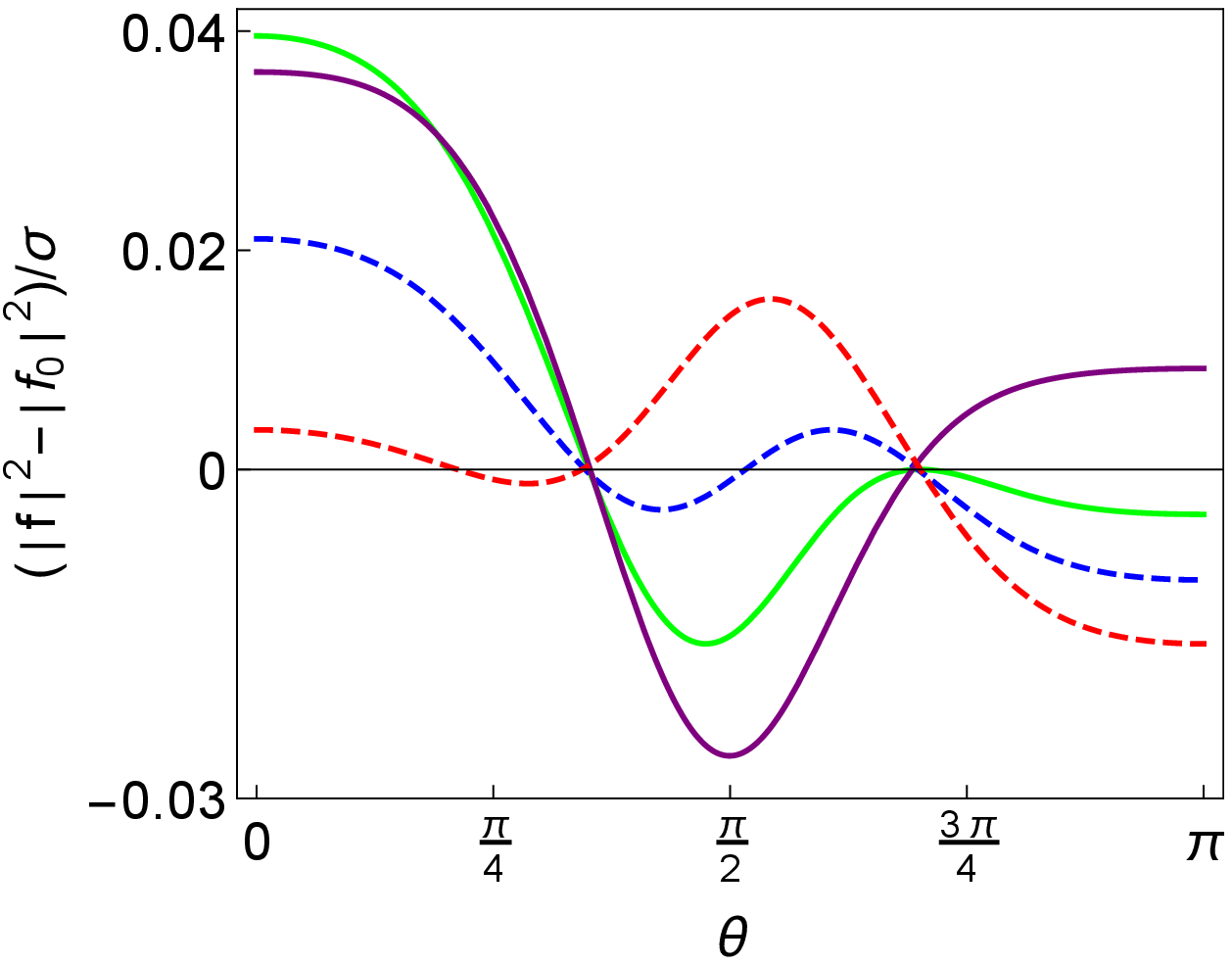}
	\caption{Plots of $|f(\boldsymbol{k}',\boldsymbol{k})|^2/\sigma$ and
	$\left[|f(\boldsymbol{k}',\boldsymbol{k})|^2-|f_0(\boldsymbol{k}',	
	\boldsymbol{k})|^2\right]/\sigma$ for a Gaussian bump (\ref{G-bump}) with a pair of distant defects located at $\bba_1=(-3\sigma,0)$ and $\bba_2=(3\sigma,0)$ as functions of $\theta$ for $\theta_0=0$, $\eta=0.1$, $\tilde\xi=\hbar^2/2m$, $k\sigma=1$, and different values of $\lambda_1$ and $\lambda_2$, namely $\lambda_1=-\lambda_2=1/2$ (green), $\lambda_1=0$ and $\lambda_2=-1/2$ (dashed blue), $\lambda_1=1/2$ and $\lambda_2=0$ (purple), and $\lambda_1=\lambda_2=1/2$ (dashed red).}
	\label{fig5b}
	\end{center}
	\end{figure}

Figures~\ref{fig5a} and  \ref{fig5b} 
show the plots of the differential cross section $\ff(\bk',\bk)|^2$ and its difference with $|\ff_0(\bk',\bk)|^2$ for a Gaussian bump in the presence of a pair of distant defects located at the points $(\pm3\sigma,0)$. Placing the defects at $(0,\pm 3\sigma)$, we obtain similar graphs that we do not include here for lack of space. 

\begin{figure}[!ht]
 	\begin{center}
 	\includegraphics[scale=.6]{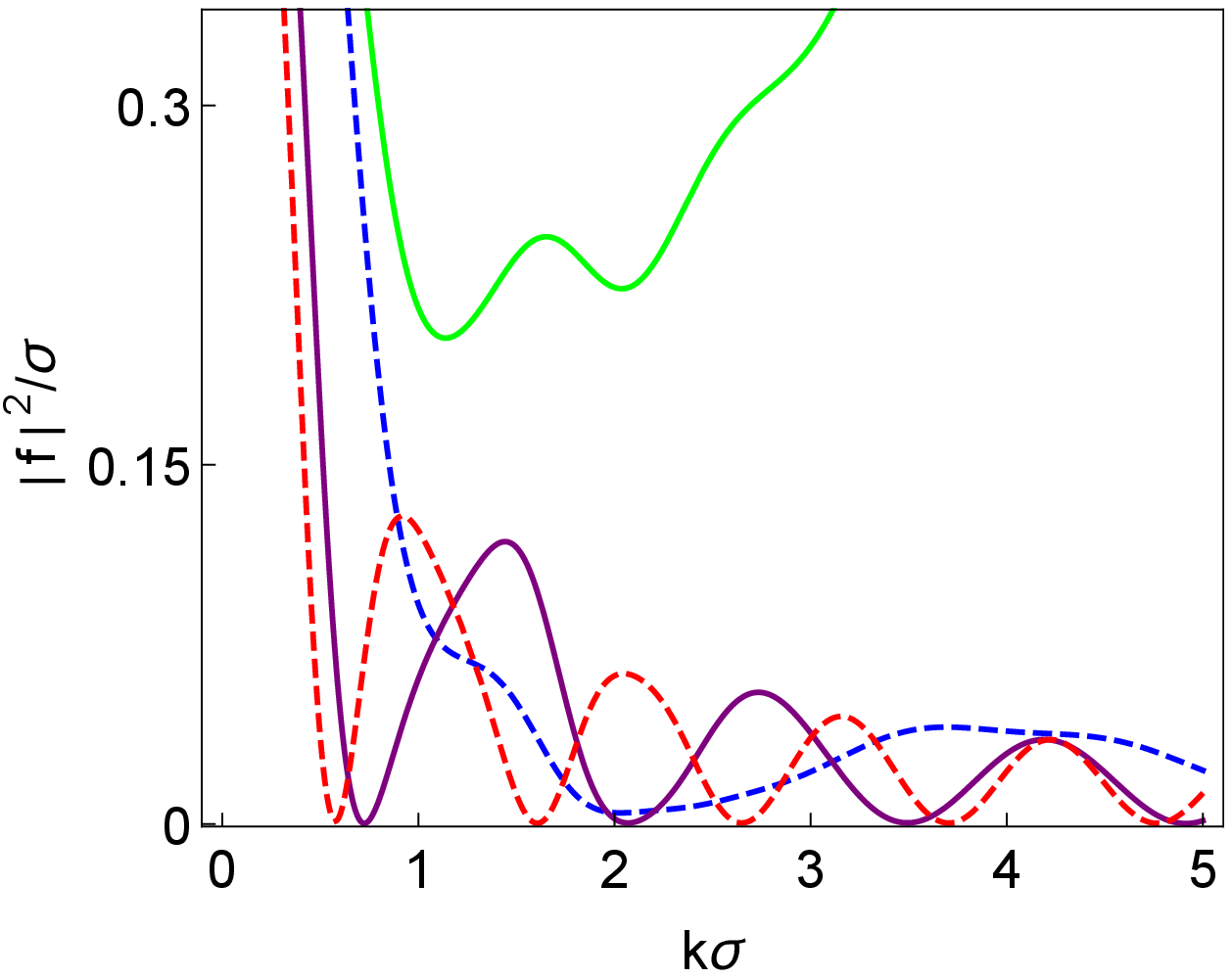}~~~~~~~~
 	\includegraphics[scale=.6]{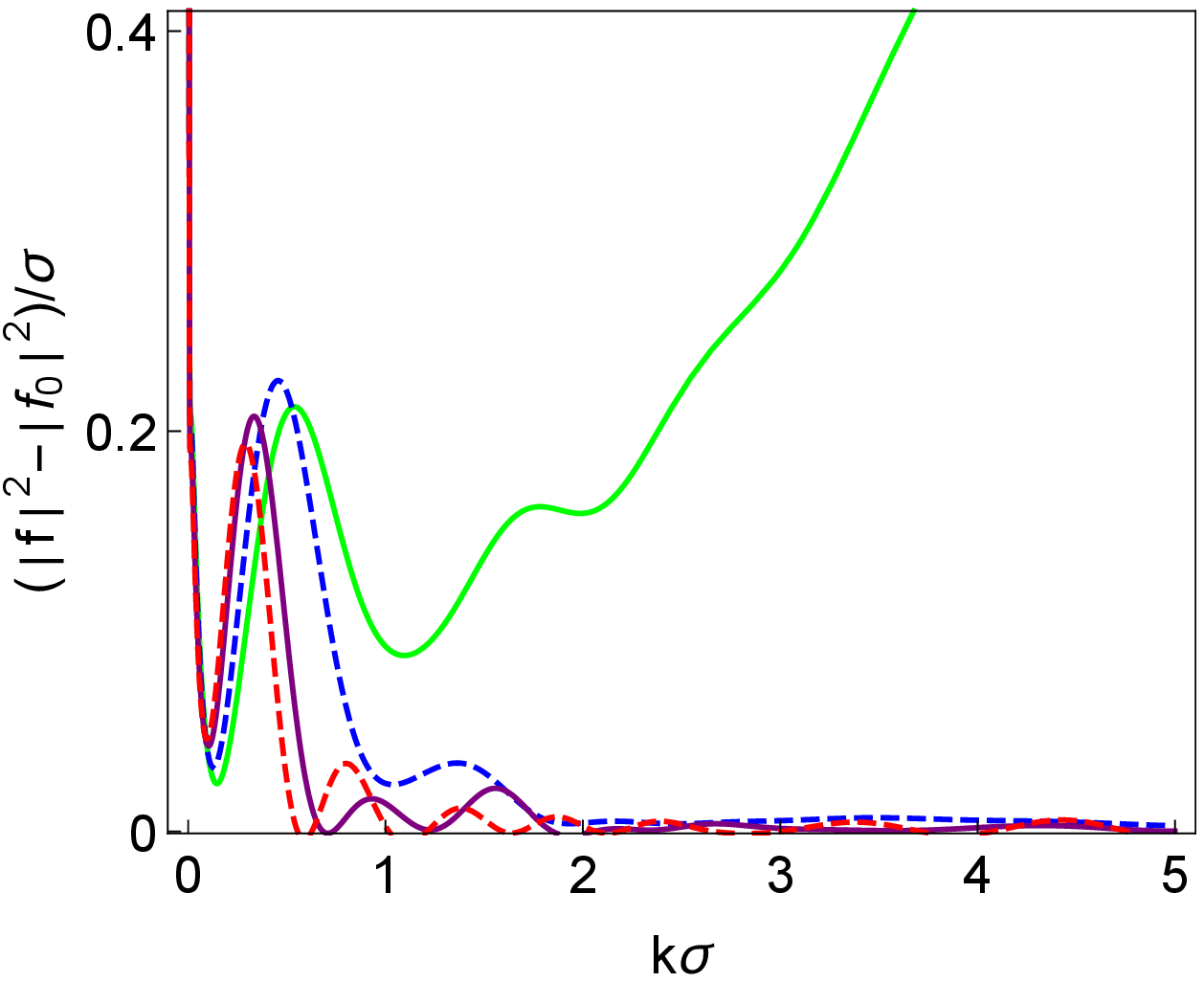}
 	\caption{Plots of $|f(\boldsymbol{k}',\boldsymbol{k})|^2/\sigma$ and $\left[|f(\boldsymbol{k}',\boldsymbol{k})|^2-|f_0(\boldsymbol{k}',\boldsymbol{k})|^2\right]/\sigma$ for a Gaussian bump (\ref{G-bump}) with a central defect and a distant defect located at $\bba=(3\sigma,0)$ as functions of $k\sigma$ for $	\theta_0=0$, $\eta=0.1$, $\tilde\xi=\hbar^2/2m$, and $\lambda_1=-\lambda_2 =1/2$, and different values of $\theta$, namely $\theta=0$ (green), $\pi/3$ (dashed blue), $2\pi/3$ (purple), and $\pi$ (dashed red).}
	\label{fig7a}
 	\end{center}
 	\end{figure}

\begin{figure}[!ht]
 	\begin{center}
 	\includegraphics[scale=.6]{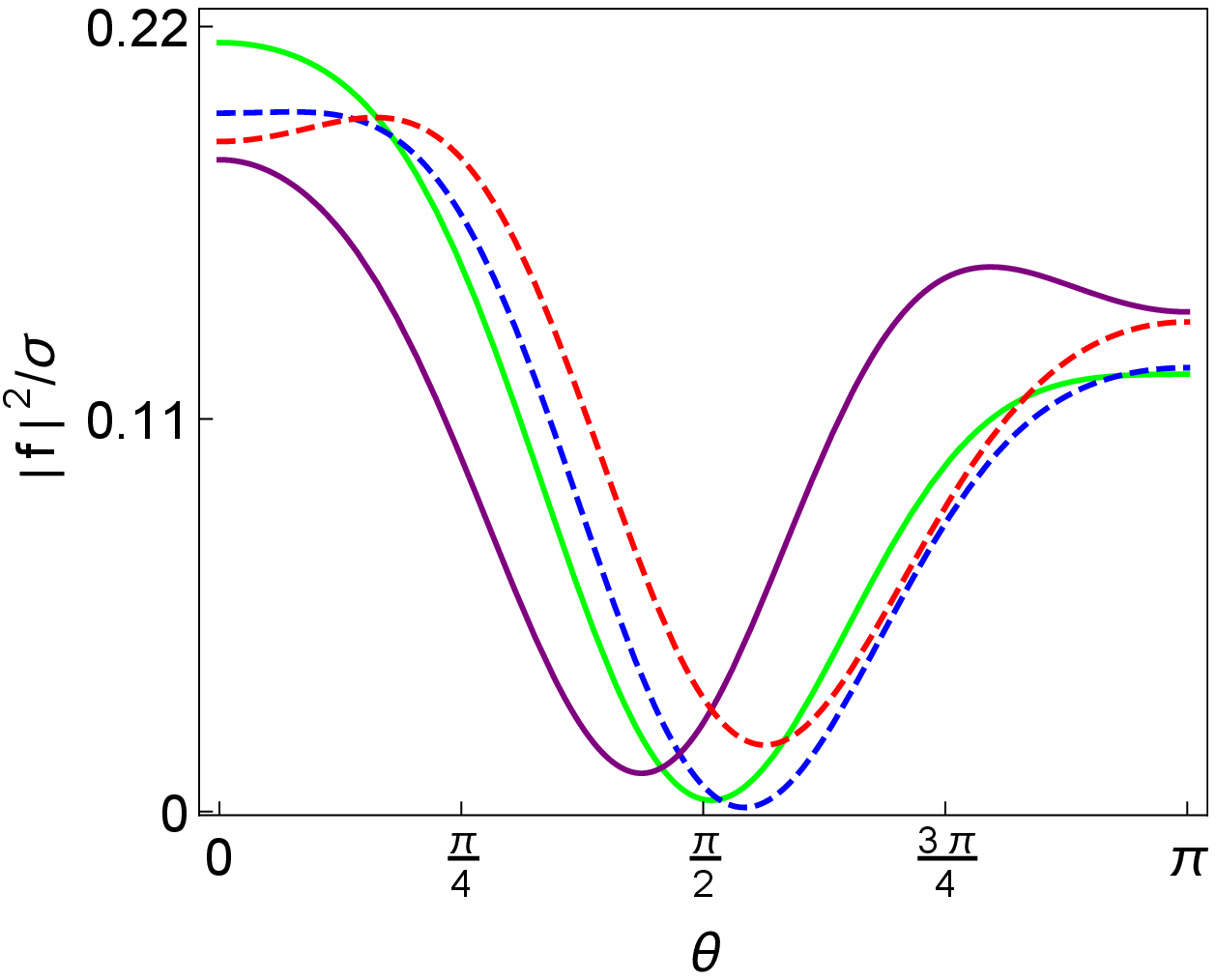}~~~~~~~~
 	\includegraphics[scale=.62]{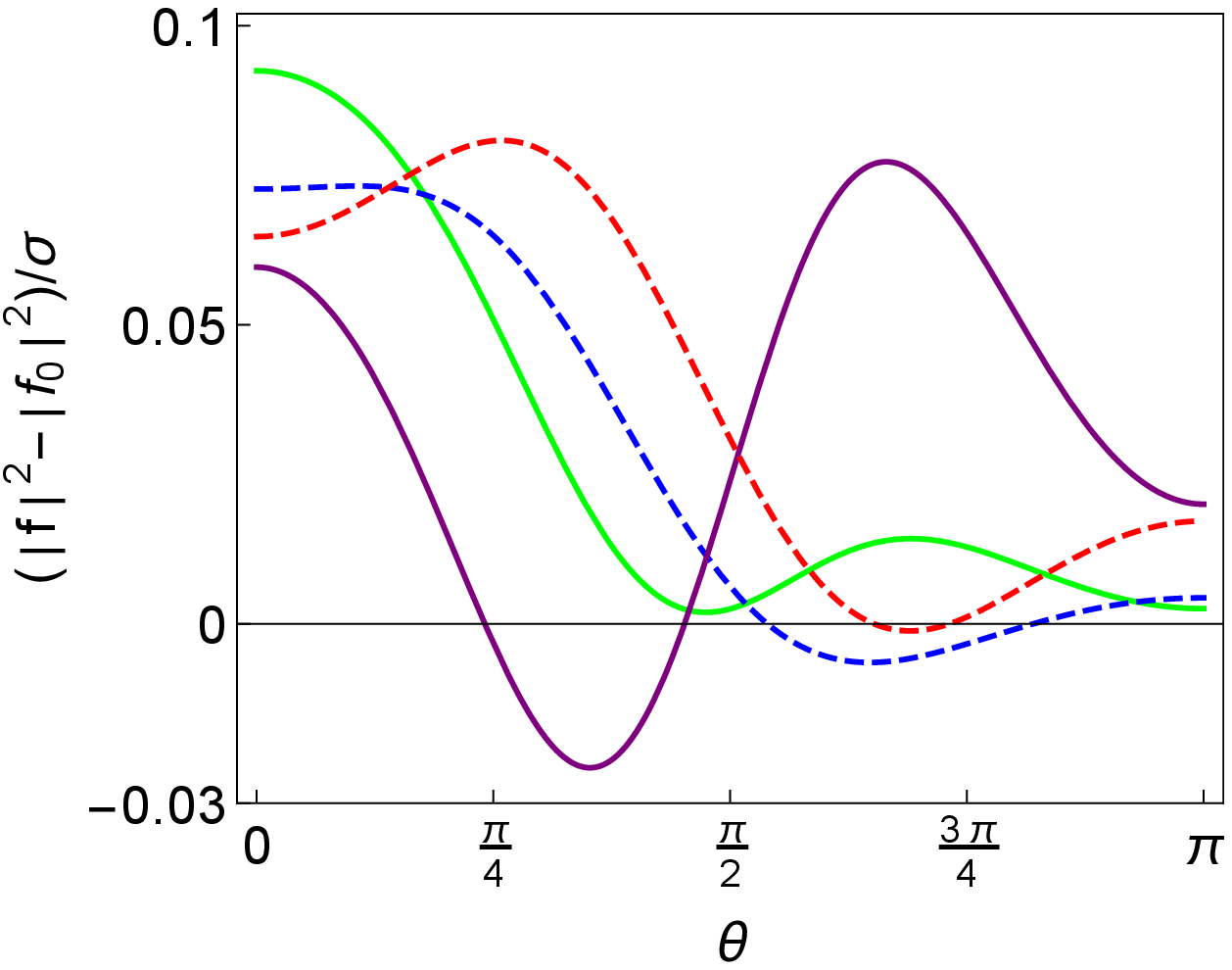}
 	\caption{Plots of $|f(\boldsymbol{k}',\boldsymbol{k})|^2/\sigma$ and $\left[|f(\boldsymbol{k}',\boldsymbol{k})|^2-|f_0(\boldsymbol{k}',\boldsymbol{k})|^2\right]/\sigma$ for a Gaussian bump (\ref{G-bump}) with a central defect and a distant defect located at $\bba=(3\sigma,0)$ as functions of $\theta$ for $\theta_0=0$, $\eta=0.1$, $\tilde\xi=\hbar^2/2m$, $k\sigma=1$, and different values of $\lambda_1$ and $\lambda_2$, namely $\lambda_1=-\lambda_2=1/2$     
	(green), $\lambda_1=0$ and $\lambda_2=-1/2$ (dashed blue), $\lambda_1=1/2$ and $\lambda_2=0$ (purple), and $\lambda_1=\lambda_2=1/2$ (dashed red).}
 	\label{fig7b}
 	\end{center}
 	\end{figure}

Next, consider a Gaussian bump (\ref{G-bump}) with a central defect and $N$ distant defects. Because we employ the first Born approximation to determine the effect of the geometry of the surface, the scattering amplitude for this setup has the form
   	\be
	\ff(\bk',\bk)\approx
	\ff_0(\bk',\bk)+\zeta\, \ff_1^{(c)}(\bk',\bk)+\zeta\, \ff_1^{(d)}(\bk',\bk),
	\label{f1=cd}
	\ee
where $\ff_0(\bk',\bk)$ is the scattering amplitude for the same defects when they are placed on a Euclidean plane, as give by (\ref{f-for-delta=}), and $\zeta\, \ff_1^{(c)}(\bk',\bk)$ and $\zeta\, \ff_1^{(d)}(\bk',\bk)$ give the geometric contribution to the scattering amplitude in the absence of the distant and central defects, respectively. Figures~\ref{fig7a} and \ref{fig7b} show the plots of the differential cross section $|\ff(\bk',\bk)|^2$ and its difference with $|\ff_0(\bk',\bk)|^2$ for a Gaussian bump in the presence of a central defect and a distant defect.
	
\begin{figure}[!ht]
	\begin{center}
	\includegraphics[scale=.6]{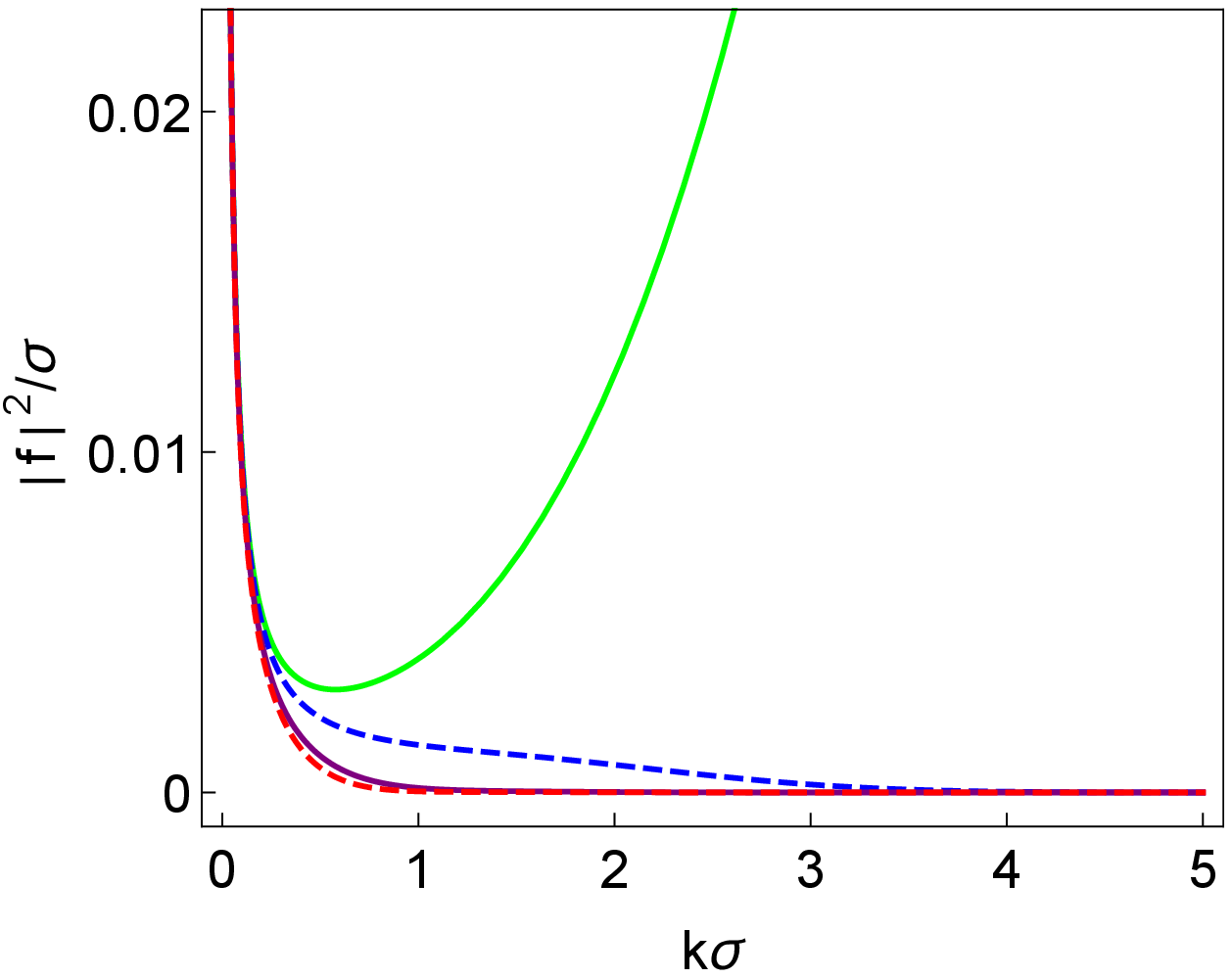}
	\includegraphics[scale=.6]{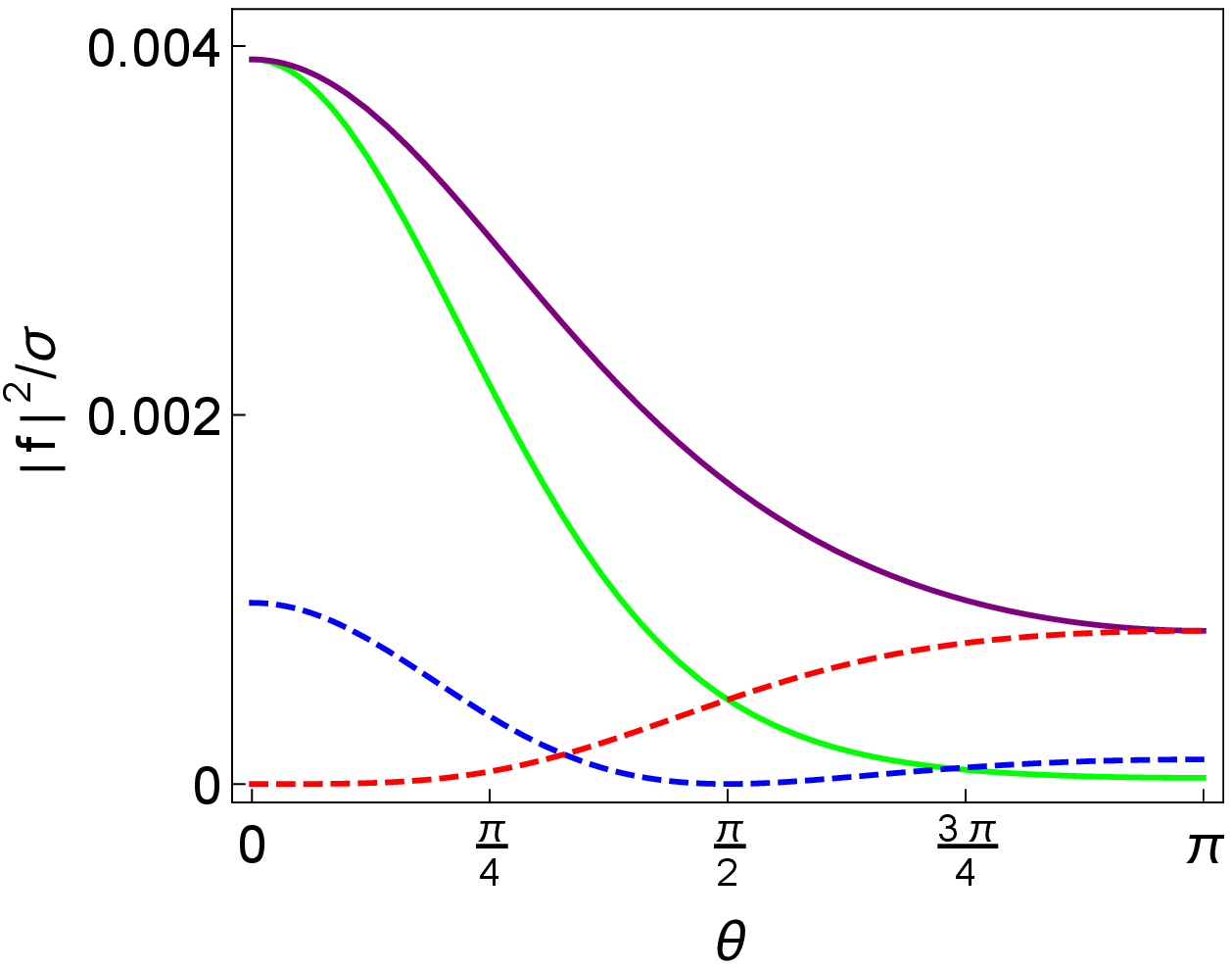}
	\caption{Plots of the differential cross-section for the geometric scattering due to the Gaussian bump  (\ref{G-bump}) as function of $k\sigma$ (on the left) for $\theta_0=0$, $\eta=0.1$, $\tilde\xi=\hbar^2/2m$, and $\lambda_1=-\lambda_2 =1/2$, and different values of $\theta$, namely $\theta=0$ (green), $\pi/3$ (dashed blue), $2\pi/3$ (purple), and $\pi$ (dashed red), and  as functions of $\theta$  (on the right)  for $\theta_0=0$, $\eta=0.1$, $\tilde\xi=\hbar^2/2m$, $k\sigma=1$, and different values of $\lambda_1$ and $\lambda_2$, namely $\lambda_1=-\lambda_2=1/2$ (green), $\lambda_1=0$ and $\lambda_2=-1/2$ (dashed blue), $\lambda_1=1/2$ and $\lambda_2=0$ (purple), and $\lambda_1=\lambda_2=1/2$ (dashed red).}
	\label{fig12}
	\end{center}
	\end{figure}

In order to understand the impact of the point defects on the contribution of the nontrivial geometry of the surface to the differential cross-section, we have compared $|\ff(\bk',\bk)|^2-|\ff_0(\bk',\bk)|^2$ with the differential cross-section for the same surface in the absence of the defects. Figure.~\ref{fig12} shows the graphs of the latter. 
Comparing the magnitude of the variations of $|\ff(\bk',\bk)|^2-|\ff_0(\bk',\bk)|^2$ as given in the right-hand graphs in Figs.~\ref{fig2a} -- \ref{fig7b} with the those shown in Fig.~\ref{fig12}, we see that the presence of the point defects enhances the contribution of the nontrivial geometry of the surface to the differential cross-section.
	
\begin{figure}[!ht]
 	\begin{center}
 	\includegraphics[scale=.6]{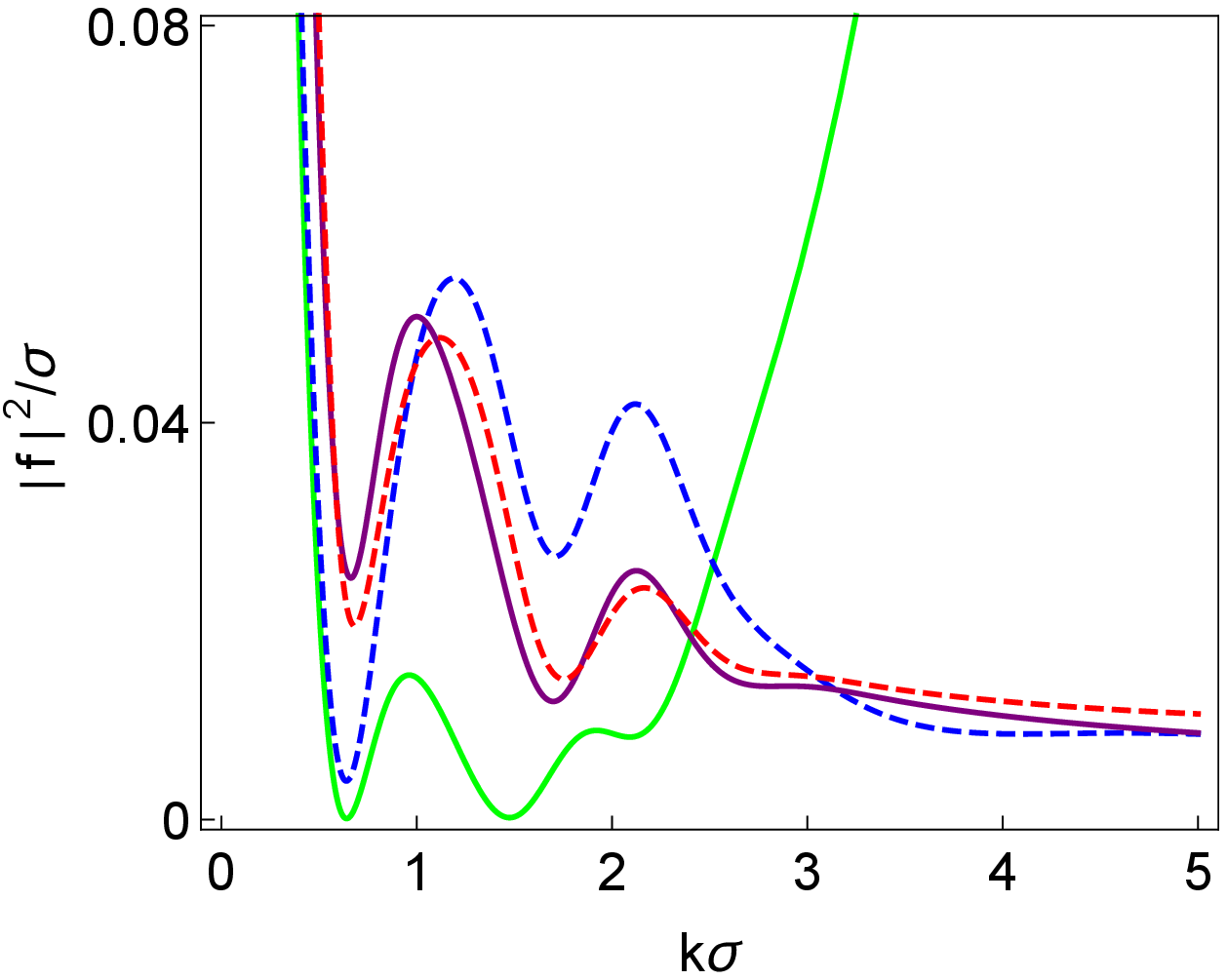}~~~~~~~~
 	\includegraphics[scale=.61]{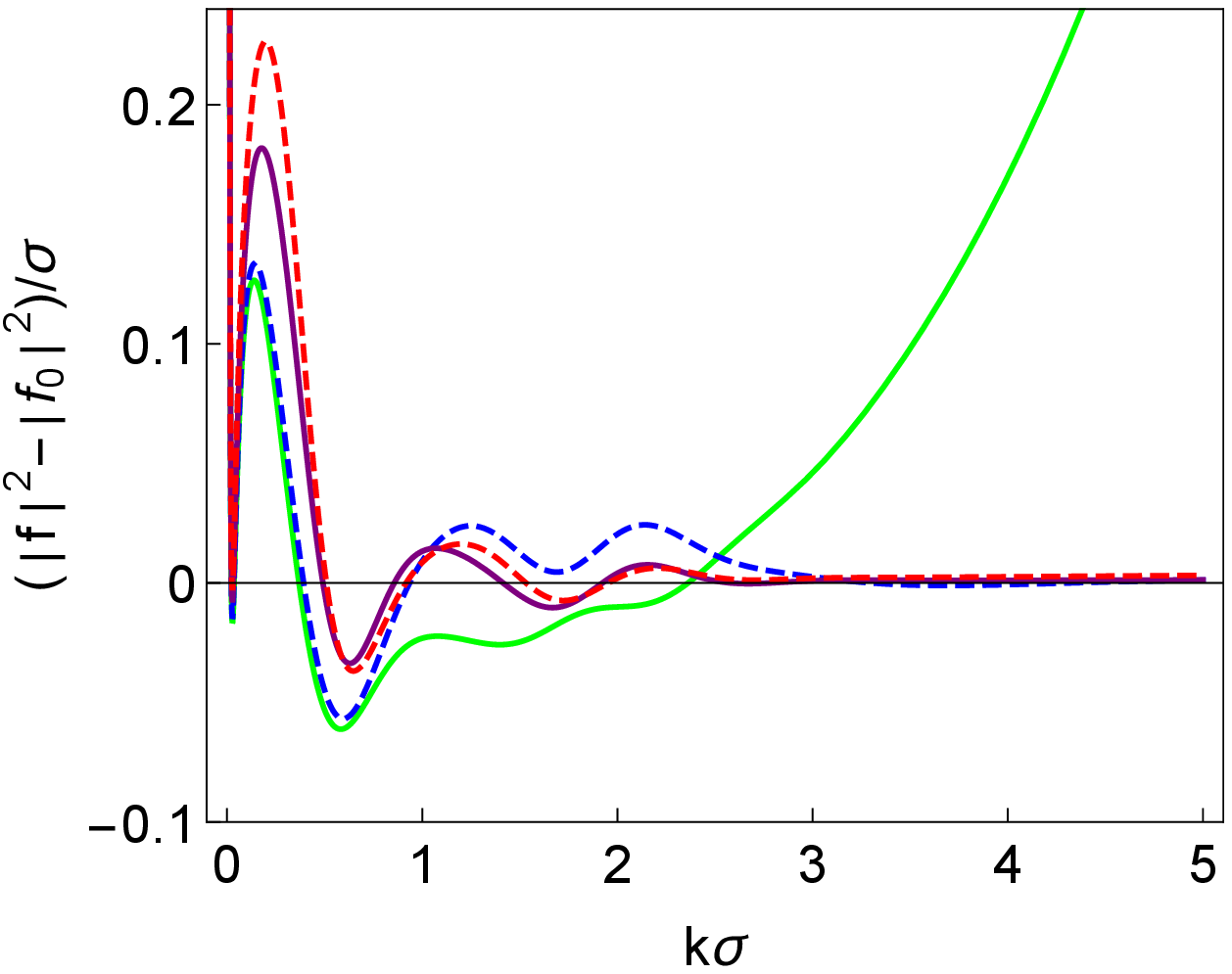}
 	\caption{Plots of $|f(\boldsymbol{k}',\boldsymbol{k})|^2/\sigma$ and $\left[|f(\boldsymbol{k}',\boldsymbol{k})|^2-|f_0(\boldsymbol{k}',\boldsymbol{k})|^2\right]/\sigma$ for a pair of identical Gaussian bumps located at $\bc_1=(-3\sigma,0)$ and $\bc_2=(3\sigma,0)$
with a point defect at the origin $(0,0)$ as functions of $k\sigma$ for $\theta_0=0$, $\eta=0.1$, $\tilde\xi=\hbar^2/2m$, and $\lambda_1=-\lambda_2 =1/2$, and different values of $\theta$, namely $\theta=0$ (green), $\pi/3$ (dashed blue), $2\pi/3$ (purple), and $\pi$ (dashed red).}
 	\label{fig8a}
 	\end{center}
 	\end{figure}

\begin{figure}[!ht]
 	\begin{center}
 	\includegraphics[scale=.6]{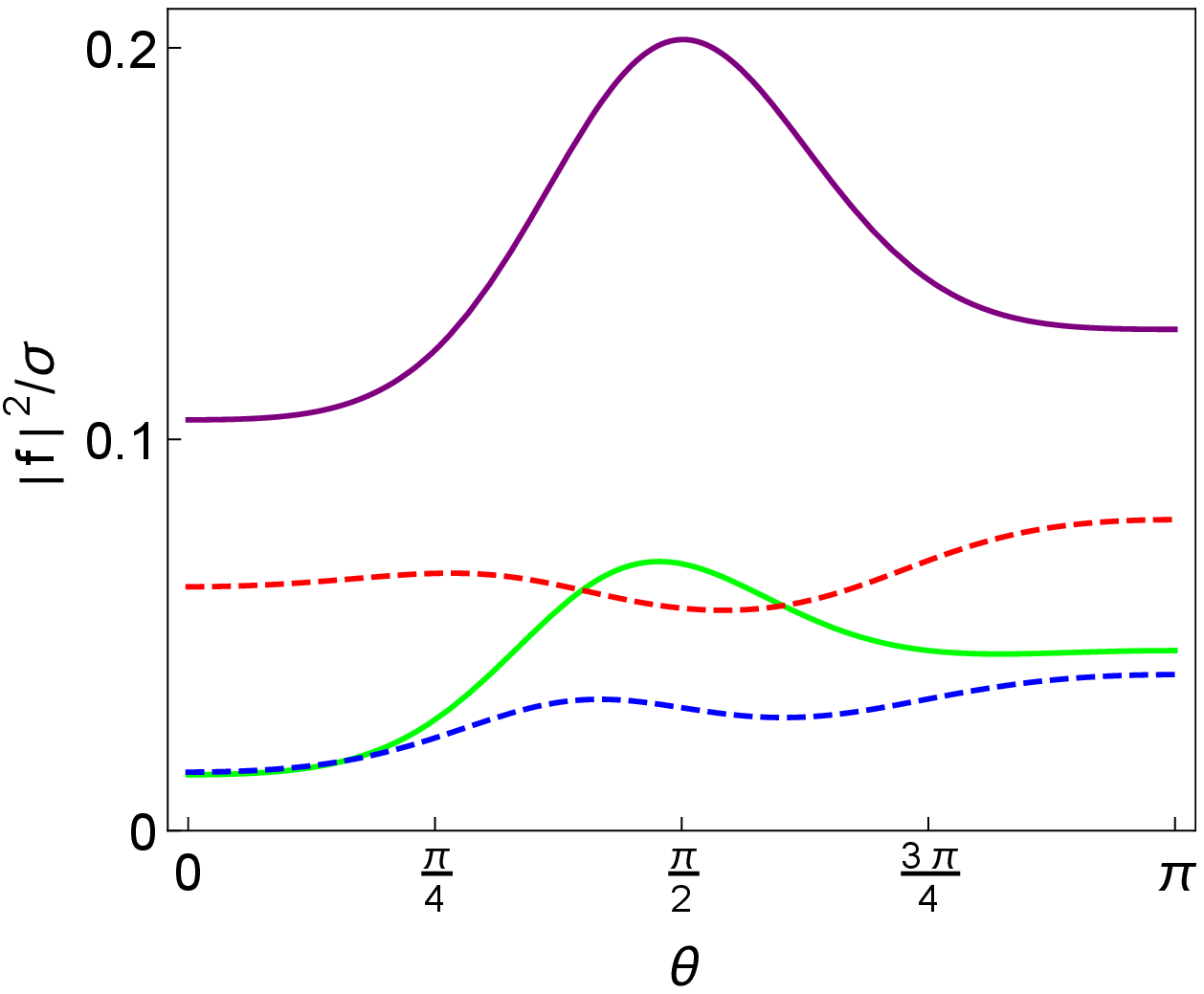}~~~~~~~~
 	\includegraphics[scale=.64]{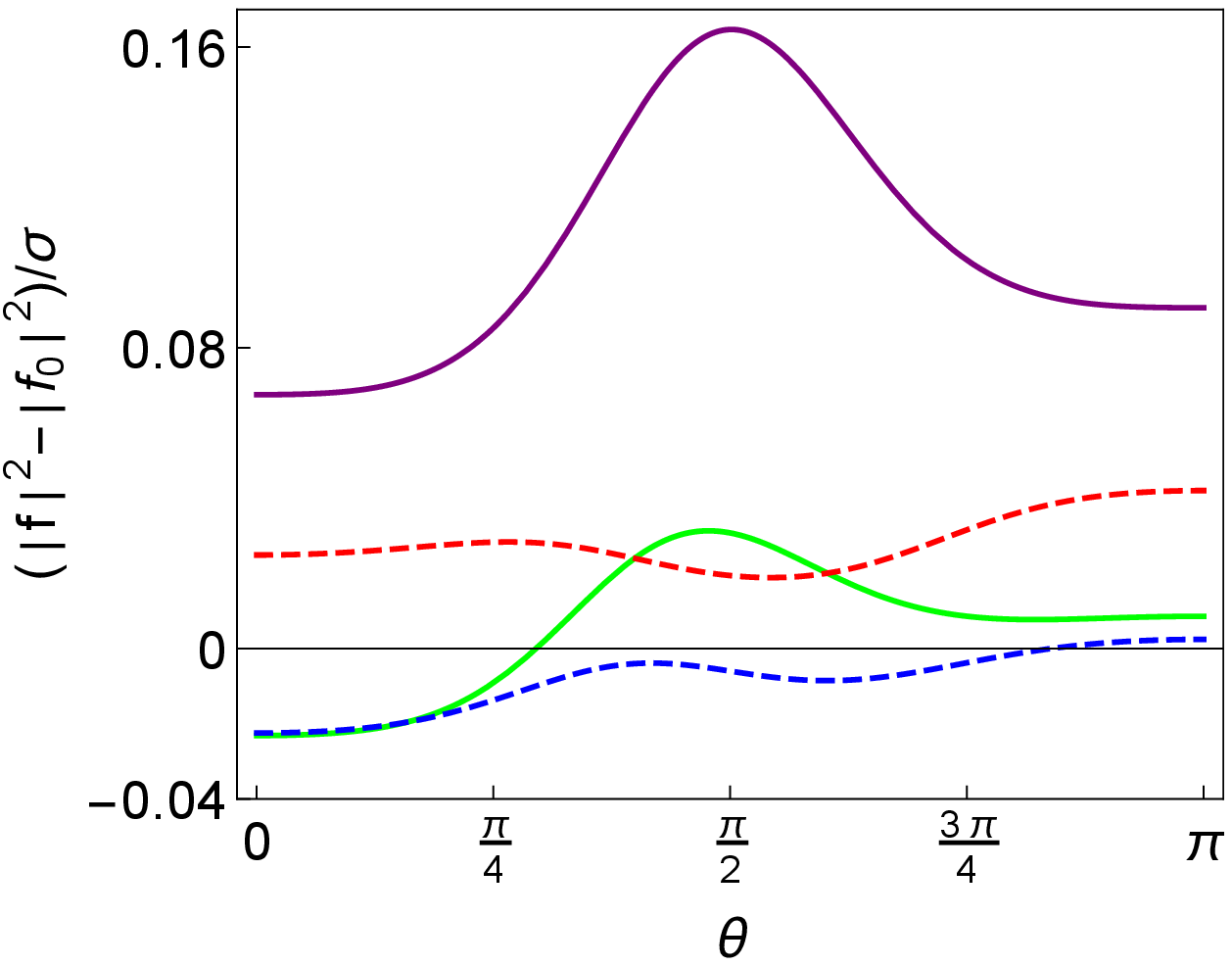}
 	\caption{Plots of $|f(\boldsymbol{k}',\boldsymbol{k})|^2/\sigma$ and $\left[|f(\boldsymbol{k}',\boldsymbol{k})|^2-|f_0(\boldsymbol{k}',\boldsymbol{k})|^2\right]/\sigma$ for a pair of identical Gaussian bumps located at $\bc_1=(-3\sigma,0)$ and $\bc_2=(3\sigma,0)$
with a point defect at the origin $(0,0)$ as functions of $\theta$ for $\theta_0=0$, $\eta=0.1$, $\tilde\xi=\hbar^2/2m$, $k\sigma=1$, and different values of $\lambda_1$ and $\lambda_2$, namely $\lambda_1=-\lambda_2=1/2$ (green), $\lambda_1=0$ and $\lambda_2=-1/2$ (dashed blue), $\lambda_1=1/2$ and $\lambda_2=0$ (purple), and $\lambda_1=\lambda_2=1/2$ (dashed red).}
 	\label{fig8b}
 	\end{center}
 	\end{figure}

\begin{figure}[!ht]
 	\begin{center}
 	\includegraphics[scale=.6]{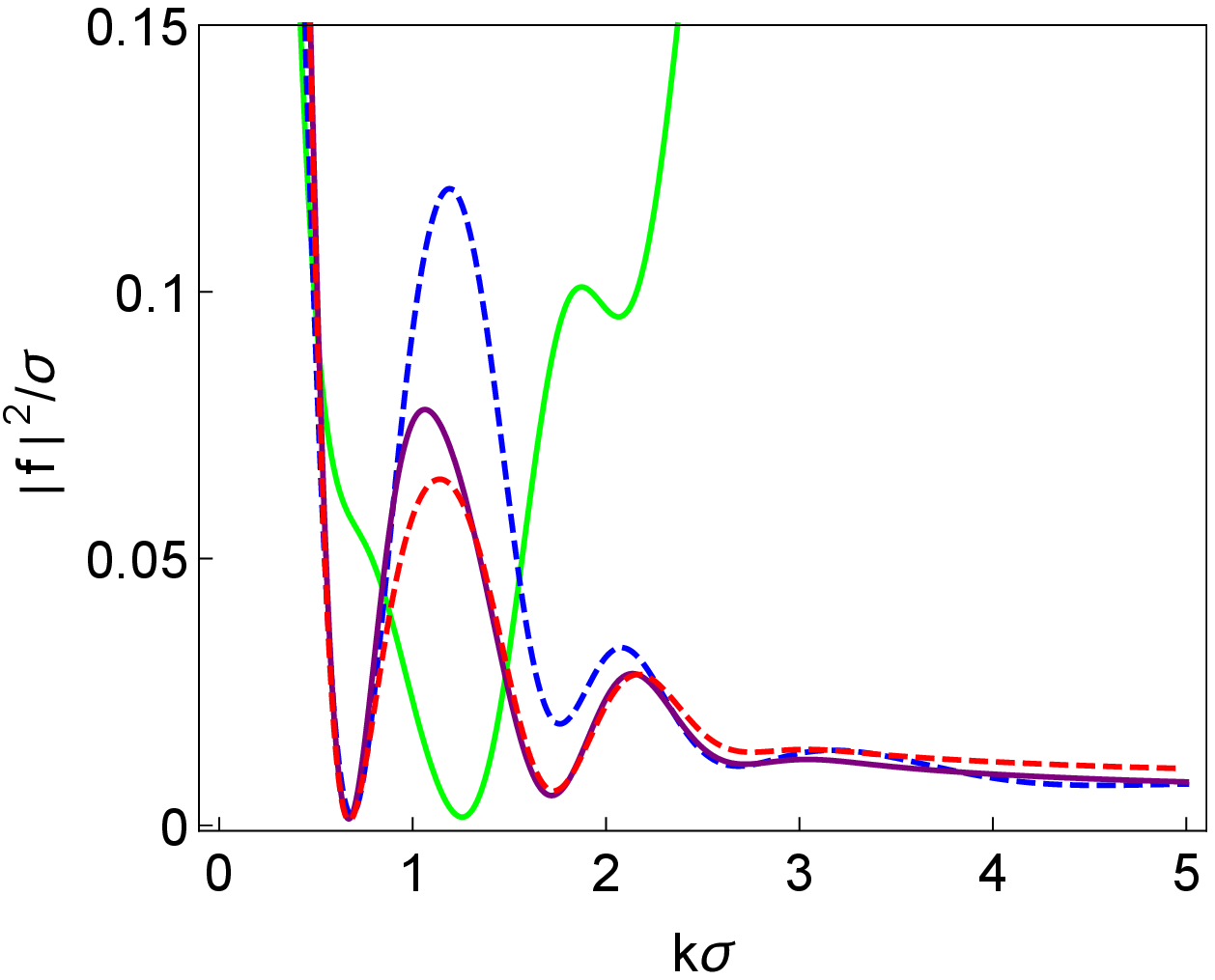}~~~~~~~~
 	\includegraphics[scale=.62]{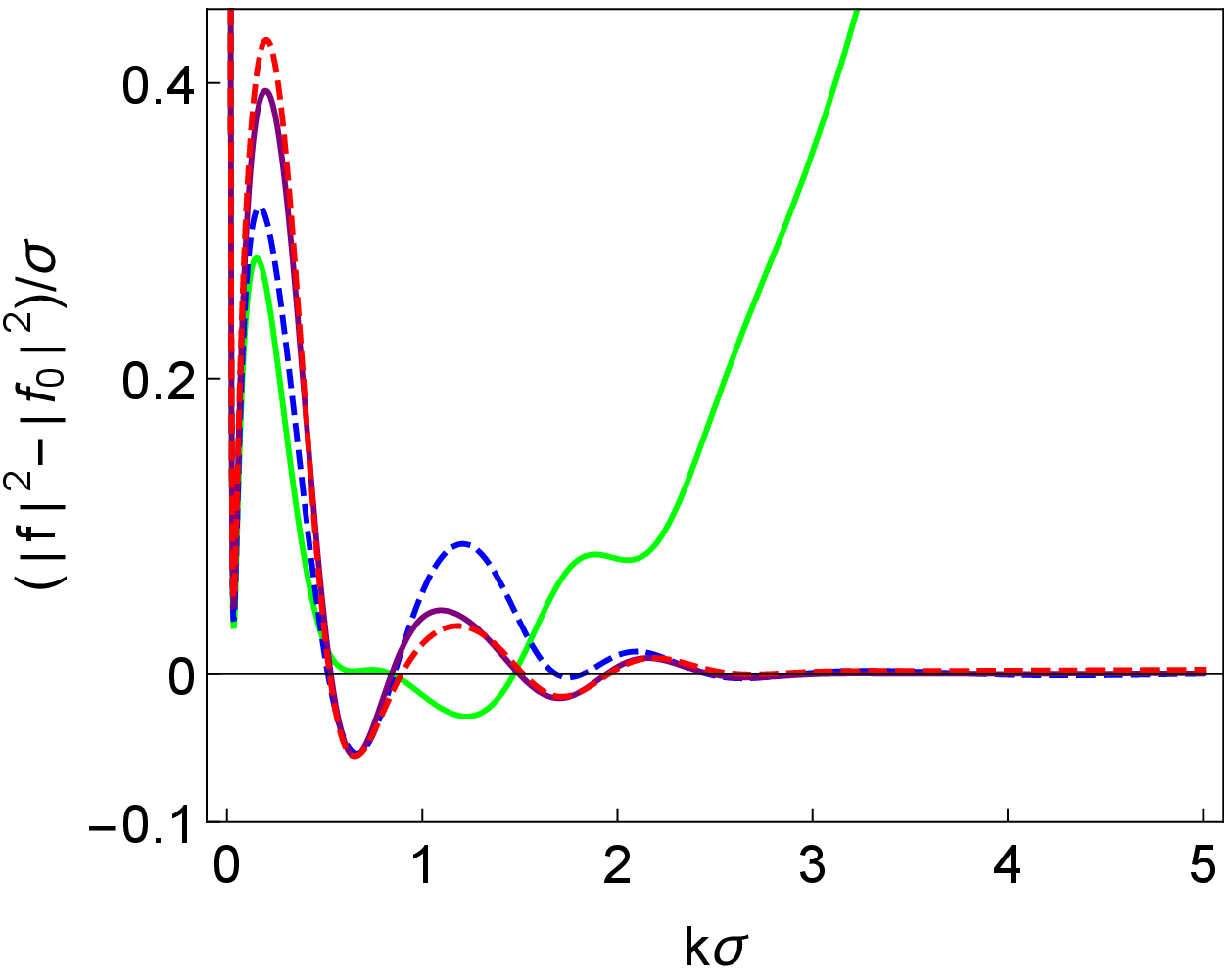}
 	\caption{Plots of $|f(\boldsymbol{k}',\boldsymbol{k})|^2/\sigma$ and $\left[|f(\boldsymbol{k}',\boldsymbol{k})|^2-|f_0(\boldsymbol{k}',\boldsymbol{k})|^2\right]/\sigma$ for four identical Gaussian bumps located at $\bc_1=(-3\sigma,0)$, $\bc_2=(3\sigma,0)$, $\bc_3=(0,-3\sigma)$, $\bc_2=(0,3\sigma)$ with a point defect at the origin $(0,0)$ as functions of $k\sigma$ for $	\theta_0=0$, $\eta=0.1$, $\tilde\xi=\hbar^2/2m$, and $\lambda_1=-\lambda_2 =1/2$, and different values of $\theta$, namely $\theta=0$ (green), $\pi/3$ (dashed blue), $2\pi/3$ (purple), and $\pi$ (dashed red).}
 	\label{fig10a}
 	\end{center}
	\end{figure}

\begin{figure}[!ht]
 	\begin{center}
 		\includegraphics[scale=.6]{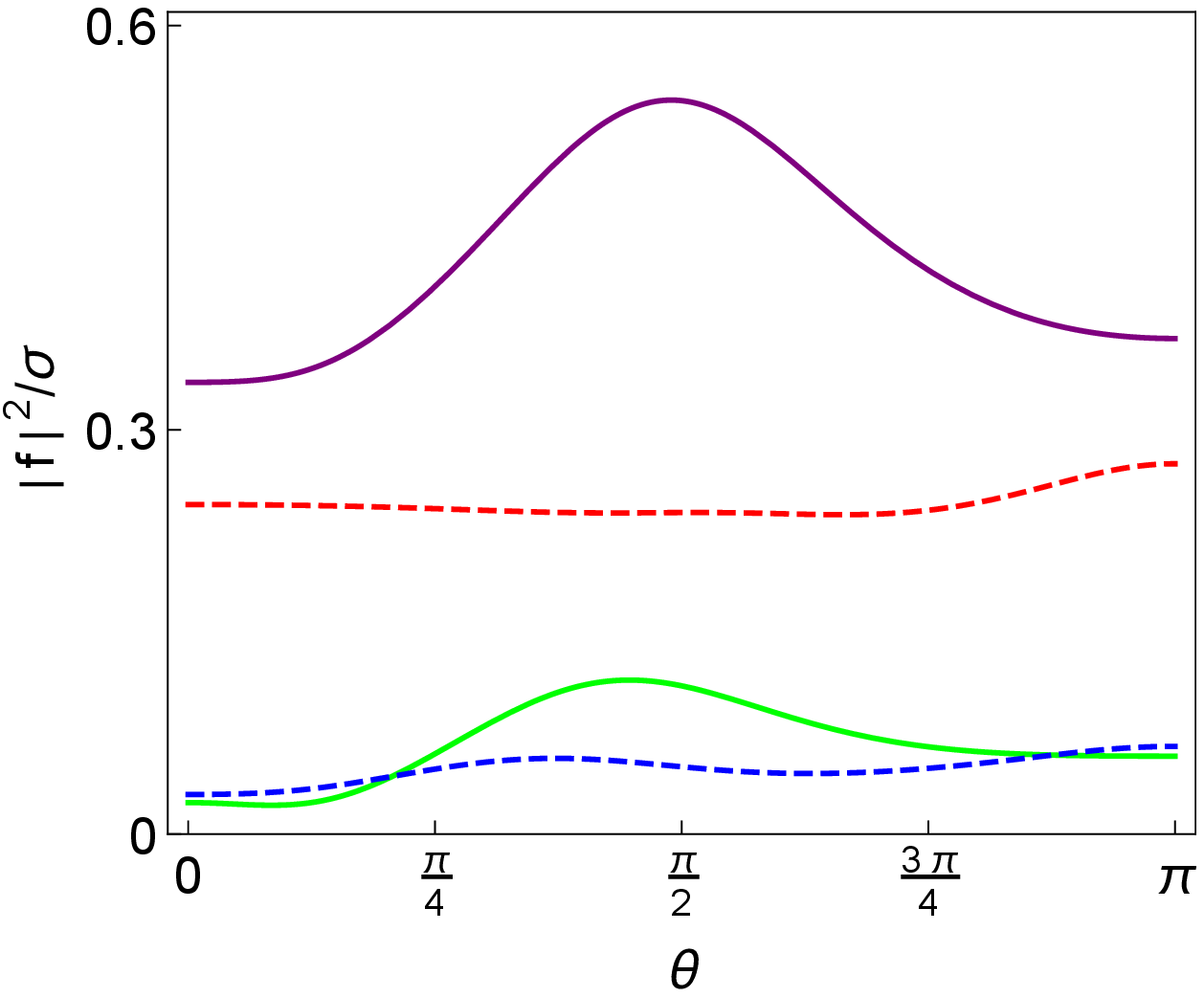}~~~~~~~~
 		\includegraphics[scale=.62]{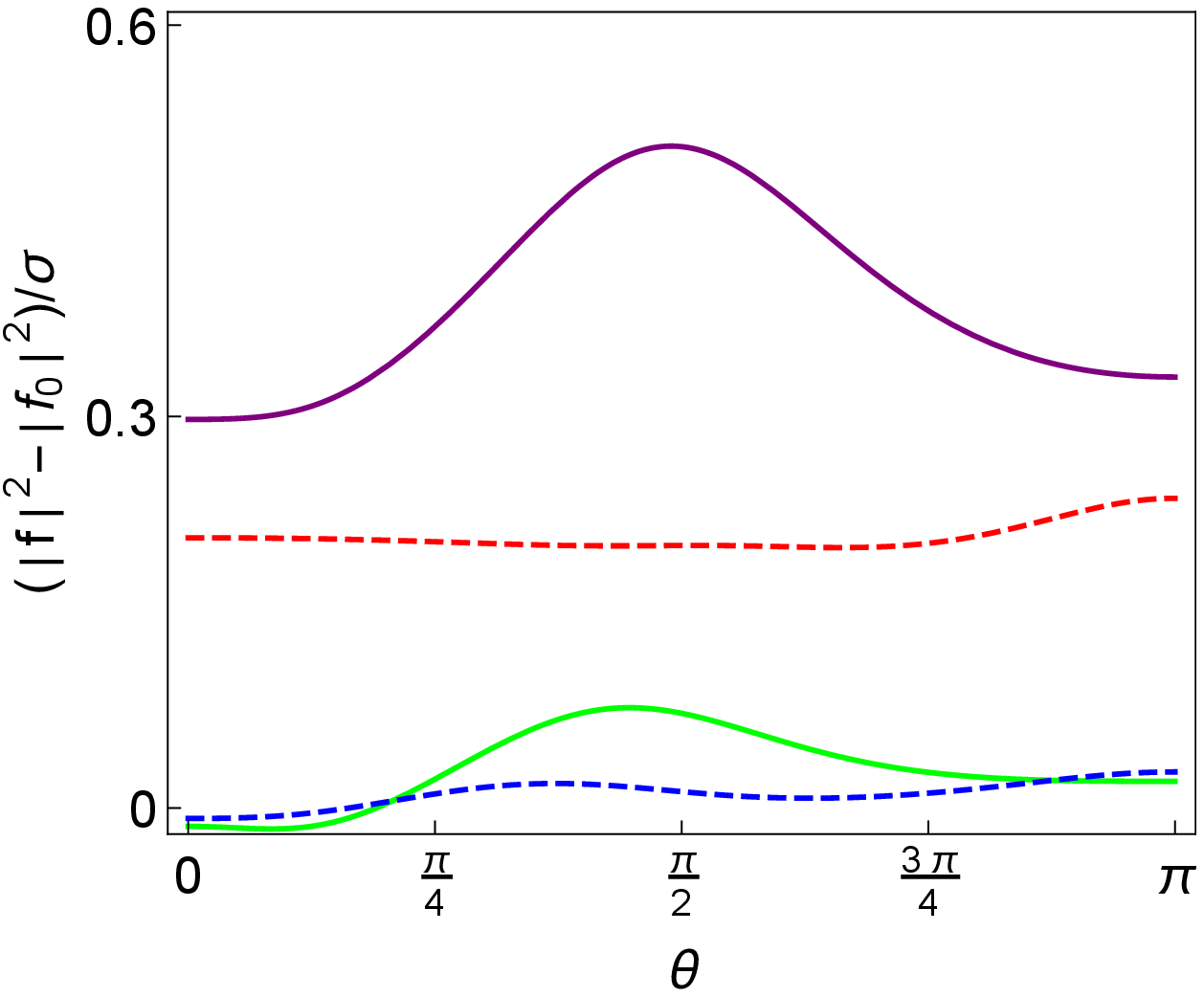}
 		\caption{Plots of $|f(\boldsymbol{k}',\boldsymbol{k})|^2/\sigma$ and $\left[|f(\boldsymbol{k}',\boldsymbol{k})|^2-|f_0(\boldsymbol{k}',\boldsymbol{k})|^2\right]/\sigma$ for four identical Gaussian bumps located at $\bc_1=(-3\sigma,0)$, $\bc_2=(3\sigma,0)$, $\bc_3=(0,-3\sigma)$, $\bc_2=(0,3\sigma)$ with a point defect at the origin $(0,0)$ as functions of $\theta$ for $\theta_0=0$, $\eta=0.1$, $\tilde\xi=\hbar^2/2m$, $k\sigma=1$, and different values of $\lambda_1$ and $\lambda_2$, namely $\lambda_1=-\lambda_2=1/2$ (green), $\lambda_1=0$ and $\lambda_2=-1/2$ (dashed blue), $\lambda_1=1/2$ and $\lambda_2=0$ (purple), and $\lambda_1=\lambda_2=1/2$ (dashed red).}
 	\label{fig10b}
 	\end{center}
	\end{figure}

We can extend the above results to cases where the surface ${S}$ is made of several Gaussian bumps that are sufficiently far from one another \cite{pra-2018}. To do this, first we recall that under a translation, $\bx\to\bx-\bc$, the scattering amplitude transforms according to $f(\bk',\bk)\to e^{i(\bk-\bk')\cdot\bc}f(\bk',\bk)$. Now, suppose that ${S}$ is the surface defined by 
	\begin{equation}
	z=f(\bx):=\sum_{m=1}^M\delta_m\, e^{-(\bx-\bc_m)^2/2\sigma_m^2},
	\label{G-bump-m}
	\end{equation}
  where $\delta_m$ and $\sigma_m$ are nonzero real parameters, $\sigma_m>0$, and $\bc_m\in\R^2$. If for all $m,m'=1,2,\cdots M$, $|\bc_m-\bc_{m'}|\gg \sigma_m+\sigma_{m'}$, then employing the first Born approximation we find that the contribution of the nontrivial geometry of ${S}$ to the scattering amplitude is given by
  	\be
	\zeta\, \ff_1(\bk',\bk)=\sum_{m=1}^M e^{i(\bk-\bk')\cdot\bc_m} \zeta\,
	\ff_{1,m}(\bk',\bk),
	\label{f1m=}
	\ee
where $\zeta\,\ff_{1,m}(\bk',\bk)$ is the geometric scattering amplitude (\ref{f1=Is}) for a Gaussian bump (\ref{G-bump}) with $\delta=\delta_m$ and $\sigma=\sigma_m$. Again, the total scattering amplitude $\ff(\bk',\bk)$ is the sum of (\ref{f1m=}) and the contribution of the defects in the absence of the geometric effects, i.e., (\ref{f-for-delta=}). Figures~\ref{fig8a}, \ref{fig8b}, \ref{fig10a}, and \ref{fig10b} provide graphical demonstrations of the outcome of this calculation for two and four identical Gaussian bumps with a single distant point defect located at the origin of the coordinate system. Here we use $\sigma$ and $\delta$ to label the common value of $\sigma_m$ and $\delta_m$, respectively.
   
\begin{figure}[!ht]
 	\begin{center}
 	\includegraphics[scale=.6]{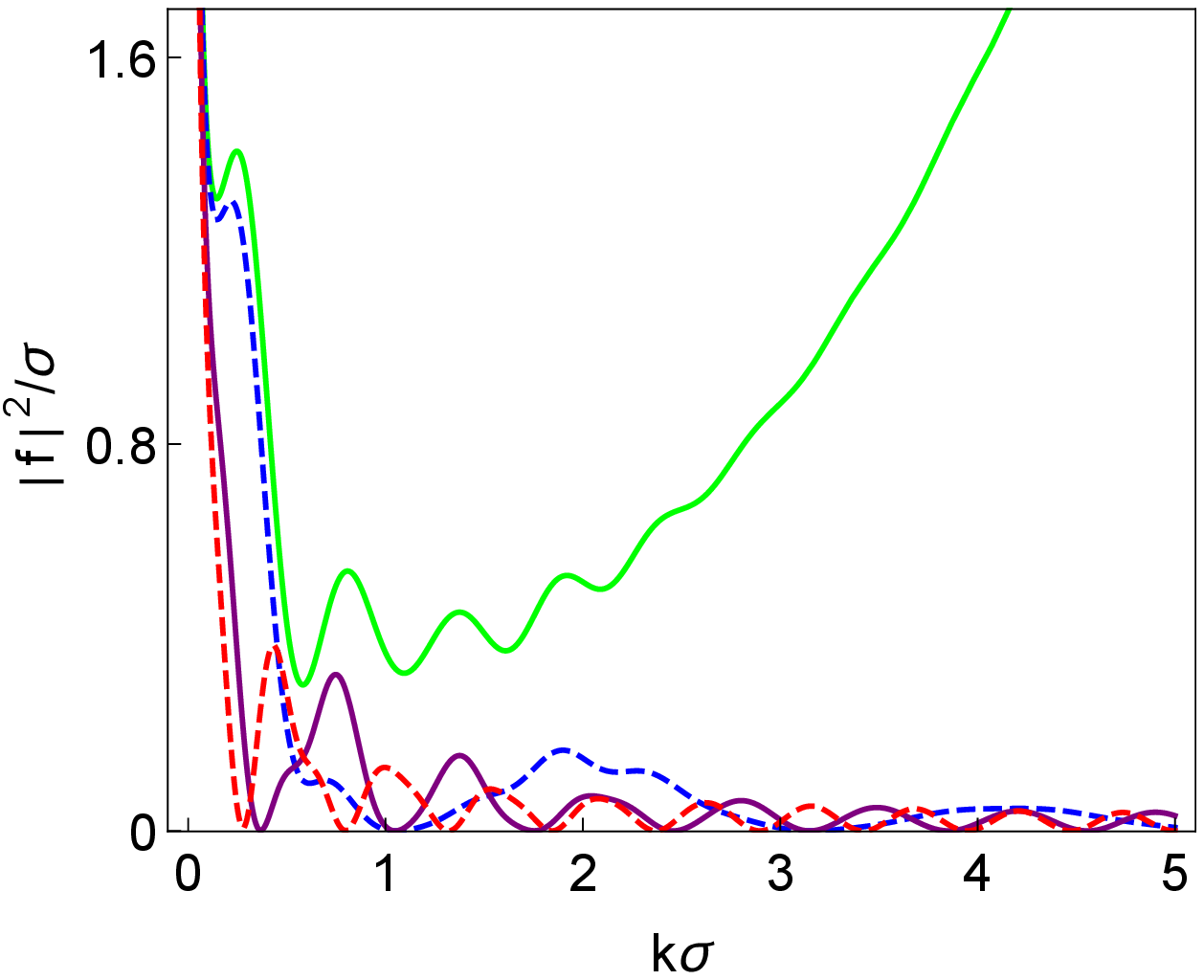}~~~~~~~~
 	\includegraphics[scale=.6]{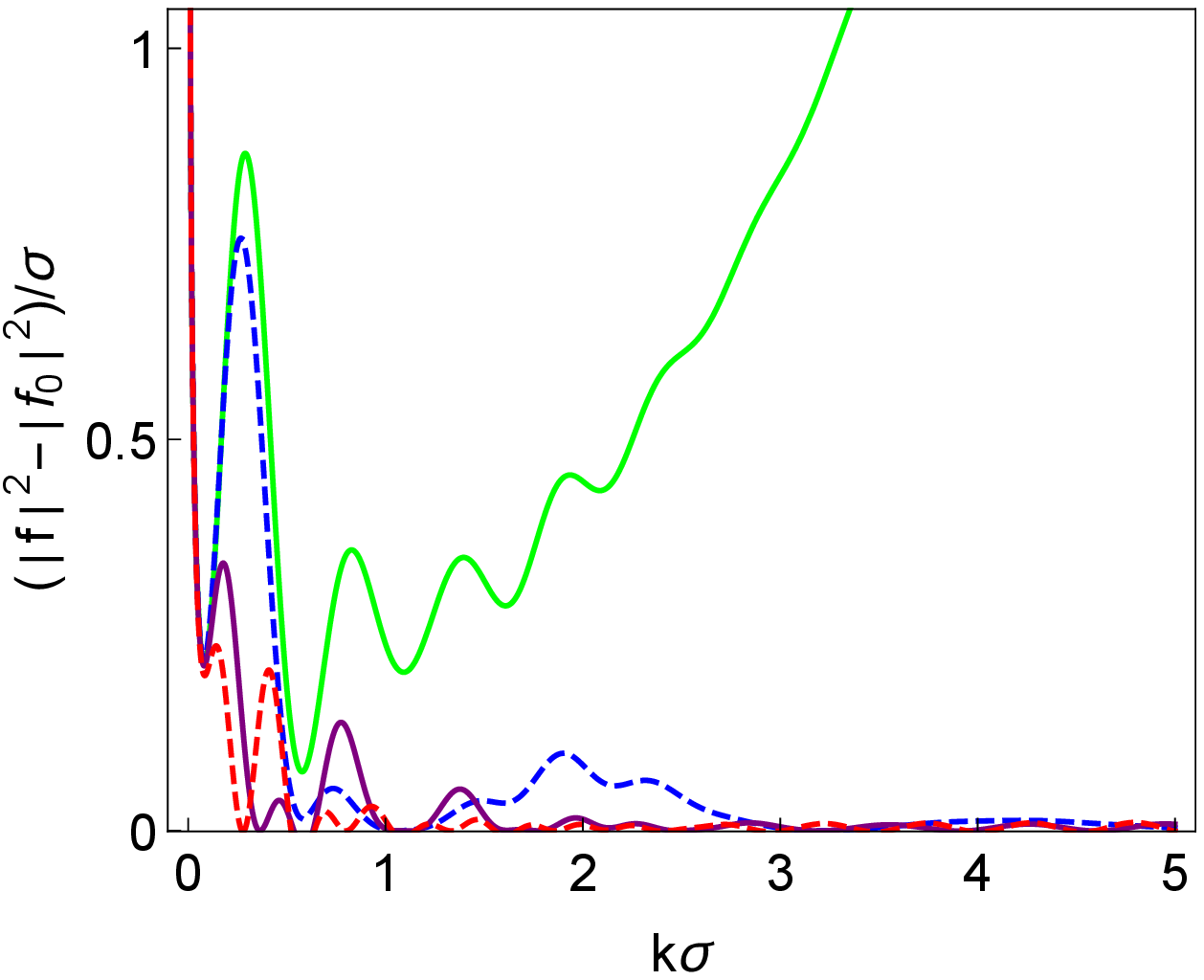}
 	\caption{Plots of $|f(\boldsymbol{k}',\boldsymbol{k})|^2/\sigma$ and $\left[|f(\boldsymbol{k}',\boldsymbol{k})|^2-|f_0(\boldsymbol{k}',\boldsymbol{k})|^2\right]/\sigma$ for a pair of identical Gaussian bumps with central defects located at $\bc_1=(-3\sigma,0)$ and $\bc_2=(3\sigma,0)$ as functions of $k\sigma$ for $\theta_0=0$, $\eta=0.1$, $\tilde\xi=\hbar^2/2m$, and $\lambda_1=-\lambda_2 =1/2$, and different values of $\theta$, namely $\theta=0$ (green), $\pi/3$ (dashed blue), $2\pi/3$ (purple), and $\pi$ (dashed red).}
 	\label{fig11a}
 	\end{center}
	\end{figure}

\begin{figure}[!ht]
 	\begin{center}
 	\includegraphics[scale=.6]{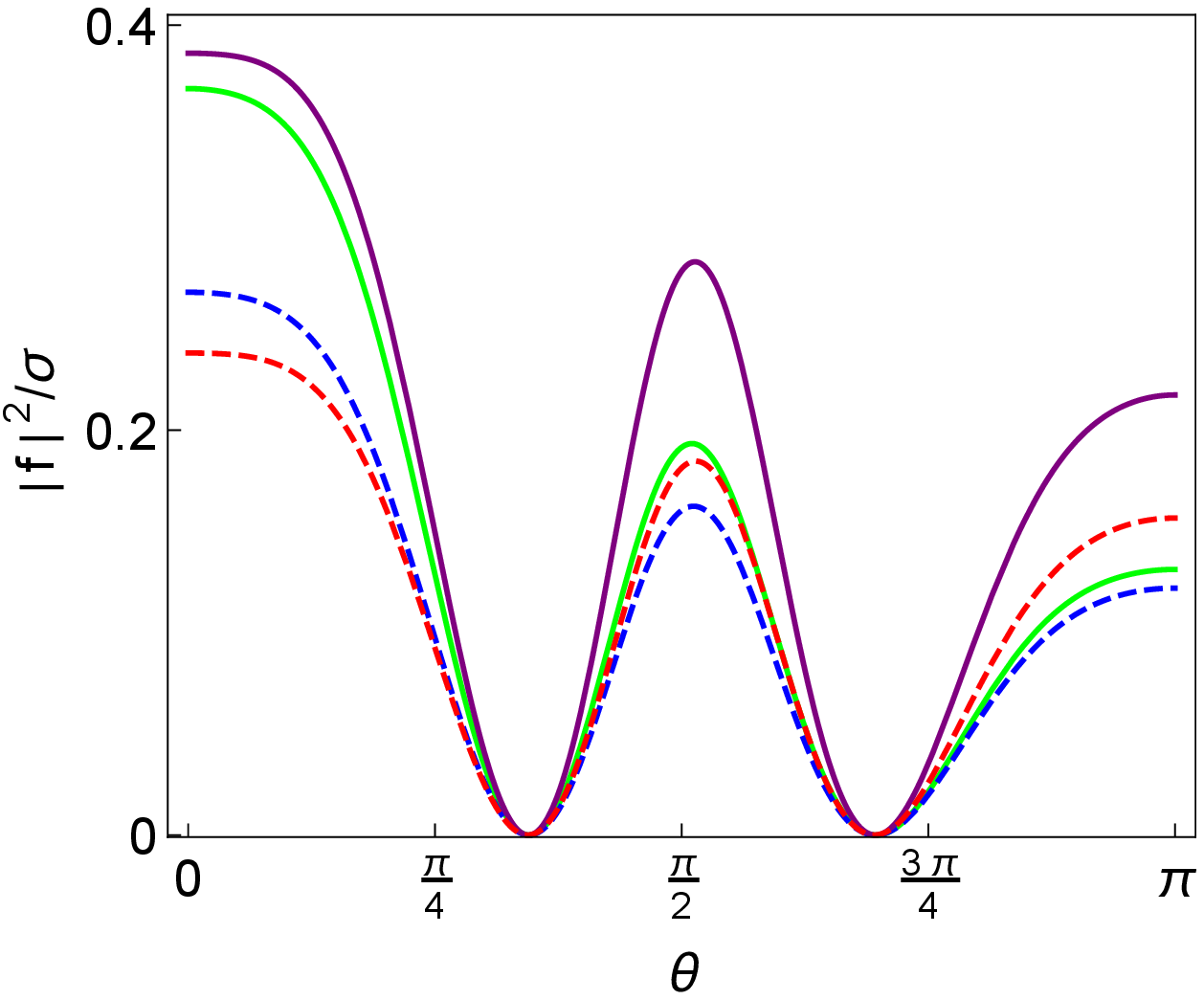}~~~~~~~~
 	\includegraphics[scale=.62]{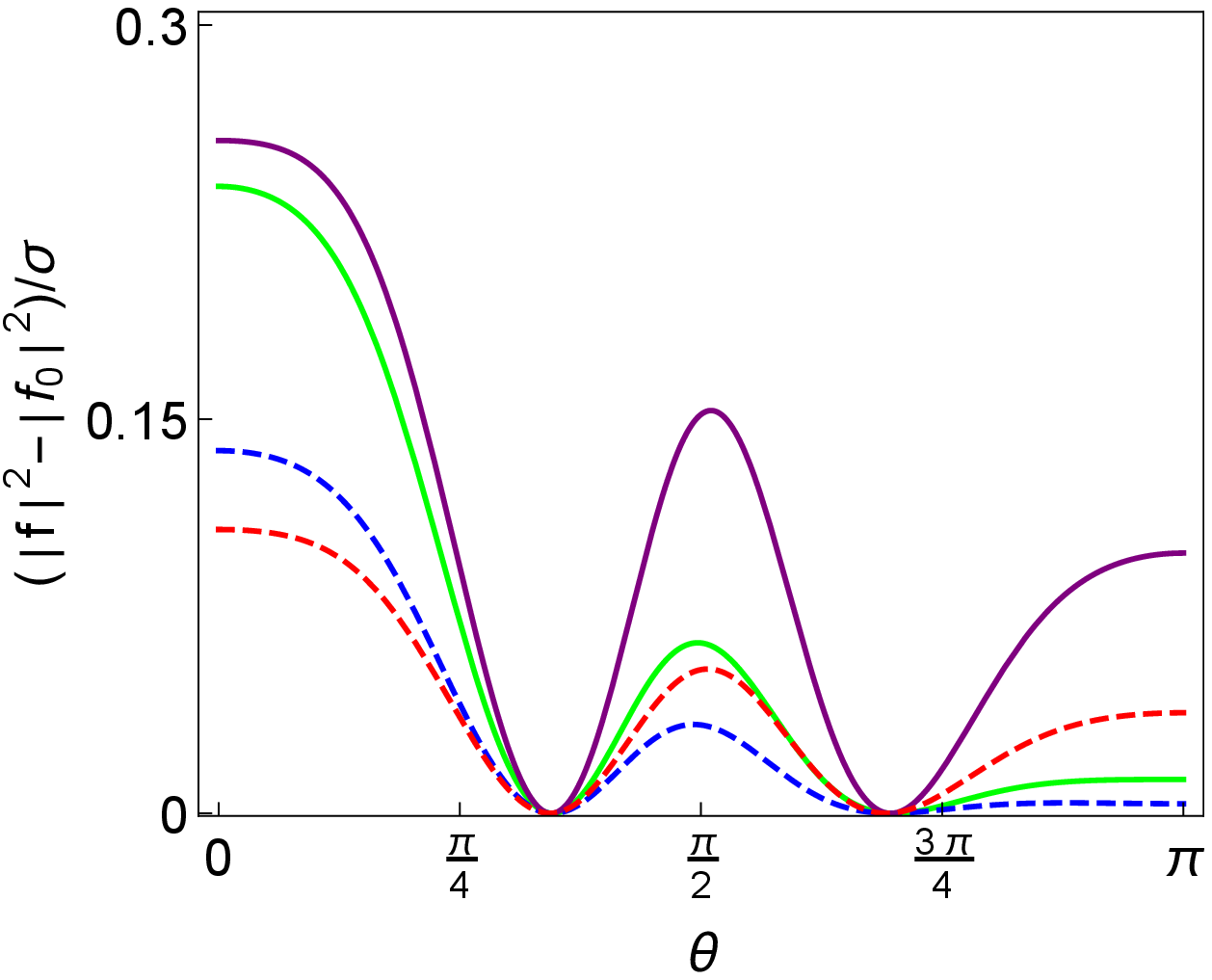}
 	\caption{Plots of $|f(\boldsymbol{k}',\boldsymbol{k})|^2/\sigma$ and $\left[|f(\boldsymbol{k}',\boldsymbol{k})|^2-|f_0(\boldsymbol{k}',\boldsymbol{k})|^2\right]/\sigma$ for a pair of identical Gaussian bumps with central defects located at $\bc_1=(-3\sigma,0)$ and $\bc_2=(3\sigma,0)$ as functions of $\theta$ for $\theta_0=0$, $\eta=0.1$, $\tilde\xi=\hbar^2/2m$, $k\sigma=1$, and different values of $\lambda_1$ and $\lambda_2$, namely $\lambda_1=-\lambda_2=1/2$ (green), $\lambda_1=0$ and $\lambda_2=-1/2$ (dashed blue), $\lambda_1=1/2$ and $\lambda_2=0$ (purple), and $\lambda_1=\lambda_2=1/2$ (dashed red).}
 	\label{fig11b}
 	\end{center}
 	\end{figure}

Next, consider the surface made of two identical Gaussian bumps with central defects at $\bc_1=(3\sigma,0)$ and $\bc_2=(-3\sigma,0)$. Then $\ff_{1,m}(\bk',\bk)$ is the geometric contribution corresponding to a Gaussian bump centered at the origin with a pair of defects placed at $\bc_1-\bc_m$ and $\bc_2-\bc_m$. Notice that one of these defects is a central defect and the other is a distant defect. Therefore, we can compute $\ff_{1,m}(\bk',\bk)$ using our results on central and distant defects. Figures~\ref{fig11a} and \ref{fig11b} show the plots of the differential cross section $|\ff(\bk',\bk)|^2$ for this arrangement of Gaussian bumps with central defects. 
 
 Comparing the graphs of $|f(\boldsymbol{k}',\boldsymbol{k})|^2-|f_0(\boldsymbol{k}',\boldsymbol{k})|^2$ given in Figs.~\ref{fig8a} --  \ref{fig11b} with those of the differential cross section for the same surface in the absence of point defects (not given here for lack of space) confirms our previous observation regarding the amplification of the geometric scattering due to the presence of the defects.

\section{Summary and concluding remarks}
\label{sec7}

When a particle is confined to move on an asymptotically flat curved surface the geometry of the latter affects its motion. If the particle is made to enter and leave the region of the surface whose geometry deviates from that of a plane, it undergoes a geometric scattering. In this paper we have studied this phenomenon for situations that the surface includes one or more point defects with known locations. We have identified these with point scatterers modeled using delta-function potentials. 

A major problem one encounters in treating delta-function point scatterers in two and higher dimensions is that the standard approach to potential scattering leads to divergent terms. This problem has been studied in detail for the case of a single delta-function potential in the plane. To regularize the unwanted divergences in dealing with this problem we have performed a renormalization scheme with a simple physical justification to reproduce the known result for a single delta function. Using this scheme we have calculated the scattering amplitude for $N$ delta-function defects with arbitrary locations in the plane. 

Next, we have extended our results for the plane to curved surfaces. Here, we included the effect of the geometry of the surface as a perturbation of the scattering system defined by the delta-function point defects placed on a plane. This leads to highly complicated formulas that we could study in some detail for surfaces made out of finitely many distant Gaussian bumps with defects placed in their peak or at large distances to the peaks. An extensive graphical examination of our analytical results shows that the presence of the point defects amplifies the geometric scattering effect substantially. A heuristic justification for this observation is that the defects cause the particle to spend more time in the curved regions of the surface.  A numerical implementation of our analytical results should allow for the study of geometric scattering effects in the presence of randomly distributed point defects.\vspace{6pt} 

\noindent{\bf Acknowledgements.} We are indebted to Farhang Loran and Teoman Turgut for their invaluable comments and suggestions. This work has been supported by the Scientific  and Technological Research Council of Turkey (T\"UB{$\dot{\rm I}$}TAK) in the framework of the Project
No.~117F108 and by the Turkish Academy of Sciences (T\"UBA).

\ed
\begin{thebibliography}{99}

\bibitem{dewitt-1952} B.~S.~DeWitt, Phys.\ Rev.~{\bf 85}, 653 (1952).

\bibitem{dewitt-1957}  B.~S.~DeWitt, Rev.\ Mod.\ Phys.~\textbf{29}, 377 (1957).

\bibitem{golovnev} A. V. Golovnev, Rep. Math. Phys. 64, 59 (2009).

\bibitem{dacosta} R.~C.~T.~da Costa, Phys.\ Rev.~A {\bf 23}, 1982 (1981).

\bibitem{kaplan} L.~Kaplan, N.~T.~Maitra, and E.~J.~Heller, Phys.\ Rev.~A {\bf 56}, 2592 (1997).

\bibitem{pra-1996} A.~Mostafazadeh, Phys.\ Rev.~A  \textbf{54}, 1165-1170 (1996).

\bibitem{pra-2018} N.~Oflaz, A.~Mostafazadeh, and M.~Ahmady, Phys.\ Rev.\ A \textbf{98}, 022126 (2018).

\bibitem{Ono-2010} S.~Ono and H.~Shima, Physica E {\bf 42} 1224 (2010).

\bibitem{Onoe-2012} J.~Onoe, T.~Ito, H.~Shima, H.~Yoshioka, S.-I.~Kimura, EPL {\bf 98}, 27001 (2012).

\bibitem{Ortix-2011} C.~Ortix, S.~Kiravittaya, O.~G.~Schmidt, and J.~ van~den~Brink, Phys.\ Rev.~B {\bf 84}, 045438 (2011).

\bibitem{Silva-2013} K.~V.~R.~A.~Silva, C.~F.~de~Freitas, and C.~Filgueiras, Eur.\ Phys.\ J.~B  \textbf{86}, 147 (2013).

\bibitem{Vadakkumbatt-2014} V.~Vadakkumbatt, E.~Joseph, A.~Pal, and A. Ghosh, Nature Commu.\ {\bf 5}, 4571 (2014).

\bibitem{Pahlavani-2015} H.~Pahlavani and M.~Botchekananfard, Physca~B {\bf 459}, 88 (2015).

\bibitem{mead-1991} L.~R.~Mead, and J.~Godines, Am.\ J.~Phys.~{\bf 59}, 935 (1991).

\bibitem{manuel} C.~Manuel and R.~Tarrach, Phys.\ Lett.~B {\bf 328} 113 (1994).

\bibitem{Adhikari1} S.~Adhikari and T.~ Frederico, Phys.\ Rev.\ Lett.~\textbf{74}, 4572 (1995).

\bibitem{Adhikari2} S.~Adhikari, T.~Frederico, and  R.~M.~Marinho,  J. Phys. A~\textbf{29}, 7157 (1996).

\bibitem{Henderson} R. J. Henderson and S. G. Rajeev, J.~Math.\ Phys.\ \textbf{38}, 2171 (1997).

\bibitem{Mitra} I.~ Mitra, A.~ DasGupta, and B.~ Dutta-Roy, Am. J. Phys. \textbf{66}, 1101 (1998).

\bibitem{Nyeo} S.~Nyeo, Am.\ J.~Phys.~\textbf{68},  571 (2000).

\bibitem{Camblong} H. E. Camblong and  C. R. Ord\'{o}n\~{e}z, Phys. Rev. A \textbf{65}, 052123 (2002).

\bibitem{teo} F.~Erman and O.~T.~Turgut, J.~Phys.~A {\bf 43}, 335204 (2010).

\bibitem{albaverio} S.~Albeverio, F.~Gesztesy, R.~Hoegh-Krohn, and H.~Holden, Solvable Models in Quantum Mechanics (American Mathematical Society, Providence, RI, 2005).

\bibitem{pra-2016} F.~Loran and A.~Mostafazadeh, Phys.\ Rev.~A \textbf{93}, 042707 (2016).

\bibitem{jpa-2018} F.~Loran and A.~Mostafazadeh, J.~Phys.~A {\bf 51}, 335302 (2018).

\bibitem{sakurai} J.~J.~Sakurai, \emph{Modern Quantum Mechanics} (Addison-Wesley, New York, 1985.)

\bibitem{watson} G.~N.Watson, {\em A Treatise on the Theory of Bessel Functions,} Cambridge University Press, Cambridge, 1944.

\bibitem{GR-table} I.~S.~Gradshteyn and I.~M.~Ryzhik, {\em Table of Integrals, Series, and Products,} Academic Press, Burlington, MA, 2007.


\end{thebibliography}
